\documentclass[a4paper,12pt]{article}
\pdfoutput=1
\usepackage{epsfig,graphicx,xcolor,amsbsy,amssymb,latexsym,amsfonts,amsmath,,tcolorbox,setspace}
\usepackage{pstricks}
\usepackage{color}
\usepackage{soul}
\usepackage[numbers,square,comma, compress]{natbib}
\usepackage{placeins}
\usepackage{wrapfig,framed,caption}
\usepackage[makeroom]{cancel}

\usepackage{multirow}

\definecolor{shadecolor}{gray}{0.925}

\usepackage{eurosym}
\usepackage[a4paper]{geometry}
\usepackage{hyperref}
\usepackage{graphicx}
\usepackage{tikz}
\usetikzlibrary{decorations.pathreplacing,decorations.markings,snakes}
\usetikzlibrary{decorations.pathmorphing}
\usetikzlibrary{patterns}
\usetikzlibrary{calc}
\usepackage{xcolor}
\usepackage{amsmath,amssymb,amsfonts,pstricks,setspace}
\usepackage[enableskew]{youngtab}

\numberwithin{equation}{section}

\newcommand{\bea}{\begin{eqnarray}\displaystyle}
\newcommand{\eea}{\end{eqnarray}}

\newcommand{\figref}[1]{Fig.~\protect\ref{#1}}

\newcommand{\buildH}[2]{H^{{(#1),\{0\}}}_{(#2)}}
\newcommand{\buildW}[2]{W^{{(#1)}}_{(#2)}}
\newcommand{\coup}[2]{\mathcal{O}^{(#1),#2}}

\title{
\begin{flushright}{\vspace{-2.5cm}\small LYCEN 2020-08\\}\end{flushright}
\vspace{2.3cm}
{\bf Diagrammatic Expansion of Non-Perturbative Little String Free Energies}\\[40pt]}

\author{\large \textsc{Stefan~Hohenegger\footnote{\tt s.hohenegger@ipnl.in2p3.fr}}}
\date{}

\begin{document}

\maketitle
\thispagestyle{empty}
\begin{center}
\renewcommand{\thefootnote}{\fnsymbol{footnote}}\vspace{-0.5cm}
${}^{\footnotemark[1]}$ Univ Lyon, Univ Claude Bernard Lyon 1, CNRS/IN2P3, IP2I Lyon,\\ UMR 5822, F-69622, Villeurbanne, France\\[2.5cm]
\end{center}

\begin{abstract}
In \href{https://arxiv.org/abs/1911.08172}{arXiv:1911.08172} we have studied the single-particle free energy of a class of Little String Theories of A-type, which are engineered by $N$ parallel M5-branes on a circle. To leading instanton order (from the perspective of the low energy $U(N)$ gauge theory) and partially also to higher order, a decomposition was observed, which resembles a Feynman diagrammatic expansion: external states are given by expansion coefficients of the $N=1$ BPS free energy and a quasi-Jacobi form that governs the BPS-counting of an M5-brane coupling to two M2-branes. The effective coupling functions were written as infinite series and similarities to modular graph functions were remarked. In the current work we continue and extend this study: Working with the full non-perturbative BPS free energy, we analyse in detail the cases $N=2,3$ and $4$. We argue that in these cases to leading instanton order all coupling functions can be written as a simple combination of two-point functions of a single free scalar field on the torus. We provide closed form expressions, which we conjecture to hold for generic $N$. To higher instanton order, we observe that a decomposition of the free energy in terms of higher point functions with the same external states is still possible but a priori not unique. We nevertheless provide evidence that tentative coupling functions are still combinations of scalar Greens functions, which are decorated with derivatives or multiplied with holomorphic Eisenstein series. We interpret these decorations as corrections of the leading order effective couplings and in particular link the latter to dihedral graph functions with bivalent vertices, which suggests an interpretation in terms of disconnected graphs. 

\end{abstract}

\newpage

\tableofcontents

\onehalfspacing

\vskip1cm

\section{Introduction}
Perturbative methods are one of the cornerstones in modern high energy physics. Besides their phenomenological applications, the study of scattering amplitudes and correlation functions in perturbation theory has revealed many interesting structures and symmetries. In particular in the context of supersymmetric field theories and string theory, very efficient computational tools have been devised, which have led to very interesting new insights into these theories. Many of these tools have been inspired by a better understanding of underlying mathematical structures, notably number theoretical concepts. Non-perturbative aspects of field theories are more difficult to tackle directly. However, in the case of supersymmetric field theories in various dimensions, their intimate connection to string theory, along with dualities of the latter, have opened up other approaches. This has equally led to very interesting results in recent years, which have revealed further interesting structures and dualities of the underlying theories. In this paper we shall discuss an instance in which  perturbative and non-perturbative aspects can be combined in a specific class of supersymmetric field theories in a somewhat unexpected fashion. This leads to a situation in which many of the above mentioned computational tools can be used at the same time to analyse these theories. To make the discussion concrete, we shall focus on a class of supersymmetric gauge theories that arise in the low energy regime of Little String Theories of A-type. Indeed, we shall elaborate on and largely extend a recent observation in \cite{Hohenegger:2019tii} on the structure of their non-perturbative BPS partition function.

 Little String Theories (LSTs) \cite{Witten:1995zh,Aspinwall:1996vc,Aspinwall:1997ye,Seiberg:1997zk,Intriligator:1997dh,Hanany:1997gh,Brunner:1997gf} in general are quantum theories in six dimensions, whose spectrum includes extended (string-like) degrees of freedom in the UV. Such theories can be constructed from (type II) string theory via particular limits that decouple the gravitational sector while keeping the string length finite. On the one hand, this connection to full-fledged (type II) string theory (and in particular its various dual descriptions) opens up a very powerful means to explicitly study these theories. On the other hand, since LSTs can be viewed as a 'simplified version' of string theory, in which symmetries and structures may be easier to access, this may also teach us new lessons about string theory (or its dual descriptions) in return. Accordingly, various different constructions of LSTs have been explored, whose low energy field theory descriptions exhibit different gauge- and matter contents: LSTs follow an ADE classification and recent work \cite{Bhardwaj:2015oru,Bhardwaj:2019hhd} has focused on exploring the landscape of these theories in more depth, using similar methods from classifying superconformal field theories in 6 or less dimensions \cite{Heckman:2013pva,Heckman:2015bfa,Xie:2015rpa,Jefferson:2017ahm,Jefferson:2018irk,Caorsi:2018zsq,Bhardwaj:2018vuu,Apruzzi:2019opn,Bhardwaj:2019jtr,Martone:2020nsy,Argyres:2020wmq}.

A class of LSTs of A-type can be constructed in M-theory via so-called BPS M-brane webs: these consist of $N$ parallel M5-branes arranged on a circle\footnote{Here $\rho$ and $\tau$ denote the circumferences of the two circles (which subsequently are complexified). Both are measured in units of the radius of the compact time direction, which throughout this paper is implicitly set equal to $1$. We refer to \cite{Hohenegger:2015cba,Hohenegger:2015btj,Hohenegger:2016eqy} for more details about the precise brane setup.}  $S_\rho^1$ and compactified on $S^1_\tau$ while probing a flat transverse space, with M2-branes stretched between them. These theories allow various different low energy limits \cite{Bastian:2017ary,Bastian:2018dfu,Bastian:2018fba} that give rise to supersymmetric gauge theories with different gauge structure. In the current work, we shall exclusively be concerned with the region in the parameter space, that describes a $U(N)$ gauge theory with matter in the adjoint representation. The BPS states of the M-brane system can be counted from the perspective of the one-dimensional intersections of the M2- and M5-branes (called M-strings) \cite{Haghighat:2013gba,Haghighat:2013tka,Hohenegger:2013ala}. Indeed, the BPS partition function $\mathcal{Z}_{N,1}$ can be calculated as the equivariant elliptic genus \cite{GritsenkoEllGen} of an $\mathcal{N}=(0,2)$ supersymmetric sigma model. Furthermore, there exist various dual descriptions \cite{Hohenegger:2013ala} of these M5-brane configurations which allow for very efficient ways of calculating the BPS counting function $\mathcal{Z}_{N,1}$. A very useful description in this regards is F-theory compactified on a class of toric Calabi-Yau threefolds $X_{N,1}$ \cite{Kanazawa:2016tnt}. The partition function $\mathcal{Z}_{N,1}$ is captured by the topological string partition function on $X_{N,1}$ \cite{Haghighat:2013gba,Haghighat:2013tka,Hohenegger:2013ala,Hohenegger:2015cba,Hohenegger:2015btj,Hohenegger:2016eqy}. With the help of the toric diagram of $X_{N,1}$ (which can directly be inferred from the M-brane web), the latter can be computed in an algorithmic fashion using the (refined) topological vertex~\cite{Aganagic:2003db,Iqbal:2007ii}. 

This intrinsically geometric description, together with the very explicit form in which $\mathcal{Z}_{N,1}$ (or the related free energy $\mathcal{F}_{N,1}$) can be presented, have proven very fruitful and have aided in unravelling numerous interesting and surprising structures and symmetries \cite{Hohenegger:2016eqy,Bastian:2017jje,Bastian:2017ary,Bastian:2018dfu,Hohenegger:2016yuv,Ahmed:2017hfr,Bastian:2017ing,Bastian:2018jlf,Bastian:2019hpx,Bastian:2019wpx,IqbalHohenegger}. Here we shall not present all the observations made in recent studies (see \cite{IqbalHohenegger} for a more complete review), but only recount those that are relevant for the current work: the Calabi-Yau manifolds $X_{N,1}$ are only a subset of a two-parameter class of manifolds, labelled $X_{N,M}$, which can be used to describe orbifolds of the M5-brane configurations mentioned above \cite{Hohenegger:2016eqy}. In \cite{Hohenegger:2016yuv} it was argued that $X_{N,M}$ is dual to $X_{N',M'}$ if $NM=N'M'$ and $\text{gcd}(N,M)=\text{gcd}(N',M')$, in the sense that the K\"ahler cones of both these manifolds are part of a larger common extended moduli space. Within this space, $X_{N,M}$ and $X_{N',M'}$ are related through a combination of flop- and other symmetry transformations. The duality map induced by these symmetry transformations was conjectured\footnote{This conjecture was subsequently proven for $\text{gcd}(N,M)=1$ in \cite{Bastian:2017ing} and for generic $(N,M)$ (but in a particular limit of the regularisation parameters that are needed to render $\mathcal{Z}_{N,1}$ well defined) in \cite{Haghighat:2018gqf}.} in \cite{Hohenegger:2016yuv} to leave the corresponding partition functions invariant, \emph{i.e.} $\mathcal{Z}_{N,M}(\omega)=\mathcal{Z}_{N',M'}(\omega')$, where $\omega$ and $\omega'$ denote the dependence on the K\"ahler parameters. Together with the triality \cite{Bastian:2017ary} of gauge theories engineered from a single $X_{N,M}$ (\emph{i.e.} the fact that generically the K\"ahler moduli space of a given $X_{N,M}$ allows for 3 regions that engineer low energy gauge theories), this leads to a large web of dual supersymmetric gauge theories: these theories are expected to all share the same partition function (as has been checked explicitly in numerous examples \cite{Bastian:2018dfu}) and their instanton series correspond to different (but equivalent) series expansions of $\mathcal{Z}_{N,M}$. It was furthermore argued in \cite{Bastian:2018jlf} that this web of dual gauge theories in fact also implies highly non-trivial (and intrinsically non-perturbative) symmetries for individual theories: focusing on the case of the $U(N)$ gauge theories engineered by $X_{N,1}$, invariance of the free energy $\mathcal{F}_{N,1}$ was shown under a particular dihedral symmetry group.\footnote{We refer the reader to \cite{Bastian:2018jlf} for the details.} The implications of this symmetry were further explored in \cite{Bastian:2019hpx,Bastian:2019wpx}: by studying series expansions of the single-particle\footnote{Starting from the partition function $\mathcal{Z}_{N,1}$, the free energy is defined as $\mathcal{F}_{N,1}=\ln \,\mathcal{Z}_{N,1}$. In contrast to that, $\mathcal{F}_{N,1}^{\text{plet}}=\text{Plog}\,\mathcal{Z}_{N,1}$ is defined with the help of the plethystic logarithm $\text{Plog}(f(x))=\sum_{n=1}^\infty \frac{\mu(n)}{n}\,\ln f(nx)$, where $\mu$ is the M\"obius function. Physically, $\mathcal{F}_{N,1}^{\text{plet}}$ receives contributions only from single-particle states.} free energy $\mathcal{F}^{\text{plet}}_{N,1}$ for $N=2,3$ and partially $4$, characteristic patterns were observed, which together with the above mentioned symmetries allowed to conjecture a resummation of the former in an intriguing fashion: although being based on a limited (and a priori formal) series expansion, this allowed to write different contributions of $\mathcal{F}_{N,1}^{\text{plet}}$ in terms of generating functions of multiple divisor sums introduced in \cite{Bachmann:2013wba} (see appendix~\ref{App:DivisorSums} for their definitions). In the case of $N=2$ this was further shown to be equivalent to using generalised Eisenstein series as described in \cite{Weil}. Despite being, as mentioned, based on a limited, a priori formal series expansion, this form of writing $\mathcal{F}^{\text{plet}}_{N,1}$ is fully compatible with all the expected symmetries, notably modular transformations (with modular parameter $\rho$) as well as the non-perturbative symmetries discovered in \cite{Bastian:2018jlf}. Moreover, as discussed in \cite{IqbalHohenegger} this result also exhibits the correct pole-structure as a function of the gauge parameters of the $U(N)$ low energy gauge theory that can abstractly be inferred from the partition function $\mathcal{Z}_{N,1}$.

Analysing the form of $\mathcal{F}_{N,1}^{\text{plet}}$ proposed in \cite{Bastian:2019hpx,Bastian:2019wpx} in the so-called unrefined limit (\emph{i.e.} the limit in which $\mathcal{Z}_{N,1}$ captures the unrefined topological string partition function of $X_{N,1}$) it was further observed in \cite{Hohenegger:2019tii} that it can be re-written in a fashion that strongly resembles a Feynman diagrammatic decomposition. Focusing again on the examples $N=2,3$ and partially 4, it was argued that to leading instanton order (from the perspective of the low energy $U(N)$ gauge theory), $\mathcal{F}_{N,1}^{\text{plet}}$ can be written in a way resembling $N$-point functions: the external states were given either by (coefficients of) the BPS counting function of the LST with $N=1$ or a (quasi) Jacobi form that governs the BPS-counting of a single M5-brane coupling to two M2-branes. Furthermore, it was remarked that the effective couplings appearing in this decomposition were akin to modular graph functions (or more generally modular graph forms -- see appendix~\ref{App:GraphFunctions} for a very brief review) \cite{Broedel:2015hia,DHoker:2015wxz,DHoker:2016mwo,DHoker:2017pvk,Zerbini:2018sox,Zerbini:2018hgs,Gerken:2018jrq,Mafra:2019ddf,Mafra:2019xms,Gerken:2019cxz,Gerken:2020yii}. In particular, it was observed that the first non-trivial such coupling in the case of $N=2$ is related to the (second derivative of the) Greens function of a free scalar field on a torus. Higher orders in the instanton expansion of $\mathcal{F}_{N,1}^{\text{plet}}$ exhibit a similar pattern, however, new elements appear and no concrete pattern was put forward in \cite{Hohenegger:2019tii}.

The observation of \cite{Hohenegger:2019tii} has linked an intrinsically non-perturbative quantity (the instanton one-particle free energy $\mathcal{F}_{N,1}^{\text{plet}}$ of a supersymmetric gauge theory with $U(N)$ gauge theory) to a very simple scalar two-point function that is a fundamental building block in perturbative (in this case one-loop) scattering amplitudes of string theory. The purpose of this paper is to elaborate on this connection and in particular to give evidence for the complete structure at leading instanton order. Instead\footnote{This difference only plays a role to higher instanton orders, since to leading order there is in fact no difference between $\mathcal{F}_{N,1}^{\text{plet}}$ and $\mathcal{F}_{N,1}$.} of $\mathcal{F}_{N,1}^{\text{plet}}$ we shall work with the (unrefined limit of the) full free energy $\mathcal{F}_{N,1}$. By analysing in more detail the effective couplings (which have for the most part been written as infinite series in \cite{Hohenegger:2019tii}) for $N=2,3$ and computing them for $N=4$, we find that they can be written as combinations of (derivatives of) $N$ scalar two-point functions. Based on these results, we propose a simple closed  form for generic $N$, that fits all results that are available in the literature. From a physics perspective, we therefore provide strong evidence that the leading instanton contribution to the free energy of a class of $U(N)$ gauge theories with adjoint matter is entirely determined by two-point correlators of free scalar fields on a torus, as well as the leading instanton contribution of the free energy for $N=1$, for which in fact a closed form expansion can be presented. Higher orders in the instanton expansion of $\mathcal{F}_{N,1}$ can be presented in a fashion that resembles higher (\emph{i.e.} $>N$) point functions, whose external states are still the same building blocks as to leading order. Furthermore, for many examples we present evidence that the tentative coupling functions are still composed of scalar Greens functions: however, they now appear 'decorated', either through derivative operators or multiplied by holomorphic modular forms. For all the examples we have studied we show that the latter can in fact be re-written as a class of modular graph forms \cite{DHoker:2015wxz,DHoker:2016mwo} with bivalent vertices. While the structures we observe point towards a unified picture, the fact that the fundamental building blocks appearing in the external states form an overcomplete basis (and thus introduce an intrinsic ambiguity in the decomposition of the 'Feynman graphs'), prevents us from conjecturing a closed form for higher instanton orders at this point.  

The rest of the paper is organised as follows: in Section~\ref{Sect:Review} we review the free energy of LSTs of A-type as well as the results of \cite{Hohenegger:2019tii}. Due to the technical nature of the explicit computations, we present a detailed summary of our results in Section~\ref{Sect:SummaryResults}. Sections~\ref{Sect:ExN2}, \ref{Sect:ExN3} and \ref{Sect:ExN4} present a detailed study of the examples $N=2,3$ and 4 respectively. Finally, Section~\ref{Sect:Conclusions} contains our conclusions and an interpretation of our results. Details and definitions of modular objects, the precise definitions of the fundamental building blocks (along with their explicit expressions to low orders) that appear as external states, as well as technical details of some computations that have been deemed too long for the main body of the paper have been relegated to three~appendices.

\section{Review: LST Free Energies and Graphs}\label{Sect:Review}
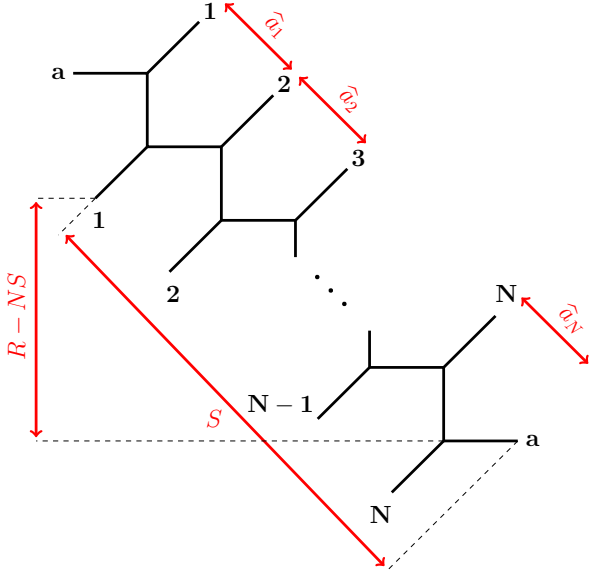
\begin{wrapfigure}{l}{0.48\textwidth}
\begin{center}
\vspace{-0.5cm}
\scalebox{0.65}{\parbox{12.1cm}{\begin{tikzpicture}[scale = 1.50]
\draw[ultra thick] (-1,0) -- (0,0) -- (0,-1) -- (1,-1) -- (1,-2) -- (2,-2) -- (2,-2.5);
\node[rotate=315] at (2.5,-3) {\Huge $\cdots$};
\draw[ultra thick] (3,-3.5) -- (3,-4) -- (4,-4) -- (4,-5) -- (5,-5);
\draw[ultra thick] (0,0) -- (0.7,0.7);
\draw[ultra thick] (1,-1) -- (1.7,-0.3);
\draw[ultra thick] (2,-2) -- (2.7,-1.3);
\draw[ultra thick] (4,-4) -- (4.7,-3.3);
\draw[ultra thick] (0,-1) -- (-0.7,-1.7);
\draw[ultra thick] (1,-2) -- (0.3,-2.7);
\draw[ultra thick] (3,-4) -- (2.3,-4.7);
\draw[ultra thick] (4,-5) -- (3.3,-5.7);
\node at (-1.2,0) {\large {\bf $\mathbf a$}};
\node at (5.2,-5) {\large {\bf $\mathbf a$}};
\node at (0.85,0.85) {\large {$\mathbf 1$}};
\node at (1.85,-0.15) {\large {$\mathbf 2$}};
\node at (2.85,-1.15) {\large {$\mathbf 3$}};
\node at (4.85,-3) {\large {$\mathbf{N}$}};
\node at (-0.65,-2) {\large {$\mathbf{1}$}};
\node at (0.35,-3) {\large {$\mathbf{2}$}};
\node at (1.8,-4.5) {\large {$\mathbf{N-1}$}};
\node at (3.15,-6) {\large {$\mathbf{N}$}};
\draw[ultra thick,red,<->] (1.05,0.95) -- (1.95,0.05);
\node[red,rotate=315] at (1.75,0.65) {{\large {\bf {$\widehat{a}_1$}}}};
\draw[ultra thick,red,<->] (2.05,-0.05) -- (2.95,-0.95);
\node[red,rotate=315] at (2.75,-0.35) {{\large {\bf {$\widehat{a}_2$}}}};
\draw[ultra thick,red,<->] (5.05,-3.05) -- (5.95,-3.95);
\node[red,rotate=315] at (5.75,-3.35) {{\large {\bf {$\widehat{a}_N$}}}};
\draw[dashed] (-0.7,-1.7) -- (-1.2,-2.2);
\draw[dashed] (5,-5) -- (3.25,-6.75);
\draw[ultra thick,red,<->] (3.2,-6.7) -- (-1.1,-2.2);
\node[red] at (0.9,-4.7) {{\large {\bf {$S$}}}};
\draw[dashed] (4,-5) -- (-1.5,-5);
\draw[dashed] (-0.7,-1.7) -- (-1.5,-1.7);
\draw[ultra thick,red,<->] (-1.5,-4.95) -- (-1.5,-1.75);
\node[red,rotate=90] at (-1.75,-3.3) {{\large {\bf{$R-NS$}}}};
\end{tikzpicture}}}
\caption{\sl Web diagram of $X_{N,1}$ with the parameters $(\widehat{a}_{1},\ldots,\widehat{a}_{N}, S, R)$.}
\label{Fig:GenNweb}
\end{center}
${}$\\[-1.7cm]
\end{wrapfigure}

In this paper we mostly follow the notation of \cite{IqbalHohenegger} (see also \cite{Hohenegger:2019tii,Bastian:2019wpx}). Little String~Theories (LSTs) of type $A_N$  can be described via F-theory compactified on a class of toric Calabi-Yau threefolds $X_{N,1}$, whose web diagram is schematically shown in \figref{Fig:GenNweb}. This diagram is doubly periodic, \emph{i.e.} the horizontal lines labelled $\mathbf{a}$ as well as the diagonal lines labelled $(\mathbf{1},\ldots,\mathbf{N})$ are pairwise identified. The manifold $X_{N,1}$ is parametrised by a total of $N+2$ K\"ahler parameters. While various different bases for the latter can be chosen (see \cite{Hohenegger:2015btj,Hohenegger:2016eqy,Hohenegger:2016yuv}), in this paper we shall follow \cite{Bastian:2017ing,Bastian:2017ary,Bastian:2018dfu} and use the parameters $(\widehat{a}_1,\ldots,\widehat{a}_N,S,R)$, which are schematically shown in \figref{Fig:GenNweb} as the (sum of) areas of certain curves of $X_{N,1}$ represented by lines in the web diagram.

Starting from the web diagram in~\figref{Fig:GenNweb}, the non-perturbative BPS partition function $\mathcal{Z}_{N,1}(\widehat{a}_{1,\ldots,N},S,R;\epsilon_{1,2})$ of the LST is captured by the partition function of the topological string on $X_{N,1}$. The latter in turn can be computed in an algorithmic fashion using the (refined) topological vertex formalism. From this point of view, the parameters $\epsilon_{1,2}$ appearing in $\mathcal{Z}_{N,1}$~are

\noindent
related to the coupling constant and the refinement of the topological string.\footnote{One way to define the refined A-model of a Calabi-Yau manifold $X$ is via M-theory compactified on $X \times  S^1 \times\text{TN}$, where the Taub-NUT space TN is twisted along $S^1$ thus giving rise to the two parameters $\epsilon_{1,2}$. The topological partition function can then be directly computed by counting BPS-states in M-theory \cite{GV1,GV2,Hollowood:2003cv}. See also \cite{Antoniadis:2010iq,Antoniadis:2013bja,Antoniadis:2013mna,Antoniadis:2015spa} for a world-sheet definition of the refined topological string and an interpretation of $\epsilon_{1,2}$.} From the perspective of the LST, they can be thought of as regularisation parameters that are necessary to render the non-perturbative partition function well defined. Indeed, in the context of the low-energy gauge theory description, they are identified with the deformation parameters of Nekrasov's $\Omega$-background \cite{Moore:1997dj,Lossev:1997bz,Nekrasov:2002qd}.

Explicit expressions for $\mathcal{Z}_{N,1}$ for general $N$ in terms of Jacobi theta functions have been given in \cite{Hohenegger:2015cba,Hohenegger:2015btj,Hohenegger:2016eqy,Bastian:2017ing}. In this paper, we will mostly be concerned with the free energy, which is directly related to $\mathcal{Z}_{N,1}$ in the following way
\begin{align}
&\mathcal{F}_{N,1}(\widehat{a}_1,\ldots,\widehat{a}_N,S,R;\epsilon_{1,2})
=\ln \mathcal{Z}_{N,1}(\widehat{a}_1,\ldots,\widehat{a}_N,S,R;\epsilon_{1,2})\,.\label{FreeEnergy}
\end{align}
Notice, following \cite{Ahmed:2017hfr,IqbalHohenegger} (but unlike \cite{Hohenegger:2019tii}) we consider the full free energy which is defined as the logarithm of $\mathcal{Z}_{N,1}$. Our strategy in analysing the LST in this paper is to consider the series expansions\footnote{Although a prior only a formal expansion, it was pointed out in \cite{IqbalHohenegger} that each term in this series is made up from (well-defined) quotients of (derivatives of) Jacobi theta functions.} of $\mathcal{F}_{N,1}$ in the unrefined limit $\epsilon_1=-\epsilon_2=\epsilon$. To this end, we define
\begin{align}
&\mathcal{F}_{N,1}(\widehat{a}_1,\ldots,\widehat{a}_N,S,R;\epsilon,-\epsilon)=\sum_r Q_R^r\,P_N^{(r)}(\widehat{a}_1,\ldots,\widehat{a}_N,S,\epsilon)\,,&&\text{with} &&Q_R=e^{2\pi iR}\,.
\end{align}
From the perspective of the low-energy $U(N)$ gauge theory, where $R$ is related to the coupling constant, this corresponds to an instanton expansion and $P_N^{(r)}$ can be interpreted as the $r$-instanton free energy. For further convenience, we also define the $\epsilon$-expansion\footnote{To make contact with the coefficients $P^{(r)}_{N,(2s_1,2s_2)}$ defined in \cite{Hohenegger:2019tii}, we remark that 
\begin{align}
P^{(r)}_{N,(s)}(\widehat{a}_1,\ldots,\widehat{a}_N,S)=\sum_{s_1+s_2=s}(-1)^{s_2}\,P^{(r)}_{N,(2s_1,2s_2)}(\widehat{a}_1,\ldots,\widehat{a}_N,S)\,.
\end{align}
} 
\begin{align}
P_N^{(r)}(\widehat{a}_1,\ldots,\widehat{a}_N,S,\epsilon)=\sum_{s=0}^\infty\,\epsilon^{2s-2}\,P^{(r)}_{N,(s)}(\widehat{a}_1,\ldots,\widehat{a}_N,S)\,.
\end{align}
Furthermore, we can also define a Fourier series with respect to the parameters $\widehat{a}_{1,\ldots,N}$
\begin{align}
P^{(r)}_{N,(s)}(\widehat{a}_{1,\ldots,N},S)=\sum_{n_1,\ldots,n_N}Q_{\widehat{a}_1}^{n_1}\ldots Q_{\widehat{a}_N}^{n_N}\,P^{(r),\{n_1,\ldots,n_N\}}_{N,(s)}(S)\,,&&\text{with} &&Q_{\widehat{a}_i}=e^{2\pi i \widehat{a}_i}\,.
\label{FourierModesFreeEnergy}
\end{align}
For ease of writing we shall also use the shorthand notation $\underline{n}=\{n_1,\ldots,n_N\}$. In \cite{Bastian:2019wpx} (see also \cite{IqbalHohenegger}) the following decomposition has been introduced
\begin{align}
P^{(r)}_{N,(s)}(\widehat{a}_{1,\ldots, N},S)=H^{(r),\{0,\ldots,0\}}_{(s)}(\rho,S)+\sum_{\underline{n}}^\prime H^{(r),\underline{n}}_{(s)}(\rho,S)\,Q_{\widehat{a}_1}^{n_1}\ldots Q_{\widehat{a}_N}^{n_N}\,,&&\text{with} &&\rho=\sum_{i=1}^n \widehat{a}_i\,,\label{DecompositionFourierModes}
\end{align}
where the prime indicates that the summation is understood over all $\underline{n}=\{n_1,\ldots,n_N\}\in (\mathbb{N}\cup \{0\})^N$ with at least one of the $n_i=0$. The $H^{(r),\underline{n}}_{(s)}(\rho,S)$ are formally defined as \cite{Bastian:2019wpx}
\begin{align}
&H^{(r),\{n_1,\ldots,n_N\}}_{(s)}(\rho,S)=\sum_{\ell=0}^\infty Q_\rho^\ell\,P^{(r),\{n_1+\ell,n_2+\ell,\ldots,n_N+\ell\}}_{N,(s)}(S)\,,&&\text{with} &&Q_\rho=e^{2\pi i \rho}=\prod_{i=1}^N Q_{\widehat{a}_i}\,.
\end{align}
Since they will be important building blocks in many computations in this paper, we discuss the functions $\buildH{r}{s}(\rho,S)$ in detail in appendix~\ref{App:BuildingN1}.

In \cite{Hohenegger:2019tii} non-trivial evidence has been provided that the $P^{(r=1)}_{N,(s)}(\widehat{a}_{1,\ldots, N},S)$ introduced in (\ref{DecompositionFourierModes}) afford a decomposition in terms of rather simple building blocks with a very suggestive graphical presentation that resembles in some way an amplitude expansion. Indeed, in~\cite{Hohenegger:2019tii} the following decomposition  was proposed\footnote{As remarked before, the proposition in \cite{Hohenegger:2019tii} was strictly speaking for a reduced free energy, in which the logarithm in (\ref{FreeEnergy}) is replaced by the plethystic logarithm
\begin{align}
\mathcal{F}_{N,1}^{\text{plet}}(\widehat{a}_{1,\ldots,N},S,R;\epsilon_{1,2})=\text{Plog}\mathcal{Z}_{N,1}(\widehat{a}_{1,\ldots,N},S,R;\epsilon_{1,2})=\sum_{n=1}^\infty \frac{\mu(n)}{n}\,\ln\mathcal{Z}_{N,1}(n \widehat{a}_{1,\ldots,N},n S, nR;n\epsilon_{1,2})\,,\label{DefPletLog}
\end{align}
where $\mu(n)$ is the M\"obius function. To order $\mathcal{O}(Q_R)$, however, only the term $n=1$ can contribute. Since, $\mu(1)=1$, this implies that to this order $\mathcal{F}_{N,1}^{\text{plet}}$ and $\mathcal{F}_{N,1}$ are in fact identical, such that the results of \cite{Hohenegger:2019tii} for $r=1$ directly carry over to our current setup. As we shall see below, this is no longer the case for $r>1$.
} 
\begin{align}
P^{(r=1)}_{N,(s)}(\widehat{a}_{1,\ldots,N},S)=\buildH{1}{s}(\rho,S)\sum_{\alpha=0}^{N-1}\left(\buildW{1}{0}(\rho,S)\right)^{N-1-\alpha}\,\left(\buildH{1}{0}(\rho,S)\right)^{\alpha}\,\coup{N}{\alpha}(\widehat{a}_{1,\ldots,N-1},\rho)\,,\label{GraphDecompositionOrder1}
\end{align}
where we have implicitly used $\rho=\sum_{i=1}^N\widehat{a}_i$. Here $\buildH{1}{s}$ and $\buildW{1}{0}$ are respectively the expansion coefficients of the free energy for $N=1$ and the function $W(\rho,\epsilon)$, which are reviewed in detail in appendix~\ref{App:BuildingBlocks} (where also explicit expressions (for some low values of $s$) are given).

\begin{wrapfigure}{r}{0.52\textwidth}
\begin{center}
\vspace{-0.3cm}
\scalebox{1}{\parbox{8.8cm}{\begin{tikzpicture}[scale = 1.50]
\draw[ultra thick] (-0.95,-1) -- (0,0);
\draw (-0.975,-1.025) circle (0.05);
\draw (-1.18,-0.77) circle (0.05);
\draw (-1.385,-0.255) circle (0.05);
\node[rotate=-65] at (-1.2,-0.5) {$\cdots$};
\draw[ultra thick] (-1.35,-0.25) -- (0,0);
\draw[ultra thick] (-1.15,-0.75) -- (0,0);
\node at (-1.2,-1.4) {$\buildW{1}{0}$};
\node at (-1.55,-0.95) {$\buildW{1}{0}$};
\node at (-1.8,-0.375) {$\buildW{1}{0}$};
\node[rotate=-58] at (-2.4,-1.2) {\parbox{2cm}{\small $(N-1-\alpha)$ \\[-6pt]${}$\hspace{0.5cm}times}};
\node at (1.4,-1.4) {$\buildH{1}{0}$};
\node at (1.8,-0.9) {$\buildH{1}{0}$};
\node at (2,-0.375) {$\buildH{1}{0}$};
\node[rotate=58] at (2.6,-1.2) {\parbox{2cm}{\small $\alpha$ times}};
\draw[ultra thick] (0.95,-1) -- (0,0);
\draw (1.385,-0.255) circle (0.05);
\draw (1.18,-0.77) circle (0.05);
\draw (0.975,-1.025) circle (0.05);
\draw[ultra thick] (1.35,-0.25) -- (0,0);
\draw[ultra thick] (1.15,-0.75) -- (0,0);
\node[rotate=65] at (1.2,-0.5) {$\cdots$};
\draw[ultra thick] (0,1.225) -- (0,0);
\draw (0,1.28) circle (0.05);
\node at (0,1.7) {$\buildH{1}{s}$};
\draw[ultra thick,<->] (1.7,-1.8) to [out=20,in=-90] (2.5,-0.55);
\draw[ultra thick,<->] (-1.5,-1.7) to [out=143,in=280] (-2.3,-0.35);
\draw[ultra thick,fill=black] (0,0) circle (0.05);
\draw[thick,fill=gray!60!white] (0,0) circle (0.44);
\node[rotate=0] at (0,0) {$\coup{N}{\alpha}$};

\end{tikzpicture}}}
\caption{\sl Diagrammatic expansion of $P^{(r=1)}_{N,(s)}$.}
\label{Fig:FeynmanR1}
\end{center}
${}$\\[-1.7cm]
\end{wrapfigure}
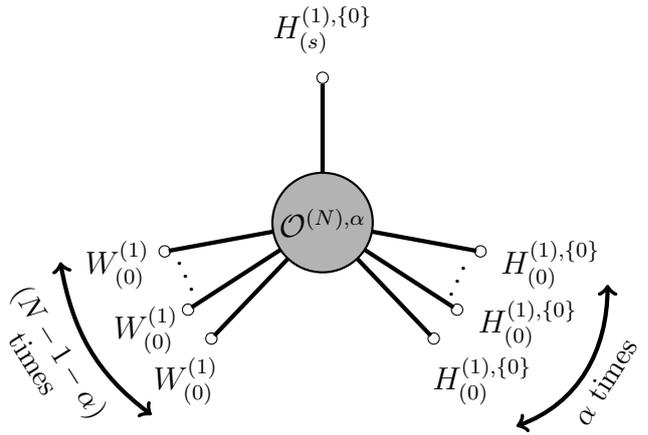 

\noindent
 Physically, the latter has appeared in previous works and was shown to be related to the BPS counting of configurations of M5-branes with single M2-branes ending on them on either side \cite{Hohenegger:2015cba,Hohenegger:2016eqy}. It has also recently appeared in \cite{IqbalHohenegger}, in the context of extracting Hecke structures in the spectrum of the LST free energy in the Nekrasov-Shatashvili limit. Finally, the functions $\coup{N}{\alpha}$ are independent of $S$ (and $s$) and encode the only dependence of $P^{(r=1)}_{N,(s)}$ on $(\widehat{a}_{1},\ldots,\widehat{a}_{N-1})$. They were schematically written in \cite{Hohenegger:2019tii} as
\begin{align}
\coup{N}{\alpha}(\widehat{a}_{1,\ldots,N})=\sum_\ell\sum_{n_1,\ldots,n_\alpha=1}^\infty\frac{p_\ell^\alpha(n_{1,\ldots,n_\alpha})\Lambda_\ell^\alpha(\widehat{a}_{1,\ldots,N},n_{1,\ldots,\alpha})}{\prod_{a=1}^\alpha\left(1-Q_\rho^{t_\ell^\alpha(n_{1,\ldots,\alpha})}\right)}\,.\label{GraphFctSchem}
\end{align}
We refer the reader to sections~\ref{Sect:ExN2}, \ref{Sect:ExN3} and \ref{Sect:ExN4} for explicit examples for $N=2,3,4$ respectively. In (\ref{GraphFctSchem}), the $p_\ell^\alpha$ denote $\ell$ homogeneous polynomials of order $\alpha$ in the summation variables $n_{1,\ldots,\alpha}$, $t_\ell^\alpha$ are linear functions in $n_{1,\ldots,\alpha}$ and the $\Lambda_\ell^\alpha$ are rational functions of the $Q_{\widehat{a}_{1,\ldots,N}}$.

Graphically, eq.~(\ref{GraphDecompositionOrder1}) was presented in \cite{Hohenegger:2019tii} in a way shown in \figref{Fig:FeynmanR1}: the building blocks $\buildH{1}{s}$, $\buildH{1}{0}$ and $\buildW{1}{0}$ are interpreted as external states in a (Feynman)diagram, where the $\coup{N}{\alpha}$ play the role of effective couplings. This interpretation was justified by the observation that the first non-trivial such function (appart from $\coup{N}{0}$, which (based on several examples) were conjectured to be equal to $N$) $\coup{2}{1}$ is related to the Green's function of a free scalar field on the torus and thus indeed represents a 2-point function. 

The free energy $\mathcal{F}_{N,1}^{\text{plet}}$ to order $Q_R^r$ with $r>1$ was also analysed in \cite{Hohenegger:2019tii} for $N=2,3$ and it was concluded that at least partially it also allows for a decomposition in the building blocks $\buildH{r}{s}$ and $\buildW{r}{s}$. However, also new structures emerge, which were difficult to interpret conceptually. In the current work, we repeat the analysis for the full free energy $\mathcal{F}_{N,1}$ to higher orders in $Q_R$.
\section{Summary of Results}\label{Sect:SummaryResults}
The goal of this paper is two-fold: On the one hand side, we analyse in more detail the coupling functions $\coup{N}{\alpha}$ that appear to leading instanton order. On the other hand, we consider the $P^{(r)}_{N,(s)}$ for $r>1$ which stem from the full free energy $\mathcal{F}_{N,1}$ (rather than $\mathcal{F}_{N,1}^{\text{plet}}$) and provide evidence in several examples (\emph{i.e.} $N=2,3$ and $4$) that they similarly exhibit a decomposition that resembles effective couplings. Since our arguments are based on series expansions of simple examples, which are rather technical to some extent, we provide in this section a brief summary of our observations.

To order $\mathcal{O}(Q_R)$ we provide evidence that the observation in \cite{Hohenegger:2019tii} (namely that $\coup{2}{1}$ is given by the derivative of the free scalar two-point function on the torus) can be generalised as follows
\begin{align}
&\coup{N}{\alpha}(\widehat{a}_{1,\ldots,{N-1}},\rho)=\frac{1}{(2\pi)^{2\alpha}}\sum_{\ell=0}^{N-1}\sum_{{\mathcal{S}\subset\{0,\ldots,N-1\}\setminus\{\ell\}}\atop {|\mathcal{S}|=\alpha}}\prod_{j\in \mathcal{S}}\left(\mathbb{G}''(\widehat{b}_\ell-\widehat{b}_j;\rho)+\frac{2\pi i}{\rho-\bar{\rho}}\right)\,,\label{StructureCouplingR1}
\end{align}
for all couplings $\alpha=0,\ldots,N-1$ and where we have introduced the new arguments
\begin{align}
\widehat{b}_0=0&&\text{and} &&\widehat{b}_j=\sum_{n=1}^j \widehat{a}_n\,,\hspace{1.5cm} \forall j=1,\ldots,N-1\,.
\end{align}
Furthermore, $\mathbb{G}(z;\rho)$ is the two-point function of a free boson on the torus and the prime de-

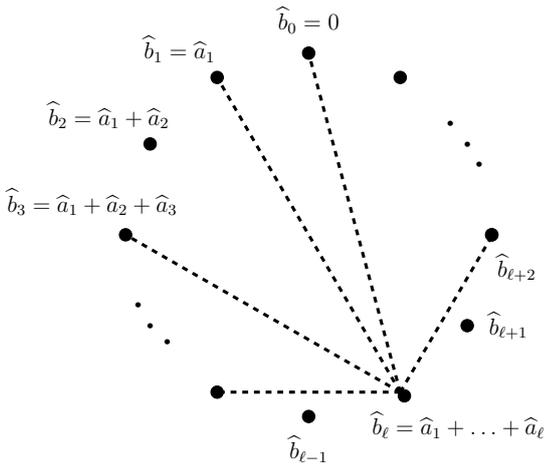
\begin{wrapfigure}{l}{0.44\textwidth}
\begin{center}
\vspace{-0.5cm}
\scalebox{0.73}{\parbox{9.9cm}{\begin{tikzpicture}[scale = 1.50]
\draw[fill=black] (2.2, 0) circle (0.075);
\draw[fill=black] (1.1, 1.905255888) circle (0.075);
\draw[fill=black] (0, 2.2) circle (0.075);
\draw[fill=black] (-1.1,  1.905255888) circle (0.075);
\draw[fill=black] (-1.905255888, 1.1) circle (0.075);
\draw[fill=black] (-2.2,   0) circle (0.075);
\draw[fill=black] (-1.1, -1.905255888) circle (0.075);
\draw[fill=black] (0., -2.2) circle (0.075);
\draw[fill=black] (1.15, -1.95) circle (0.075);
\draw[fill=black] (1.905255888, -1.1) circle (0.075);
\draw[fill=black] (2.2, 0) circle (0.075);
\node[rotate=-65] at (-1.905255888, -1.1) {\tiny $\bullet$};
\node[rotate=-65] at (-2.05, -0.85) {\tiny $\bullet$};
\node[rotate=-65] at (-1.7, -1.3) {\tiny $\bullet$};
\node[rotate=-65] at (1.905255888, 1.1) {\tiny $\bullet$};
\node[rotate=-65] at (2.05, 0.85) {\tiny $\bullet$};
\node[rotate=-65] at (1.7, 1.35) {\tiny $\bullet$};
\draw[ultra thick,dashed]  (1.1, -1.905255888) -- (0.02,2.125);
\draw[ultra thick,dashed]  (1.1, -1.905255888) -- (2.175, -0.06);
\draw[ultra thick,dashed]  (1.1, -1.905255888) -- (-1.1,  1.905255888);
\draw[ultra thick,dashed]  (1.1, -1.905255888) -- (-2.2,  0);
\draw[ultra thick,dashed]  (1.1, -1.905255888) -- (-1.025, -1.905255888);
\node at (0,2.6) {$\widehat{b}_0=0$};
\node at (-1.55,2.25) {$\widehat{b}_1=\widehat{a}_1$};
\node at (-2.4,1.45) {$\widehat{b}_2=\widehat{a}_1+\widehat{a}_2$};
\node at (-2.6,0.4) {$\widehat{b}_3=\widehat{a}_1+\widehat{a}_2+\widehat{a}_3$};
\node at (0,-2.6) {$\widehat{b}_{\ell-1}$};
\node at (1.8,-2.3) {$\widehat{b}_\ell=\widehat{a}_1+\ldots+\widehat{a}_{\ell}$};
\node at (2.4,-1.1) {$\widehat{b}_{\ell+1}$};
\node at (2.5,-0.4) {$\widehat{b}_{\ell+2}$};
\end{tikzpicture}}}
\caption{\sl Diagram of a single term in the summation over $\mathcal{S}$ (for fixed $\ell$) in eq.~(\ref{StructureCouplingR1}): The dashed lines correspond to scalar two-point functions $\mathbb{G}''(\widehat{b}_\ell-\widehat{b}_j;\rho)+\frac{2\pi i}{\rho-\bar{\rho}}$.}
\label{Fig:ScalarGenCorrelatorR1}
\end{center}
${}$\\[-1.7cm]
\end{wrapfigure}

\noindent
notes a derivative with respect to the $z$-argument (see appendix~\ref{App:GraphFunctions} for more details). Furthermore, the second summation in (\ref{StructureCouplingR1}) is over all subsets $\mathcal{S}$ of $\{1,\ldots,\bcancel{\ell},\ldots,N-1\}$ of cardinality $\alpha$ (\emph{i.e.} which have exactly $\alpha$ elements). Therefore, the product over $j$ contains precisely $\alpha$ scalar Greens functions.  

While the notation in (\ref{StructureCouplingR1}) is somewhat involved, a term for fixed $\mathcal{S}$ can be represented in a very simple graphical form, as shown in \figref{Fig:ScalarGenCorrelatorR1}. Further (more concrete) examples for $N=2,3,4$ can be found in subsequent sections. 

To order $\mathcal{O}(Q_R^r)$ for $r>1$, the structure of the free energy $P^{(r)}_{N,(s)}$ is more involved. However, nonetheless, in the examples we have studied, it still affords a decomposition similar to the $r=1$ counterpart, albeit in a more intricate fashion: analysing the cases $N=2$ and $N=3$ up to orders $\mathcal{O}(Q_R^3)$ and $\mathcal{O}(Q_R^2)$ respectively and up to $s=2$, we find that these examples can entirely be decomposed in terms of $\buildH{r}{s}$ and $\buildW{r}{s}$. However, the fact that $\buildH{r}{s}$ and $\buildW{r}{s}$ are an over-complete basis renders the decomposition ambiguous (in particular for higher values of $s$). Nevertheless we observe that tentative coupling functions appearing in such decompositions lend themselves to an interpretation as corrected couplings $\coup{N}{1}$ with additional internal points. We provide structural evidence for this interpretation by the following observations
\begin{itemize}
\item The coupling functions for $r>1$ are still combinations of (derivatives of) scalar two-point functions multiplied by modular forms. In the cases of low $s$ and $r$, where the intrinsic ambiguity is limited, we observe that the latter can be arranged in such a way as to render the couplings holomorphic
\item The modular forms that appear in this process are in fact (holomorphic) graph forms. In the case of $N=2$ and $(r,s)=(2,0)$, in which case the ambiguity of the decomposition is still under control, we can in fact show that the appearing graph form can be related through Cauchy-Riemann differential operators to a graph function, which in turn can be written as the double integral over two scalar two-point functions. Such terms would indeed be interpreted as disconnected contributions in string one-loop amplitudes. 
\end{itemize}
However, while very intriguing, the intrinsic ambiguity in the decomposition of the free energy, prevents us from making this observation more precise at the current time.

\section{Example $N=2$}\label{Sect:ExN2}
\subsection{Decomposition at Order $\mathcal{O}(Q_R)$ and Scalar Correlators}
\begin{figure}[htbp]
\begin{center}
\scalebox{1}{\parbox{16.4cm}{\begin{tikzpicture}[scale = 1.50]
\draw[ultra thick] (-1.15,0) -- (0,0);
\draw (-1.19,0) circle (0.05);
\node at (-1.6,0) {$\buildW{1}{0}$};
\draw[ultra thick] (1.15,0) -- (0,0);
\draw (1.19,0) circle (0.05);
\node at (1.8,0) {$\buildH{1}{s}$};
\draw[thick,fill=gray!60!white] (0,0) circle (0.41);
\node[rotate=0] at (0,0) {$\coup{2}{0}$};
\node at (0,-1) {\bf (a)};
\begin{scope}[xshift=6.5cm]
\draw[ultra thick] (-1.15,0) -- (0,0);
\draw (-1.19,0) circle (0.05);
\node at (-1.8,0) {$\buildH{1}{0}$};
\draw[ultra thick] (1.15,0) -- (0,0);
\draw (1.19,0) circle (0.05);
\node at (1.8,0) {$\buildH{1}{s}$};
\draw[thick,fill=gray!60!white] (0,0) circle (0.41);
\node[rotate=0] at (0,0) {$\coup{2}{1}$};
\node at (0,-1) {\bf (b)};
\end{scope}
\end{tikzpicture}}}
\caption{\sl Diagrammatic expansion of $P^{(r=1)}_{2,(s)}$ in (\ref{N2BasicGraphFunctions}) in terms of two-point functions: {\bf (a)}~$\buildH{1}{s}$ coupling to $\buildW{1}{0}$ through $\coup{2}{0}$; {\bf (b)}~$\buildH{1}{s}$ coupling to $\buildH{1}{0}$ through $\coup{2}{1}$.}
\label{Fig:FeynmanR1ExN2}
\end{center}
\end{figure}
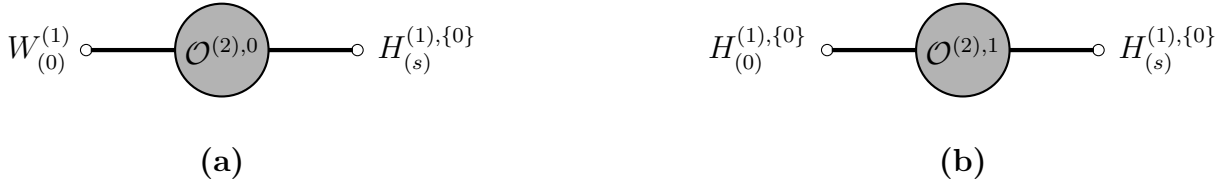

\begin{figure}[htbp]
\begin{center}
\scalebox{1}{\parbox{11.9cm}{\begin{tikzpicture}[scale = 1.50]
\draw[fill=black] (-0.75,0) circle (0.05);
\node at (-0.75,-0.4) {$\widehat{b}_0=0$};
\draw[fill=black] (0.75,0) circle (0.05);
\node at (0.75,0.4) {$\widehat{b}_1=\widehat{a}_1$};
\node at (0,-1) {\bf (a)};
\begin{scope}[xshift=5.5cm]
\draw[ultra thick,dashed] (-0.75,0) -- (0.75,0);
\draw[fill=black] (-0.75,0) circle (0.05);
\node at (-0.75,-0.4) {$\widehat{b}_0=0$};
\draw[fill=black] (0.75,0) circle (0.05);
\node at (0.75,0.4) {$\widehat{b}_1=\widehat{a}_1$};
\node at (0,-1) {\bf (b)};
\end{scope}
\end{tikzpicture}}}
\caption{\sl Diagrammatic presentation of the combinations of scalar Greens functions appearing in the summation over $\mathcal{S}$ (for fixed $\ell$) in (\ref{GraphsN2R1}): {\bf (a)} for $\coup{2}{0}$ and {\bf (b)} for $\coup{2}{1}$. Dashed lines with end-points $(\widehat{b}_i,\widehat{b}_j)$ represent factors of $\mathbb{G}''(\widehat{b}_i-\widehat{b}_j;\rho)+\frac{2\pi i}{\rho-\bar{\rho}}$.}
\label{Fig:FeynmanR1GraphFctN2}
\end{center}
\end{figure}
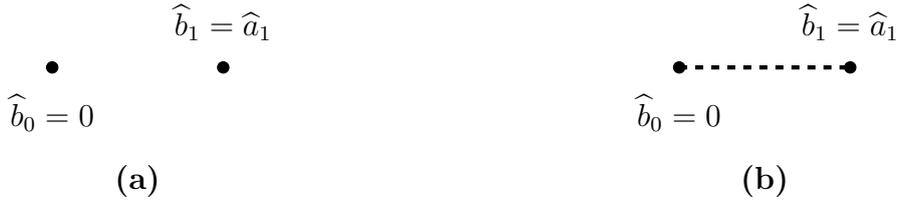

The simplest example is the case $N=2$. As was already argued in \cite{Hohenegger:2019tii} to order $\mathcal{O}(Q_R)$, the free energy\footnote{This is in fact the same term that appears at order $\mathcal{O}(Q_R)$ in $\mathcal{F}_{2,1}^{\text{plet}}$ as defined in~eq.~(\ref{DefPletLog}).} $P_{2,(s)}^{(r=1)}(\widehat{a}_1,S)$ can be written as the sum of two terms\footnote{For ease of writing we shall drop the arguments $(\rho,S)$ appearing in the building blocks $\buildH{r}{s}$ and $\buildW{r}{s}$.}
\begin{align}
P_{2,(s)}^{(r=1)}(\widehat{a}_1,\rho,S)=\buildH{1}{s}\,\buildW{1}{0}\,\coup{2}{0}+\buildH{1}{s}\,\buildH{1}{0}\,\coup{2}{1}(\widehat{a}_1,\rho)\,,\label{N2BasicGraphFunctions}
\end{align}
where the two coupling functions are given explicitly as
\begin{align}
&\coup{2}{0}=2\,,&&\text{and}&&\coup{2}{1}(\widehat{a}_1,\rho)=-\sum_{n=1}^\infty\frac{2n}{1-Q_\rho^n}\left(Q_{\widehat{a}_1}^n+\frac{Q_\rho^n}{Q_{\widehat{a}_1}^n}\right)\,.\label{N2CoupR11}
\end{align}
Graphically, the two terms appearing in (\ref{N2BasicGraphFunctions}) can be represented as shown in \figref{Fig:FeynmanR1ExN2}, in the form of two-point functions. In this presentation, $\buildH{1}{s}$, $\buildH{1}{0}$ and $\buildW{1}{0}$ are attached to the external legs, while $\coup{2}{0}$ and $\coup{2}{1}$ are (effective) couplings. 
In \cite{Hohenegger:2019tii} it was furthermore shown that $\coup{2}{1}$ can be written in the following fashion
\begin{align}
\coup{2}{1}(\widehat{a}_1,\rho)=\frac{2}{(2\pi)^2}\left[\wp(\widehat{a}_1;\rho)+\frac{\pi^2}{3}\,E_2(\rho)\right]=\frac{2}{(2\pi)^2}\left[\mathbb{G}''(\widehat{a}_1;\rho)+\frac{2\pi i}{\rho-\bar{\rho}}\right]\,.\label{N2CoupR12}
\end{align}
Here $\wp(z;\rho)$ is Weierstrass' elliptic function, $E_2$ is the second Eisenstein series (see appendix~\ref{App:JacobiForms} for the definitions) and $\mathbb{G}(z;\rho)$ is the Green's function of a free scalar field on a torus. Another way of arguing for the result (\ref{N2CoupR12}), which shall be relevant for similar computations in subsequent sections, is presented in appendix~\ref{SeriesExpansionSimCoup}. We can combine (\ref{N2CoupR11}) and (\ref{N2CoupR12}) into
{\allowdisplaybreaks
\begin{align}
\coup{2}{\alpha}(\widehat{a}_1,\rho)&=\left\{\begin{array}{lcl} 2 & \text{for} & \alpha=0\,,\\ \frac{1}{(2\pi)^2}\sum_{\ell=0}^1\sum_{j\neq \ell}\left(\mathbb{G}''(\widehat{b}_\ell-\widehat{b}_j;\rho)+\frac{2\pi i}{\rho-\bar{\rho}}\right) & \text{for} & \alpha=1\,,\end{array}\right.\nonumber\\
&=\frac{1}{(2\pi )^{2i}}\sum_{\ell=0}^1\sum_{{\mathcal{S}\subset\{0,1\}\setminus\{\ell\}}\atop{|\mathcal{S}|=i}}\prod_{j\in\mathcal{S}}\left(\mathbb{G}''(\widehat{b}_\ell-\widehat{b}_j;\rho)+\frac{2\pi i}{\rho-\bar{\rho}}\right)\,,\label{GraphsN2R1}
\end{align}}
where $\widehat{b}_0=0$ and $\widehat{b}_1=\widehat{a}_1$, while $|\mathcal{S}|$ in the last relation denotes the cardinality of the set $\mathcal{S}$. Here we have also used that $\mathbb{G}''(z;\rho)=\mathbb{G}''(-z;\rho)$. Graphically, the combinations of scalar Greens functions appearing in (\ref{GraphsN2R1}) can be represented as shown in \figref{Fig:FeynmanR1GraphFctN2}, where dashed lines represent factors of $\mathbb{G}''(\widehat{b}_i-\widehat{b}_j;\rho)+\frac{2\pi i}{\rho-\bar{\rho}}$.
\subsection{Decomposition at Order $\mathcal{O}(Q_R^r)$ for $r>1$}
To order $\mathcal{O}(Q_R^r)$ for $r>1$, the expansions of $\mathcal{F}_{2,1}$ and $\mathcal{F}^{\text{plet}}_{2,1}$ are different, such that the results of \cite{Hohenegger:2019tii} cannot be directly carried over. We therefore consider below in detail the cases $r=2,3$.
\subsubsection{Contributions at Order $\mathcal{O}(Q_R^2)$}
We have analysed $P_{2,(s)}^{(r=2)}$ as a series expansion in $Q_\rho$ to order $10$ up to $s=4$. The results we find are compatible with a decomposition of the form
{\allowdisplaybreaks
\begin{align}
P_{2,(s)}^{(2)}(\widehat{a}_1,\rho,S)=\sum_{j=0}^4&(\phi_{-2,1}(\rho,S))^{j}(\phi_{0,1}(\rho,S))^{4-j}\bigg[f^{j,(r=2)}_{(s)}(\rho,S)\nonumber\\
&+\sum_{n=1}^\infty\left(Q_{\widehat{a}_1}^n+\frac{Q_\rho^n}{Q_{\widehat{a}_1}^n}\right)\sum_{k=0}^{s+2}\frac{n^{2k+1}}{1-Q_\rho^n}\,g^{j,k,(r=2)}_{(s)}(\rho,S)\bigg]\,.\label{BasicDecompositionN2R2}
\end{align}  }
Here $f^{j,(r=2)}_{(s)}$ are quasi modular forms of weight $2(s+j-1)$ and $g^{j,k,(r=2)}_{(s)}$ are modular forms of weight $2(s+j-2-k)$. Explicit expressions for $f^{j,(2)}_{(s)}$ and $g^{j,k,(2)}_{(s)}$ can be given as combinations of Eisenstein series, for example for $s=0$ we find
{\allowdisplaybreaks
\begin{align}
&f_{s=0}^{0,(2)}=0\,,&&f_{s=0}^{1,(2)}=-\frac{1}{4608}\,,&&f_{s=0}^{2,(2)}=-\frac{E_2}{1152}\,,&&f_{s=0}^{3,(2)}=-\frac{E_4}{1152}\,,&&f_{s=0}^{3,(2)}=\frac{E_6-E_4 E_2}{144}\,,\nonumber\\
&g_{s=0}^{0,0,(2)}=0\,,&&g_{s=0}^{1,0,(2)}=0\,,&&g_{s=0}^{2,0,(2)}=-\frac{1}{96}\,,&&g_{s=0}^{3,0,(2)}=0\,,&&g_{s=0}^{3,0,(2)}=-\frac{E_2}{12}\,,\nonumber\\
&g_{s=0}^{0,1,(2)}=0\,,&&g_{s=0}^{1,1,(2)}=0\,,&&g_{s=0}^{2,1,(2)}=0\,,&&g_{s=0}^{3,1,(2)}=-\frac{1}{12}\,,&&g_{s=0}^{3,1,(2)}=0\,,\nonumber\\
&g_{s=0}^{0,2,(2)}=0\,,&&g_{s=0}^{1,2,(2)}=0\,,&&g_{s=0}^{2,2,(2)}=0\,,&&g_{s=0}^{3,2,(2)}=0\,,&&g_{s=0}^{3,2,(2)}=-\frac{1}{24}\,.
\end{align}}
For $s=0$, these explicit expressions afford the following decomposition in terms of the basic building blocks $\buildH{2}{0}$:\footnote{This decomposition is unique if we assume only terms involving $\buildH{2}{0}$, but not $\buildW{2}{0}$.}
\begin{align}
&P^{(2)}_{2,(0)}(\widehat{a}_1,\rho,S)=\frac{2}{3}\,\buildH{2}{0}\,\buildW{2}{0}\,\coup{2}{0}+\frac{4}{3}\,\buildH{2}{0}\,\buildH{2}{0}\,\coup{2}{1}(\widehat{a}_1,\rho)\nonumber\\
&\hspace{0.5cm}-\frac{1}{48}\left(\buildH{1}{0}\right)^4\left[\mathfrak{d}E_4(\rho)+4E_4(\rho)\,\mathcal{I}_0(\rho,\widehat{a}_1)+2\mathcal{I}_2(\rho,\widehat{a}_1)\right]+\frac{4}{3}\left(\buildH{1}{0}\right)^2\,\buildH{2}{0}\,\mathcal{I}_1(\rho,\widehat{a}_1)\,,\label{ExpansionN2R2s0}
\end{align}
 where we defined the differential operator $\mathfrak{d}:=Q_\rho\frac{d}{dQ_\rho}$ and $\mathcal{I}_k$ is defined in (\ref{DefIalpha}). For generic $k\in\mathbb{N}\cup\{0\}$, the latter can be written as derivatives of $\mathcal{I}_0$ (see (\ref{I0Generating})), which in turn is related to the (derivative of the) scalar Greens function $\mathbb{G}(\rho,\widehat{a}_1)$ (\ref{RelI0Greens}). In view of the discussion of the free energy to order $\mathcal{O}(Q_R)$, it is tempting to give the following interpretation of the various terms appearing in (\ref{ExpansionN2R2s0}):
 \begin{enumerate}
 \item Up to different numerical prefactors, the terms in the first line of (\ref{ExpansionN2R2s0}) correspond schematically to the same two-point functions as in \figref{Fig:FeynmanR1ExN2}, except that the external states have been replaced by their counterparts for $r=2$. 
 \item The first term in the second line of (\ref{ExpansionN2R2s0}) has the appearance of a 4-point function (since it contains 4 powers of $\buildH{1}{0}$). However, the tentative coupling-function (\emph{i.e.} without the four 'external states' $\buildH{1}{0}$)
\begin{align}
\mathcal{O}^{(2),\text{4-pt}}_{r=2,s=0}(\widehat{a}_1,\rho)=-\frac{1}{48}\left[\mathfrak{d}E_4(\rho)+4E_4(\rho)\,\mathcal{I}_0(\rho,\widehat{a}_1)+2\mathcal{I}_2(\rho,\widehat{a}_1)\right]\,,\label{N24ptCoupling}
\end{align}
depends (apart from $\rho$) only on a single 'position', namely $\widehat{a}_1$. Following the interpretation of the coupling functions appearing at order $\mathcal{O}(Q_R)$ as schematically shown in \figref{Fig:FeynmanR1GraphFctN2}, this would suggest that of the 4 states, two are external (and are inserted at positions $\widehat{b}_0=0$ and $\widehat{b}_1=\widehat{a}_1$), while the positions of the remaining two states are internal (and could be integrated over). Two graphs of this type, which could give rise to the various different terms appearing in the coupling $\mathcal{O}^{(2),\text{4-pt}}_{r=2,s=0}$, are schematically shown in \figref{Fig:FeynmanR24pt}.
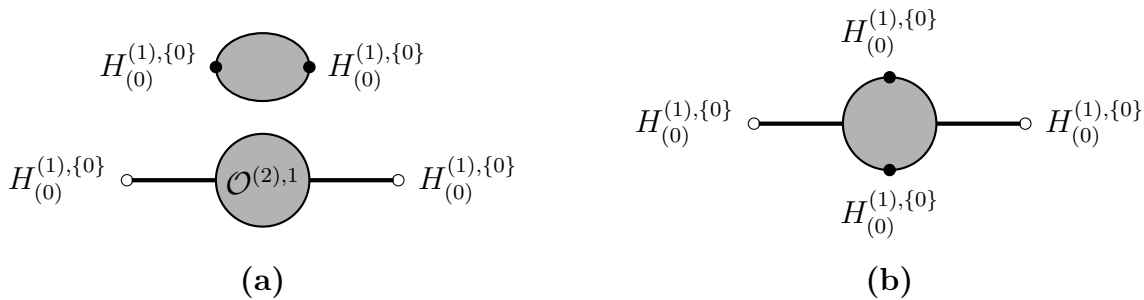
\begin{figure}[htbp]
\begin{center}
\scalebox{1}{\parbox{15.2cm}{\begin{tikzpicture}[scale = 1.50]
\node at (-1,0) {$\buildH{1}{0}$};
\node at (1,0) {$\buildH{1}{0}$};
\draw[thick,fill=gray!60!white] (0,0) ellipse (0.41cm and 0.3cm);
\draw[fill=black] (0.41,0) circle (0.05);
\draw[fill=black] (-0.41,0) circle (0.05);
\draw (-1.19,-1) circle (0.05);
\draw (1.19,-1) circle (0.05);
\draw[ultra thick] (-1.15,-1) -- (1.15,-1);
\node at (-1.8,-1) {$\buildH{1}{0}$};
\node at (1.8,-1) {$\buildH{1}{0}$};
\draw[thick,fill=gray!60!white] (0,-1) circle (0.41);
\node[rotate=0] at (0,-1) {$\coup{2}{1}$};
\node at (0,-1.9) {\bf (a)};
\begin{scope}[xshift=5.5cm,yshift=-0.5cm]
\draw (-1.19,0) circle (0.05);
\draw (1.19,0) circle (0.05);
\draw[ultra thick] (-1.15,0) -- (1.15,0);
\node at (-1.8,0) {$\buildH{1}{0}$};
\node at (1.8,0) {$\buildH{1}{0}$};
\draw[thick,fill=gray!60!white] (0,0) circle (0.41);
\draw[fill=black] (0,0.41) circle (0.05);
\draw[fill=black] (0,-0.41) circle (0.05);
\node at (0,0.8) {$\buildH{1}{0}$};
\node at (0,-0.8) {$\buildH{1}{0}$};
\node at (0,-1.4) {\bf (b)};
\end{scope}

\end{tikzpicture}}}
\caption{\sl Diagrammatic presentation of the 4-point functions appearing in the first term of the second line in (\ref{ExpansionN2R2s0}) and which contribute to the coupling (\ref{N24ptCoupling}). The nodes represented by $\circ$ correspond to external states (with insertion points $\widehat{b}_0=0$ and $\widehat{b}_1$ respectively), while the nodes represented by $\bullet$ correspond to internal states. Following the decomposition (\ref{CorrelatorsR2N2}), diagram {\bf (a)} leads to the coupling $\mathcal{O}^{(2),\text{4-pt}}_{r=2,s=0}(\widehat{a}_1,\rho)$ before the subtraction of the quasi-holomorphic contribution, while diagram {\bf (b)} leads to $\mathcal{O}^{(2),\text{4-pt},1}_{r=2,s=0}(\widehat{a}_1,\rho)$.}
\label{Fig:FeynmanR24pt}
\end{center}
\end{figure}
These are two distinct corrections where the two internal states modify the two-point function, either in the form of a disconnected diagram as in \figref{Fig:FeynmanR24pt}~{\bf (a)} (from which still the $E_2$-dependent contributions need to be subtracted) or through a direct insertion as in \figref{Fig:FeynmanR24pt}~{\bf (b)}.

This tentative interpretation is supported by two further (somewhat circumstantial) pieces of evidence:
\begin{itemize}
\item[\emph{(i)}] The effective coupling in (\ref{N24ptCoupling}) naturally decomposes into the sum of two terms
\begin{align}
\mathcal{O}^{(2),\text{4-pt}}_{r=2,s=0}(\widehat{a}_1,\rho)=\mathcal{O}^{(2),\text{4-pt},1}_{r=2,s=0}+\mathcal{O}^{(2),\text{4-pt},2}_{r=2,s=0}\,,
\end{align}
where each term stems from one of the two diagrams in \figref{Fig:FeynmanR24pt} and where we defined
{\allowdisplaybreaks
\begin{align}
\mathcal{O}^{(2),\text{4-pt},1}_{r=2,s=0}(\widehat{a}_1,\rho)&=-\frac{1}{48}\left[\mathfrak{d}E_4(\rho)+4E_4(\rho)\,\mathcal{I}_0(\rho,\widehat{a}_1)\right]=\frac{E_4}{24}\,\coup{2}{1}(\widehat{a}_1,\rho)-\frac{1}{48}\,\mathfrak{d}E_4\nonumber\\
&=\frac{1}{24}\left[E_4\,\coup{2}{1}(\widehat{a}_1,\rho)-\text{quasi-holomorphic}\right]\,,\nonumber\\
\mathcal{O}^{(2),\text{4-pt},1}_{r=2,s=0}(\widehat{a}_1,\rho)&=-\frac{1}{24}\,\mathcal{I}_2(\rho,\widehat{a}_1)=\frac{1}{48}\,D_{\widehat{a}_1}^4\,\coup{2}{1}(\widehat{a}_1,\rho)\,.\label{CorrelatorsR2N2}
\end{align}}
Here we used the shorthand notation $D_{\widehat{a}_1}=\frac{1}{2\pi i}\,\frac{\partial}{\partial\widehat{a}_1}$. We note in particular, that these couplings are still composed of (combinations of) scalar two-point functions $\mathbb{G}''(\widehat{b}_i-\widehat{b}_j;\rho)+\frac{2\pi i}{\rho-\bar{\rho}}$, which are 'decorated' either through multiplication with the modular form $E_4$ (and a subsequent removal of non-holomorphic contributions) or through the action of the derivative operators $D_{\widehat{a}_1}$. Indeed, following the presentation of \figref{Fig:FeynmanR1GraphFctN2} for $r=1$, the two objects in (\ref{CorrelatorsR2N2}) can graphically be presented as in \figref{Fig:FeynmanR2GraphFctN24pt}:
\begin{figure}[htbp]
\begin{center}
\scalebox{1}{\parbox{13.5cm}{\begin{tikzpicture}[scale = 1.50]
\draw[thick] (0,0.5) ellipse (0.38cm and 0.2cm);
\node at (0,0.5) {\footnotesize $E_4$};
\draw[ultra thick,dashed] (-0.75,0) -- (0.75,0);
\draw[fill=black] (-0.75,0) circle (0.05);
\node at (-1,-0.4) {$\widehat{b}_0=0$};
\draw[fill=black] (0.75,0) circle (0.05);
\node at (1.1,-0.4) {$\widehat{b}_1=\widehat{a}_1$};
\node at (0,-1) {\bf (a)};
\begin{scope}[xshift=5.5cm]
\draw[ultra thick,dashed] (-0.75,0) -- coordinate[pos=0.3] (A2) coordinate[pos=0.69] (B2) (0.75,0);
\draw[fill=black] (-0.75,0) circle (0.05);
\node at (-1.4,0.05) {$\widehat{b}_0=0$};
\draw[fill=black] (0.75,0) circle (0.05);
\node at (1.5,0.05) {$\widehat{b}_1=\widehat{a}_1$};
\fill[black] (A2) circle (1.5pt);
\draw[thick] ($ (A2) + (0.1,0.1) $) -- ($ (A2) + (-0.1,-0.1) $);
\draw[thick] ($ (A2) + (0.1,-0.1) $) -- ($ (A2) + (-0.1,0.1) $);
\fill[black] (B2) circle (1.5pt);
\draw[thick] ($ (B2) + (0.1,0.1) $) -- ($ (B2) + (-0.1,-0.1) $);
\draw[thick] ($ (B2) + (0.1,-0.1) $) -- ($ (B2) + (-0.1,0.1) $);
\node at (0,-1) {\bf (b)};
\end{scope}
\end{tikzpicture}}}
\caption{\sl Diagrammatic presentation of the couplings in (\ref{CorrelatorsR2N2}). Dashed lines with end-points $(\widehat{b}_0,\widehat{b}_1)$ represent factors of $\mathbb{G}''(\widehat{a}_1;\rho)+\frac{2\pi i}{\rho-\bar{\rho}}$, while crosses indicate the action of the derivative operator $D^2_{\widehat{a}_1}$: {\bf (a)} the coupling $\mathcal{O}^{(2),\text{4-pt},1}_{r=2,s=0}(\widehat{a}_1,\rho)$ (where the subtraction of the quasi-holomorphic contributions is understood); {\bf (b)} the coupling $\mathcal{O}^{(2),\text{4-pt},1}_{r=2,s=0}$.}
\label{Fig:FeynmanR2GraphFctN24pt}
\end{center}
\end{figure}
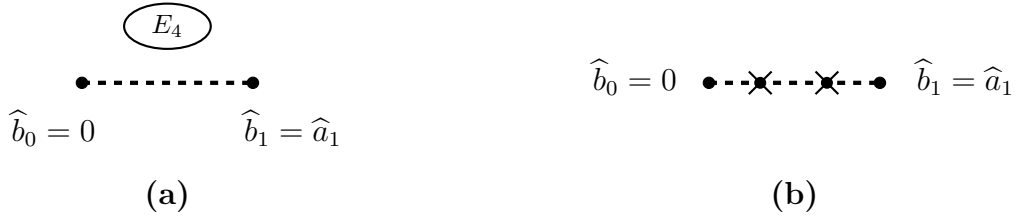
while as before dashed lines represent the scalar Greens function, crosses indicate the action of the operator $D^2_{\widehat{a}_1}$. Furthermore, the term 'quasi-holomorphic' in (\ref{CorrelatorsR2N2}) denotes the subtraction of all terms containing the quasi-holomorphic form $E_2$ appearing in an expansion of $E_4\,\coup{2}{1}$ (which is implicitly assumed in \figref{Fig:FeynmanR2GraphFctN24pt}~{\bf (a)}). Indeed, using (\ref{DerivativeE4}) as well as (\ref{N2CoupR12}) along with the expansion (\ref{WeierstrassDef}) of the Weierstrass function we find
\begin{align}
\mathcal{O}^{(2),\text{4-pt},1}_{r=2,s=0}(\widehat{a}_1,\rho)&=\frac{E_4(\rho)}{48\pi^2\widehat{a}_1^2}+\frac{E_6(\rho)}{144}+\frac{E_4(\rho)}{48\pi^2}\sum_{k=1}^\infty 2(2k+1)\zeta(2k+2)E_{2k+2}(\rho)\,\widehat{a}_1^{2k}\,,\label{ExpandE4I0}
\end{align}
which only depends on the holomorphic Eisenstein series $E_{2k}$ (with $k>1$), but not~$E_2$.

\item[\emph{(ii)}] The holomorphic Eisenstein series $E_4$ appearing in the bubble in \figref{Fig:FeynmanR2GraphFctN24pt}~{\bf (a)} (and which tentatively corresponds to the disconnected contribution in \figref{Fig:FeynmanR24pt}~{\bf (a)}) is in fact (proportional to) a (holomorphic) graph form \cite{DHoker:2016mwo} that is related to a graph with two bivalent vertices
\begin{align}
&\mathcal{C}\big[{}^{4\,0}_{0\,0}\big](\rho)=2\,\zeta(4)\,E_4(\rho)&&\parbox{3cm}{\begin{tikzpicture}[scale = 1.60]
\node at (0,0) {$\bullet$};
\node at (1.5,0) {$\bullet$};
\begin{scope}[ultra thick,decoration={markings,mark=at position 0.8 with {\arrow{>}}}] 
\draw[thin,postaction={decorate}] (0,0) to [out=80,in=180] (0.75,0.4) to [out=0,in=100](1.5,0);
\draw[thin,postaction={decorate}] (0,0) to [out=280,in=180] (0.75,-0.4) to [out=0,in=260](1.5,0);
\end{scope}
\node[fill=white] at (0.75,0.4) {\parbox{0.55cm}{\tiny $(4,0)$}};
\node[fill=white] at (0.75,-0.4) {\parbox{0.55cm}{\tiny $(0,0)$}};
\end{tikzpicture}}\label{GraphFormPicture}
\end{align}
We refer to appendix~\ref{App:GraphFunctions} as well as to \cite{DHoker:2015wxz,DHoker:2016mwo} (see also \cite{Broedel:2015hia,DHoker:2017pvk,Zerbini:2018sox,Zerbini:2018hgs,Gerken:2018jrq,Mafra:2019ddf,Mafra:2019xms,Gerken:2019cxz}) for the definition and our conventions and notation. As was explained in \cite{DHoker:2016mwo}, graph forms of the type $\mathcal{C}\big[{}^{2k\,\,0}_{\,\,\,0\,\,0}\big]$ in (\ref{GraphFormPicture}) can be related to $\mathcal{C}\big[{}^{k\,0}_{k\,0}\big]$ through the action of Cauchy-Riemann differential operators by using the general relation (\ref{GraphFunctionSplit})
\begin{align}
\mathcal{C}\big[{}^{4\,0}_{0\,0}\big](\rho)=\frac{1}{3!\,(\text{Im}\rho)^4}\,\nabla^2\,(\text{Im}(\rho))^2\,\mathcal{C}\big[{}^{2\,0}_{2\,0}\big](\rho)\,,&&\text{with} &&\nabla=2i\,(\text{Im}\rho)^2\,\frac{\partial}{\partial\rho}\,.
\end{align}
Using furthermore (\ref{GraphFunctionShift}) we have
\begin{align}
\mathcal{C}\big[{}^{2\,0}_{2\,0}\big](\rho)=\mathcal{C}\big[{}^{1\,1}_{1\,1}\big](\rho)=\int_{\Sigma}\frac{d^2z_1}{\text{Im}\rho}\int_{\Sigma}\frac{d^2z_2}{\text{Im}\rho}\,\left(\frac{\pi}{(\text{Im}\rho)}\,\mathbb{G}(z_1-z_2;\rho)\right)^2\,.
\end{align}
Here $\mathcal{C}\big[{}^{1\,1}_{1\,1}\big](\rho)$ is a dihedral modular graph function that can be written as the torus integral over two insertion points and whose integrand is proportional to two powers of the scalar two-point function. Therefore indeed, the bubble in \figref{Fig:FeynmanR24pt}~{\bf (a)} can be interpreted as (a differential operator acting on) a disconnected diagram with two integrated positions which are contracted through scalar Greens functions.
\end{itemize}
%
 \item The last term in the second line of (\ref{ExpansionN2R2s0}) has the appearance of a 3-point function (since it contains 3 powers of $\buildH{1}{0}$). Graphical representations of such terms are shown in \figref{Fig:FeynmanR23pt}, which differ by a choice of external and internal points.
\begin{figure}[htbp]
\begin{center}
\scalebox{1}{\parbox{15.2cm}{\begin{tikzpicture}[scale = 1.50]
\draw (-1.19,0) circle (0.05);
\draw (1.19,0) circle (0.05);
\draw[ultra thick] (-1.15,0) -- (1.15,0);
\node at (-1.8,0) {$\buildH{1}{0}$};
\node at (1.8,0) {$\buildH{1}{0}$};
\draw[thick,fill=gray!60!white] (0,0) circle (0.41);
\draw[fill=black] (0,0.41) circle (0.05);
\node at (0,0.8) {$\buildH{2}{0}$};
\node at (0,-1) {\bf (a)};
\begin{scope}[xshift=5.5cm]
\draw (-1.19,0) circle (0.05);
\draw (1.19,0) circle (0.05);
\draw[ultra thick] (-1.15,0) -- (1.15,0);
\node at (-1.8,0) {$\buildH{1}{0}$};
\node at (1.8,0) {$\buildH{2}{0}$};
\draw[thick,fill=gray!60!white] (0,0) circle (0.41);
\draw[fill=black] (0,0.41) circle (0.05);
\node at (0,0.8) {$\buildH{1}{0}$};
\node at (0,-1) {\bf (b)};
\end{scope}
\end{tikzpicture}}}
\caption{\sl Diagrammatical representation of the apparent 3-point function $\mathcal{O}^{(2),\text{3-pt}}_{r=2,s=0}$ in (\ref{CorrelatorN2R23pt}) as a two-point function (with external states marked with $\circ$) corrected by one internal state (marked with $\bullet$). {\bf (a)} internal state is given by $\buildH{2}{0}$; {\bf (b)} internal state is given by $\buildH{1}{0}$}
\label{Fig:FeynmanR23pt}
\end{center}
\end{figure}
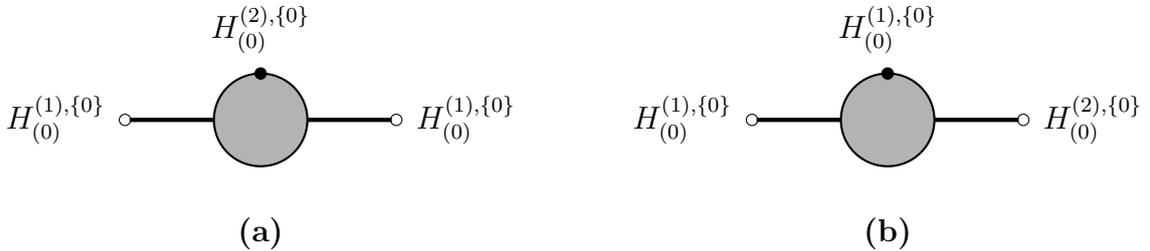 
Just as in the discussion of the 4-point functions, the internal states simply modify the tentative coupling functions 
 \begin{align}
 \mathcal{O}^{(2),\text{3-pt}}_{r=2,s=0}(\widehat{a}_1,\rho)=\frac{4}{3}\mathcal{I}_1(\rho,\widehat{a}_1)=-\frac{8}{3}\,D_{\widehat{a}_1}^2\,\coup{2}{1}(\widehat{a}_1,\rho)\,.\label{CorrelatorN2R23pt}
 \end{align}
 Graphically, both terms in \figref{Fig:FeynmanR23pt} contribute to the this coupling, which itself can be presented as in \figref{Fig:ScalarCouplingRepR23pt}. As before, also this coupling can be written as a $D_{\widehat{a}_1}^2$ derivative (represented by the cross in \figref{Fig:ScalarCouplingRepR23pt}) of the scalar 2-point function on the torus.

 \end{enumerate}
 
 \noindent
While it is quite interesting from the perspective of 

\setlength{\columnsep}{30pt}%
\begin{wrapfigure}{r}{0.375\textwidth}
\begin{center}
\vspace{-1.5cm}
\scalebox{1}{\parbox{5.8cm}{\begin{tikzpicture}[scale = 1.50]
\draw[ultra thick,dashed] (-0.75,0) -- coordinate[pos=0.5] (A) (0.75,0);
\draw[fill=black] (-0.75,0) circle (0.05);
\node at (-1.4,0.05) {$\widehat{b}_0=0$};
\draw[fill=black] (0.75,0) circle (0.05);
\node at (1.5,0.05) {$\widehat{b}_1=\widehat{a}_1$};
\fill[black] (A) circle (1.5pt);
\draw[thick] ($ (A) + (0.1,0.1) $) -- ($ (A) + (-0.1,-0.1) $);
\draw[thick] ($ (A) + (0.1,-0.1) $) -- ($ (A) + (-0.1,0.1) $);
\end{tikzpicture}}}
\caption{\sl Diagrammatic presentation of the coupling in (\ref{CorrelatorN2R23pt}). The dashed line with end-points $(\widehat{b}_0,\widehat{b}_1)$ represents a factor of $\mathbb{G}''(\widehat{a}_1;\rho)+\frac{2\pi i}{\rho-\bar{\rho}}$, while the cross indicates the action of the derivative operator $D^2_{\widehat{a}_1}$.}
\label{Fig:ScalarCouplingRepR23pt}
\end{center}
${}$\\[-1.7cm]
\end{wrapfigure} 

\noindent
the results we have obtained to order $\mathcal{O}(Q_R)$ that such a decomposition of the free energy to order $\mathcal{O}(Q_R^2)$ exists, the latter is still quite speculative. In order to add more credibility to this proposed interpretation as corrections to the two-point function it would be important to also analyse $P^{(2)}_{2,(s)}$ for higher values of $s$: while we have worked out the corresponding functions $f^{j,{(2)}}_{s}$ and $g^{j,k,{(2)}}_{s}$ appearing in (\ref{BasicDecompositionN2R2}) up to $s=4$, it turns out that the decomposition in the basic building blocks $\buildH{r}{s}$ is not unique anymore, but leaves certain ambiguities. For example to order $s=1$ we can present $P^{(2)}_{2,(1)}$ in the following fashion
{\allowdisplaybreaks
\begin{align}
&P^{(2)}_{2,(1)}(\widehat{a}_1,\rho,S)=\frac{2}{3}\,\buildH{2}{1}\,\buildW{2}{0}\,\coup{2}{0}+\frac{4}{3}\,\buildH{2}{1}\,\buildH{2}{0}\,\coup{2}{1}(\widehat{a}_1,S)\nonumber\\
&\hspace{0.5cm}+\frac{4}{3}\,\left(\buildH{1}{0}\right)^2\,\mathcal{I}_2+\left(\buildH{1}{0}\right)^4\,\left[-\frac{17}{6048}(\mathfrak{d}E_6+6 E_6\mathcal{I}_0)+\beta E_4\mathcal{I}_1+\frac{1}{630}\,\mathcal{I}_3\right]\nonumber\\
&\hspace{0.5cm}+\left(\buildH{1}{0}\right)^3\buildH{1}{1}\left[-\frac{13+360\alpha}{240}(\mathfrak{d}E_4+4 E_4\mathcal{I}_0)+\gamma \mathcal{I}_2\right]\nonumber\\
&\hspace{0.5cm}+\left(\buildH{1}{0}\right)^2\buildH{2}{0}\left[\alpha(\mathfrak{d}E_4+4 E_4\mathcal{I}_0)-\frac{17+60\gamma}{90}\,\mathcal{I}_2\right]\nonumber\\
&\hspace{0.5cm}+\left(\frac{8}{45}+32\beta\right)\buildH{1}{0}\buildH{1}{1}\buildH{2}{0}\,\mathcal{I}_1+\left(\frac{52}{45}-32\beta\right)\left(\buildH{1}{0}\right)^2\buildH{2}{1}\,\mathcal{I}_1\,,\label{DecompN3r2}
\end{align}}
where $\alpha$, $\beta$ and $\gamma$ are undetermined constants. As is evident from this expression, these latter constants not only impact the tentative modified coupling functions that appear for the various 3-point and 4-point functions, but also any potential recurring patterns. Furthermore, since the various $\buildH{r}{s}$ now also appear with $s=0$ or $s=1$, there are many more choices which of these states are assigned to be 'internal' and which are 'external'. These ambiguities make a precise statement difficult at this point. However, we point out the following observations, which to some extent support the tentative interpretation given in this subsection:
\begin{itemize}
\item We have checked up to $s=4$ that $P^{(2)}_{2,(s)}(\widehat{a}_1,\rho,S)$ can be decomposed using only $\buildH{r}{s}$ to cover the $S$-dependence.
\item We have checked up to order $s=4$ that the numerical factors can always be chosen in such a way that the only terms that would enter into potential higher-point coupling functions are of the form
\begin{align}
&\mathfrak{d}E_{2k}+2k E_{2k}\,\mathcal{I}_0\,,\hspace{0.2cm}\forall k>1&&\text{or} &&\mathcal{I}_n\,\hspace{0.2cm}\forall n>0\,,&& \text{or} &&E_{2k}\,\mathcal{I}_n\,\hspace{0.2cm}\begin{array}{l}\forall n>0\,, \\ \forall k>1\,.\end{array}\label{ModificationCoupling}
\end{align}
All the couplings appearing up to $s=4$ are holomorphic in the sense that they can be expanded in a power series in $\widehat{a}_1$, where the individual coefficients are combinations of the Eisenstein series $E_4$ and $E_6$ only (but not $E_2$). 
\item The Eisenstein series $E_{2k}$ multiplying $\mathcal{I}_n$ appearing in (\ref{DecompN3r2}) as written in (\ref{ModificationCoupling}) can (as in the case $r=1$) be interpreted as dihedral modular graph forms $\mathcal{C}\big[{}^{2k\,0}_{\,\,0\,\,0}\big](\rho)$ in (\ref{ModGraphFormHol}) of graphs with two bivalent vertices. It is therefore again tempting to once more interpret them as disconnected contributions.
\end{itemize}
\subsubsection{Contributions at Order $\mathcal{O}(Q_R^3)$}
For completeness, we also briefly discuss the decomposition of the free energy to order $\mathcal{O}(Q_R^3)$. Indeed, $P^{(3)}_{2,(0)}$ can be presented in the form
{\allowdisplaybreaks
\begin{align}
&P^{(3)}_{2,(0)}(\widehat{a}_1,\rho,S)=\frac{3}{4}\,\buildH{3}{0}\,\buildW{3}{0}\,\coup{2}{0}+\frac{9}{4}\,\buildH{3}{0}\,\buildH{3}{0}\,\coup{2}{1}(\widehat{a}_1,\rho)\nonumber\\
&\hspace{0.5cm}+\left(\buildH{1}{0}\right)^6\,\left[-\frac{\mathfrak{d} (E_4^2)+8\,E_4^2\mathcal{I}_0}{1728}+\frac{11E_6\mathcal{I}_1}{1134}+\left(\frac{23}{1080}+\frac{3\beta}{64}\right)\,E_4\,\mathcal{I}_2-\frac{\mathcal{I}_4}{7560}\right]\nonumber\\
&\hspace{0.5cm}+\left(\buildH{1}{0}\right)^4\buildH{2}{0}\,\left[-\frac{\mathfrak{d} E_6+6\,E_6\mathcal{I}_0}{81}+\left(\frac{4}{135}+\frac{3\alpha}{64}\right)E_4\,\mathcal{I}_1+\frac{8}{405}\,\mathcal{I}_3\right]\nonumber\\
&\hspace{0.5cm}+\,\left(\buildH{1}{0}\right)^2\,\left(\buildH{2}{0}\right)^2\,\left[-\frac{2(\mathfrak{d} E_4+4\,E_4\mathcal{I}_0)}{27}+\beta\,\mathcal{I}_2\right]-\,\left(\buildH{1}{0}\right)^3\,\buildH{3}{0}\,\left(1+\frac{27\beta}{16}\right)\,\mathcal{I}_2\nonumber\\
&\hspace{0.5cm}+\left(\buildH{2}{0}\right)^3\,\alpha\,\mathcal{I}_1+\buildH{1}{0}\,\buildH{2}{0}\,\buildH{3}{0}\,\left(\frac{16}{3}-\frac{27\alpha}{16}\right)\,E_6\,\mathcal{I}_1\,,\label{ExpansionN2R3s0}
\end{align}}
where $\alpha$ and $\beta$ are two undetermined constants. In the same spirit as to order $\mathcal{O}(Q_R^2)$, the terms in the second to fifth line represent corrections to two-point functions with 4, 3, 2 or 1 internal points respectively. Due to the ambiguity in the decomposition in terms of the building blocks $\buildH{r}{0}$, it is not possible to determine precisely the arising corrected couplings. However, already the fact that $P^{(3)}_{2,(0)}$ allows a decomposition in terms of only these particular building blocks, can be seen as a further argument for the picture already developed to order $\mathcal{O}(Q_R^2)$.

\section{Example $N=3$}\label{Sect:ExN3}
After the case $N=2$, we shall now discuss Little String Theories with $N=3$.
\subsection{Decomposition at Order $\mathcal{O}(Q_R)$ and Scalar Correlators}
While it was already argued in \cite{Hohenegger:2019tii} that the free energy $P_{3,(s)}^{(r=1)}$ can be decomposed using the basic building blocks $\buildH{0}{s}$ and $\buildW{0}{s}$ in the following form
\begin{align}
P_{3,(s)}^{(r=1)}(\widehat{a}_1,\widehat{a}_2,\rho,S)=&\,\buildH{1}{s}\,\left(\buildW{1}{0}\right)^2\,\coup{3}{0}+\left(\buildH{1}{s}\right)^2\,\buildH{1}{0}\,\coup{3}{1}(\widehat{a}_1,\widehat{a}_2,\rho)\nonumber\\
&+\left(\buildH{1}{s}\right)^3\,\coup{3}{2}(\widehat{a}_1,\widehat{a}_2,\rho)\,,\label{N3BasicGraphFunctions}
\end{align}
the form of the $\coup{3}{\alpha}(\widehat{a}_1,\widehat{a}_2,\rho)$ was (mostly) only given as infinite sums
{\allowdisplaybreaks\begin{align}
\coup{3}{0}&=3\,,\label{CoupN3R10}\\
\coup{3}{1}(\widehat{a}_1,\widehat{a}_2,\rho)&=\sum_{n=1}^{\infty}\frac{-2n}{1-Q_\rho^n}\left(Q_{\widehat{a}_1}^n+\frac{Q_\rho^n}{Q_{\widehat{a}_1}^n}+Q_{\widehat{a}_2}^n+\frac{Q_\rho^n}{Q_{\widehat{a}_2}^n}+(Q_{\widehat{a}_1}Q_{\widehat{a}_2})^n+\frac{Q_\rho^n}{(Q_{\widehat{a}_1}Q_{\widehat{a}_2})^n}\right)\,,\\
\coup{3}{2}(\widehat{a}_1,\widehat{a}_2,\rho)&=\sum_{n=1}^\infty\frac{n^2}{(1-Q_\rho^n)^2}\left[Q_\rho^n\left(Q_{\widehat{a}_1}^n+Q_{\widehat{a}_2}^n+\frac{Q_\rho^n}{(Q_{\widehat{a}_1}Q_{\widehat{a}_2})^n}\right)+\left((Q_{\widehat{a}_1}Q_{\widehat{a}_2})^n+\frac{Q_\rho^n}{Q_{\widehat{a}_1}^n}+\frac{Q_\rho^n}{Q_{\widehat{a}_2}^n}\right)\right]\nonumber\\
&\hspace{0.3cm}+\sum_{n_1,n_2=1}^\infty\left(\frac{n_2(2n_1+n_2)}{(1-Q_\rho^{n_1})(1-Q_\rho^{n_2})}+\frac{(n_1+n_2)(n_1-n_2)}{(1-Q_\rho^{n_1})(1-Q_\rho^{n_1+n_2})}\right)\bigg(Q_{\widehat{a}_1}^{n_1+n_2}Q_{\widehat{a}_1}^{n_1}\nonumber\\
&\hspace{1cm}+Q_{\widehat{a}_1}^{n_1}Q_{\widehat{a}_1}^{n_1+n_2}+\frac{Q_\rho^{n_1+n_2}}{Q_{\widehat{a}_1}^{n_1+n_2}Q_{\widehat{a}_2}^{n_2}}+\frac{Q_\rho^{n_1+n_2}}{Q_{\widehat{a}_1}^{n_1}Q_{\widehat{a}_2}^{n_1+n_2}}+\frac{Q_\rho^{n_1}Q_{\widehat{a}_1}^{n_2}}{Q_{\widehat{a}_2}^{n_1}}+\frac{Q_\rho^{n_1}Q_{\widehat{a}_2}^{n_2}}{Q_{\widehat{a}_1}^{n_1}}\bigg)\,.\label{N3R1CoupsPre}
\end{align}}

\noindent
The coupling $\coup{3}{0}$ can in a trivial manner be represented as in \figref{Fig:N3ScalarCouplingR1C0}: it simply corresponds

\begin{figure}[htbp]
\begin{center}
\scalebox{1}{\parbox{16.2cm}{\begin{tikzpicture}[scale = 1.50]
\draw[fill=black] (0,0.75) circle (0.05);
\draw[fill=black] (-0.65,-0.375) circle (0.05);
\draw[fill=black] (0.65,-0.375) circle (0.05);
\draw[ultra thick,dashed]  (-0.65,-0.375) -- (0,0.75);
\node at (-0.9,-0.5) {$\widehat{b}_0$};
\node at (1,-0.5) {$\widehat{b}_1$};
\node at (0,1.1) {$\widehat{b}_2$};
\node at (2,0.3) {$+$};
\begin{scope}[xshift=4cm]
\draw[fill=black] (0,0.75) circle (0.05);
\draw[fill=black] (-0.65,-0.375) circle (0.05);
\draw[fill=black] (0.65,-0.375) circle (0.05);
\draw[ultra thick,dashed]  (0.65,-0.375) -- (0,0.75);
\node at (-0.9,-0.5) {$\widehat{b}_0$};
\node at (1,-0.5) {$\widehat{b}_1$};
\node at (0,1.1) {$\widehat{b}_2$};
\end{scope}
\node at (6.25,0.3) {$+$};
\begin{scope}[xshift=8.5cm]
\draw[fill=black] (0,0.75) circle (0.05);
\draw[fill=black] (-0.65,-0.375) circle (0.05);
\draw[fill=black] (0.65,-0.375) circle (0.05);
\draw[ultra thick,dashed]  (0.65,-0.375) -- (-0.65,-0.375);
\node at (-0.9,-0.5) {$\widehat{b}_0$};
\node at (1,-0.5) {$\widehat{b}_1$};
\node at (0,1.1) {$\widehat{b}_2$};
\end{scope}
\end{tikzpicture}}}
\caption{\sl Diagrammatical representation of the coupling $\coup{3}{1}$ in (\ref{N3CoupR11}). The dashed lines represent the two-point function $\mathbb{G}''(\widehat{b}_\ell-\widehat{b}_j)+\frac{2\pi i}{\rho-\bar{\rho}}$.}
\label{Fig:N3CouplingO1}
\end{center}
\end{figure}
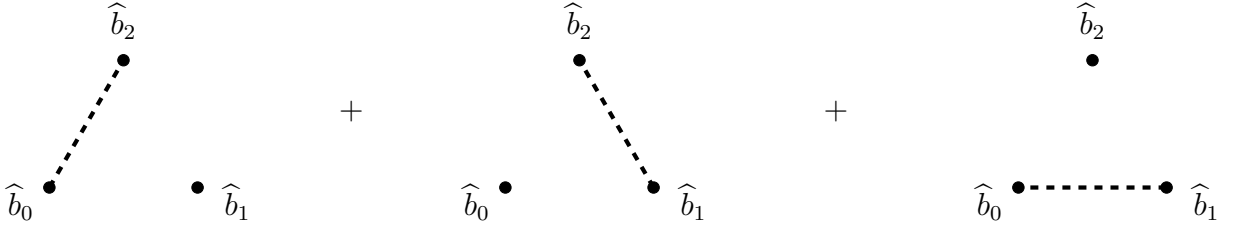

\setlength{\columnsep}{30pt}%
\begin{wrapfigure}{r}{0.27\textwidth}
\begin{center}
\vspace{-0.7cm}
\scalebox{1}{\parbox{4.8cm}{\begin{tikzpicture}[scale = 1.50]
\draw[fill=black] (0,0.65) circle (0.05);
\draw[fill=black] (-0.55,-0.275) circle (0.05);
\draw[fill=black] (0.55,-0.275) circle (0.05);
\node at (-1.1,-0.4) {$\widehat{b}_0=0$};
\node at (1.2,-0.4) {$\widehat{b}_1=\widehat{a}_1$};
\node at (0,1) {$\widehat{b}_2=\widehat{a}_1+\widehat{a}_2$};
\end{tikzpicture}}}
\caption{\sl Trivial diagrammatic presentation of the constant coupling $\coup{3}{0}$ in (\ref{N3R1CoupsPre}).}
\label{Fig:N3ScalarCouplingR1C0}
\end{center}
${}$\\[-2.2cm]
\end{wrapfigure} 

\noindent
to no correlator connecting any of the three points
\begin{align}
&\widehat{b}_0=0\,,&&\widehat{b}_1=\widehat{a}_1\,,&& \widehat{b}_2=\widehat{a}_1+\widehat{a}_2\,,
\end{align}
thus yielding a constant. Furthermore, based on the discussion of similar structures already appearing in the case of $N=2$ (see \emph{e.g.} eq.~(\ref{N2CoupR11})), we can write for $\coup{3}{2}(\widehat{a}_1,\widehat{a}_2,\rho)$\footnote{Compared to a slightly different expression in \cite{Hohenegger:2019tii}, here we have used the implicit periodicity of $\mathbb{G}''(z;\rho)$ under a shift $z\to z+\rho$ along with the symmetry $\mathbb{G}''(z;\rho)=\mathbb{G}''(-z;\rho)$}
\begin{align}
\coup{3}{1}(\widehat{a}_{1,2},\rho)=\frac{1}{(2\pi)^2}\sum_{\ell=1}^3\sum_{j\neq \ell}\left(\mathbb{G}''(\widehat{b}_\ell-\widehat{b}_j;\rho)+\frac{2\pi i}{\rho-\bar{\rho}}\right).\label{N3CoupR11}
\end{align}
Graphically, up to an overall numerical factor $\frac{2}{(2\pi)^2}$, this coupling can be represented as in~\figref{Fig:N3CouplingO1} as a single two-point function connecting any combination of the points $\{\widehat{b}_0\,,\,\widehat{b}_1\,,\widehat{b}_2\}$.

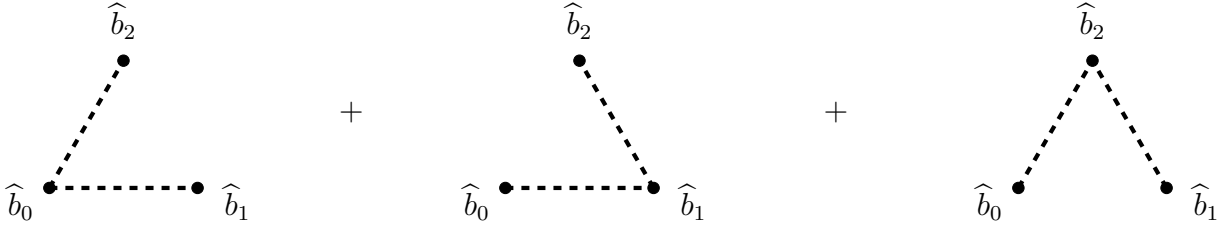
\begin{figure}[htbp]
\begin{center}
\scalebox{1}{\parbox{16.2cm}{\begin{tikzpicture}[scale = 1.50]
\draw[fill=black] (0,0.75) circle (0.05);
\draw[fill=black] (-0.65,-0.375) circle (0.05);
\draw[fill=black] (0.65,-0.375) circle (0.05);
\draw[ultra thick,dashed]  (-0.65,-0.375) -- (0,0.75);
\draw[ultra thick,dashed]  (-0.65,-0.375) -- (0.65,-0.375);
\node at (-0.9,-0.5) {$\widehat{b}_0$};
\node at (1,-0.5) {$\widehat{b}_1$};
\node at (0,1.1) {$\widehat{b}_2$};
\node at (2,0.3) {$+$};
\begin{scope}[xshift=4cm]
\draw[fill=black] (0,0.75) circle (0.05);
\draw[fill=black] (-0.65,-0.375) circle (0.05);
\draw[fill=black] (0.65,-0.375) circle (0.05);
\draw[ultra thick,dashed]  (0.65,-0.375) -- (0,0.75);
\draw[ultra thick,dashed]  (-0.65,-0.375) -- (0.65,-0.375);
\node at (-0.9,-0.5) {$\widehat{b}_0$};
\node at (1,-0.5) {$\widehat{b}_1$};
\node at (0,1.1) {$\widehat{b}_2$};
\end{scope}
\node at (6.25,0.3) {$+$};
\begin{scope}[xshift=8.5cm]
\draw[fill=black] (0,0.75) circle (0.05);
\draw[fill=black] (-0.65,-0.375) circle (0.05);
\draw[fill=black] (0.65,-0.375) circle (0.05);
\draw[ultra thick,dashed]  (0.65,-0.375) -- (0,0.75);
\draw[ultra thick,dashed]  (-0.65,-0.375) -- (0,0.75);
\node at (-0.9,-0.5) {$\widehat{b}_0$};
\node at (1,-0.5) {$\widehat{b}_1$};
\node at (0,1.1) {$\widehat{b}_2$};
\end{scope}
\end{tikzpicture}}}
\caption{\sl Diagrammatical representation of the coupling $\coup{3}{2}$ in (\ref{N3CoupR12}). The dashed lines represent the two-point function $\mathbb{G}''(\widehat{b}_\ell-\widehat{b}_j)+\frac{2\pi i}{\rho-\bar{\rho}}$.}
\label{Fig:N3CouplingO2tripple}
\end{center}
\end{figure}

Finally, the coupling function $\coup{3}{2}(\widehat{a}_1,\widehat{a}_2,\rho)$ is more complicated to analyse than the previous ones due to the presence of the double summation in the second term in (\ref{N3R1CoupsPre}). Following the discussion in \cite{Bastian:2019wpx}, however, we can represent it in terms of generating functions $T(z_1,z_2;\rho)$ (see eq.~(\ref{GenFunctionsT})) of multiple divisor sums up to length 2, which were introduced in \cite{Bachmann:2013wba}. Since these functions have a well-defined Taylor series expansion\footnote{In fact, this is the form in which the $T$ have originally been introduced in \cite{Bachmann:2013wba}.} (\ref{TaylorSeriesT}), we can compute the series expansion of $\coup{3}{2}(\widehat{a}_1,\widehat{a}_2;\rho)$ in powers of $\widehat{a}_{1,2}$. However, in order to extract the correct pole structure and also due to the complexity of (\ref{N3R1CoupsPre}), we limit ourself to study the leading powers in $\widehat{a}_{1,2}$ as a limited power series in $Q_{\rho}$ which (assuming modularity of the various contributions) can be expressed as polynomials in the Eisenstein series. The relevant details can be found in appendix~\ref{App:ExpansionCoupO32}, where we find that the series expansion matches the combination (\ref{CoupN32Calc}) of scalar Greens functions which can also be written in the form
\begin{align}
\coup{3}{2}(\widehat{a}_1,\widehat{a}_2,\rho)
=\frac{1}{(2\pi)^4}\,&\sum_{\ell=0}^{2}\prod_{j\neq \ell}\left(\mathbb{G}''(\widehat{b}_\ell-\widehat{b}_j;\rho)+\frac{2\pi i}{\rho-\bar{\rho}}\right)\,.\label{N3CoupR12}
\end{align}
Yet another way to present this is result is 
\begin{align}
\coup{3}{2}(\widehat{a}_1,\widehat{a}_2,\rho)=\frac{1}{(2\pi)^4}\,&\sum_{\ell=0}^{2}\sum_{\mathcal{S}\in\{0,1,2\}\setminus\{\ell\}\atop |\mathcal{S}|=2}\prod_{j\in\mathcal{S}}\left(\mathbb{G}''(\widehat{b}_\ell-\widehat{b}_j;\rho)+\frac{2\pi i}{\rho-\bar{\rho}}\right)\,,\label{CouplingO32R1}
\end{align}
which matches the general expression in (\ref{StructureCouplingR1}). Graphically (up to a numerical factor), this coupling can be represented as in \figref{Fig:N3CouplingO2tripple} as two scalar two-point functions connecting two distinct pairs of points $\{\widehat{b}_0,\widehat{b}_1,\widehat{b}_2\}$.

\noindent
We note that all three coupling functions $\coup{3}{0}$ in (\ref{CoupN3R10}), $\coup{3}{1}$ in (\ref{N3CoupR11}) and $\coup{3}{2}$ in (\ref{N3CoupR12}) match the general form (\ref{StructureCouplingR1}).

\subsection{Decomposition at Order $\mathcal{O}(Q_R^r)$ for $r>1$}
We can repeat the analysis of the previous section for contributions to the free energy to orders $Q_R^r$ for $r>1$. As in the case for $N=2$, however, we shall encounter the problem that a decomposition in the building blocks $\buildH{r}{s}$ and $\buildW{r}{s}$ for $r\in\{0,1\}$ is not unique since the latter form an overcomplete basis. While this ambiguity prevents us from determining the precise pattern in which $P^{(r)}_{3,(s)}$ decomposes, the general form of the result still gives further credence to the picture already advocated before. Using the same approach as in the case $N=2,$\footnote{\emph{I.e.} we study limited series expansions in $Q_\rho$, which we match to quasi-modular forms, as explained in appendix~\ref{SeriesExpansionSimCoup}.} we have matched the next-to-leading order $\mathcal{O}(Q_R^2)$ of the free energy $P^{(r=2)}_{3,(s=0)}$ to a decomposition in terms of $\buildH{r}{s}$ and $\buildW{r}{s}$ and certain (modular) coupling functions. In order to present this form despite the above mentioned ambiguities, we use a condensed notation: starting from 
\begin{align}
P_{3,(s=0)}^{(r=2)}=\sum_{a=0}^3\sum_{k}\sum_{i_1,i_2,i_3,i_4}\,c^{(a)}_{k;i_1,i_2;i_3,i_4}(\rho)\,\mathcal{K}_a^{k}(\widehat{a}_{1,2},\rho)\,\left(\buildH{1}{0}\right)^{i_1}\,\left(\buildH{2}{0}\right)^{i_2}\,\left(\buildW{1}{0}\right)^{i_3}\,\left(\buildW{2}{0}\right)^{i_4}\,,\nonumber
\end{align}
for $a=0$, $k$ can only take the value $k=1$ with the coupling $\mathcal{K}_0^{k=1}(\widehat{a}_{1,2},\rho)=1$ and the following non-vanishing coefficient functions
{\allowdisplaybreaks
\begin{align}
&c^{(0)}_{1;6,0;0,0}=\frac{(2-90 \alpha_1-81 \alpha_{19}-36 \alpha_2+18 \alpha_3)E_6}{5184}\,,\hspace{1cm} c^{(0)}_{1;4,0;2,0}=\frac{(-216 \alpha_1+72 \alpha_3+11)}{144} \,,\nonumber\\
& c^{(0)}_{1;5,0;1,0}=\frac{364-11520 \alpha_1-3 \alpha_{16}-2592 \alpha_{17}-864 \alpha_{18}+2304 \text{$\alpha$3}}{55296}\,E_4^2\,,\hspace{1cm} c^{(0)}_{1;1,2;1,0}=\alpha_{18}\,E_4\,,\nonumber\\
&c^{(0)}_{1;4,1;0,0}=\frac{7680 \alpha_1-\alpha_{16}-864 \alpha_{17}+3072 \alpha_2-1536 \alpha_3-176}{55296}\,E_4^2\,,\hspace{1cm} c^{(0)}_{1;1,0;1,2}=\frac{-3 \alpha_{16}-4}{6} \,,\nonumber\\
&c^{(0)}_{1;4,0;0,1}=\frac{ (72 \alpha_1+108 \alpha_{19}+48 \alpha_2-24 \alpha_3-1)E_6}{72}\,,\hspace{1cm} c^{(0)}_{1;3,1;1,0}=\frac{(-162 \alpha_{19}-36 \alpha_2-1)\,E_6}{54} \,,\nonumber\\
&c^{(0)}_{1;3,0;1,1}=\frac{ (576 \alpha_1+3 \alpha_{16}+1296 \alpha_{17}+432 \alpha_{18}-576 \alpha_3+4)\,E_4}{288}\,,\hspace{1cm}c^{(0)}_{1;2,2;0,0}=\alpha_{19}\,E_6\,,\nonumber\\
&c^{(0)}_{1;2,1;0,1}=\frac{ (-1152 \alpha_1+3 \alpha_{16}+1296 \alpha_{17}-2304 \alpha_2+1152 \alpha_3-16)\,E_4}{864}\,,\hspace{1cm} c^{(0)}_{1;0,3;0,0}=\alpha_{17}\,E_4\,,\nonumber\\
&c^{(0)}_{1;2,1;2,0}=\frac{ (192 \alpha_1-\alpha_{16}-864 \alpha_{17}-288 \alpha_{18}+384 \alpha_2-192
   \alpha_3-24)\,E_4}{96}\,,\hspace{1cm} c^{(0)}_{1;0,1;2,1}=\alpha_{16}\,,\nonumber\\
&c^{(0)}_{1;3,0;3,0}=-\frac{3 (8 \alpha_1-8 \alpha_3-1)\,E_4}{8}\,,\hspace{1cm}c^{(0)}_{1;0,1;0,2}=\frac{ 40-3 \alpha_{16}}{18}\,.
&\label{CoefsN3a0}
\end{align}}
For $a=1$ there are three different types of graphs $\mathcal{K}_1^{k}(\widehat{a}_{1,2},\rho)$ (\emph{i.e.} $k\in\{1,2,3\}$) and the corresponding coefficient functions $c^{(1)}_{k;i_1,i_2;i_3,i_4}(\rho)$ are tabulated in Table~\ref{Tab:Coefsa1}. Here we are using a symbolic notation for the $\mathcal{K}_1^{k}(\widehat{a}_{1,2},\rho)$: solid points represent $(\widehat{b}_0\,,\widehat{b}_1\,,\widehat{b}_2)$ and a sum over all cyclic permutations is understood. Furthermore, dashed lines with $n$ crosses between any of these points represent factors of $D_{\widehat{b}_i}^{2n} \left(\mathbb{G}''(\widehat{b}_i-\widehat{b}_j;\rho)+\frac{2\pi i}{\rho-\bar{\rho}}\right)$, such that \emph{e.g.}
\begin{align}
\mathcal{K}_1^{2}(\widehat{a}_{1,2},\rho)\hspace{0.5cm}=\hspace{0.5cm}\scalebox{1.2}{\parbox{1.4cm}{\begin{tikzpicture}
\draw[fill=black] (0,0.75) circle (0.05);
\draw[fill=black] (-0.65,-0.375) circle (0.05);
\draw[fill=black] (0.65,-0.375) circle (0.05);
\draw[thick,dashed]  (-0.65,-0.375) -- coordinate[pos=0.5] (C1) (0.65,-0.375);
\fill[black] (C1) circle (1.5pt);
\draw[thick] ($ (C1) + (0.1,0.1) $) -- ($ (C1) + (-0.1,-0.1) $);
\draw[thick] ($ (C1) + (0.1,-0.1) $) -- ($ (C1) + (-0.1,0.1) $);
\end{tikzpicture}}}\hspace{0.5cm}=\hspace{0.5cm}\mathbb{G}^{(4)}(\widehat{b}_0-\widehat{b}_1;\rho)+\mathbb{G}^{(4)}(\widehat{b}_1-\widehat{b}_2;\rho)+\mathbb{G}^{(4)}(\widehat{b}_2-\widehat{b}_0;\rho)\,.\nonumber
\end{align}
\begin{table}[htbp]
\begin{center}
\rotatebox{90}{
\scalebox{0.92}{\parbox{25.6cm}{\begin{tabular}{|c||c||c|c|c|c|}\hline
&&&&&\\[-16pt]
\multirow{5}{*}{{\bf $k$}} & \multirow{5}{*}{$\mathcal{K}_1^{k}(\widehat{a}_{1,2},\rho)$} & \parbox{1.3cm}{${}$\\[-0cm]$c^{(1)}_{k;6,0;0,0}$\\[-0.3cm]} & \parbox{1.3cm}{${}$\\[-0cm]$c^{(1)}_{k;5,0;1,0}$\\[-0.3cm]} & \parbox{1.3cm}{${}$\\[-0cm]$c^{(1)}_{k;4,1;0,0}$\\[-0.3cm]} & \parbox{1.3cm}{${}$\\[-0cm]$c^{(1)}_{k;4,0;2,0}$\\[-0.3cm]} \\\cline{3-6}
& &\parbox{1.4cm}{${}$\\[-0cm]$c^{(1)}_{k;4,0;0,1}$\\[-0.3cm]} & \parbox{1.3cm}{${}$\\[-0cm]$c^{(1)}_{k;3,1;1,0}$\\[-0.3cm]} & \parbox{1.3cm}{${}$\\[-0cm]$c^{(1)}_{k;3,0;3,0}$\\[-0.3cm]} & \parbox{1.3cm}{${}$\\[-0cm]$c^{(1)}_{k;3,0;1,1}$\\[-0.3cm]} \\ \cline{3-6}
& &\parbox{1.4cm}{${}$\\[-0cm]$c^{(1)}_{k;2,2;0,0}$\\[-0.3cm]} & \parbox{1.3cm}{${}$\\[-0cm]$c^{(1)}_{k;2,1;2,0}$\\[-0.3cm]} & \parbox{1.3cm}{${}$\\[-0cm]$c^{(1)}_{k;2,1;0,1}$\\[-0.3cm]} & \parbox{1.3cm}{${}$\\[-0cm]$c^{(1)}_{k;2,0;2,1}$\\[-0.3cm]} \\ \cline{3-6}
& &\parbox{1.4cm}{${}$\\[-0cm]$c^{(1)}_{k;2,0;0,2}$\\[-0.3cm]} & \parbox{1.3cm}{${}$\\[-0cm]$c^{(1)}_{k;1,2;1,0}$\\[-0.3cm]} & \parbox{1.3cm}{${}$\\[-0cm]$c^{(1)}_{k;1,1;3,0}$\\[-0.3cm]} & \parbox{1.3cm}{${}$\\[-0cm]$c^{(1)}_{k;1,1;1,1}$\\[-0.3cm]} \\ \cline{3-6}
& &\parbox{1.3cm}{${}$\\[-0cm]$c^{(1)}_{k;0,3;0,0}$\\[-0.3cm]} & \parbox{1.3cm}{${}$\\[-0cm]$c^{(1)}_{k;0,2;2,0}$\\[-0.3cm]} & \parbox{1.3cm}{${}$\\[-0cm]$c^{(1)}_{k;0,2;0,1}$\\[-0.3cm]} & \parbox{1.3cm}{${}$\\[-0cm]$\phantom{c^{(1)}_{k;6,0;0,0}}$\\[-0.3cm]} \\\hline\hline
&&&&&\\[-14pt]
\multirow{5}{*}{{\bf $1$}} & \multirow{5}{*}{\scalebox{1.2}{\parbox{1.4cm}{\begin{tikzpicture}
\draw[fill=black] (0,0.75) circle (0.05);
\draw[fill=black] (-0.65,-0.375) circle (0.05);
\draw[fill=black] (0.65,-0.375) circle (0.05);
\draw[thick,dashed]  (-0.65,-0.375) -- coordinate[pos=0.3] (C1) coordinate[pos=0.69] (C2) (0.65,-0.375);
\fill[black] (C1) circle (1.5pt);
\draw[thick] ($ (C1) + (0.1,0.1) $) -- ($ (C1) + (-0.1,-0.1) $);
\draw[thick] ($ (C1) + (0.1,-0.1) $) -- ($ (C1) + (-0.1,0.1) $);
\fill[black] (C2) circle (1.5pt);
\draw[thick] ($ (C2) + (0.1,0.1) $) -- ($ (C2) + (-0.1,-0.1) $);
\draw[thick] ($ (C2) + (0.1,-0.1) $) -- ($ (C2) + (-0.1,0.1) $);
\end{tikzpicture}}}} & \parbox{2.5cm}{${}$\\[-0cm]$\frac{(1-24\alpha_1-36\alpha_8)E_4}{3456(2\pi)^6}$\\[-0.3cm]} &  \parbox{0.175cm}{${}$\\[-0.2cm]$0$\\[-0.2cm]} &  \parbox{0.175cm}{${}$\\[-0.2cm]$0$\\[-0.2cm]} &  \parbox{0.8cm}{${}$\\[-0.1cm]$\frac{\alpha_1}{(2\pi)^6}$\\[-0.3cm]} \\\cline{3-6}
& & \parbox{0.8cm}{${}$\\[-0cm]$\frac{\alpha_8}{(2\pi)^6}$\\[-0.3cm]} &  \parbox{2cm}{${}$\\[-0.1cm]$\frac{1-24\alpha_1-36\alpha_8}{18(2\pi)^6}$\\[-0.3cm]} &  \parbox{0.175cm}{${}$\\[-0.1cm]$0$\\[-0.3cm]} &  \parbox{0.175cm}{${}$\\[-0.1cm]$0$\\[-0.3cm]} \\\cline{3-6}
& & \parbox{2cm}{${}$\\[-0cm]$\frac{24\alpha_1+36\alpha_8-1}{54(2\pi)^6}$\\[-0.3cm]} &  \parbox{0.175cm}{${}$\\[-0.1cm]$0$\\[-0.3cm]} &  \parbox{0.175cm}{${}$\\[-0.1cm]$0$\\[-0.3cm]} &  \parbox{0.175cm}{${}$\\[-0.1cm]$0$\\[-0.3cm]} \\\cline{3-6}
& & \parbox{0.175cm}{${}$\\[-0cm]$0$\\[-0.3cm]}  &  \parbox{0.175cm}{${}$\\[-0.1cm]$0$\\[-0.3cm]} &  \parbox{0.175cm}{${}$\\[-0.1cm]$0$\\[-0.3cm]} &  \parbox{0.175cm}{${}$\\[-0.1cm]$0$\\[-0.3cm]} \\\cline{3-6}
& & \parbox{0.175cm}{${}$\\[-0cm]$0$\\[-0.3cm]} &  \parbox{0.175cm}{${}$\\[-0.1cm]$0$\\[-0.3cm]} &  \parbox{0.175cm}{${}$\\[-0.1cm]$0$\\[-0.3cm]} &  \parbox{0.175cm}{${}$\\[-0.1cm]$\phantom{0}$\\[-0.3cm]} \\\hline\hline
&&&&&\\[-14pt]
\multirow{5}{*}{{\bf $2$}} & \multirow{5}{*}{\scalebox{1.2}{\parbox{1.4cm}{\begin{tikzpicture}
\draw[fill=black] (0,0.75) circle (0.05);
\draw[fill=black] (-0.65,-0.375) circle (0.05);
\draw[fill=black] (0.65,-0.375) circle (0.05);
\draw[thick,dashed]  (-0.65,-0.375) -- coordinate[pos=0.5] (C1) (0.65,-0.375);
\fill[black] (C1) circle (1.5pt);
\draw[thick] ($ (C1) + (0.1,0.1) $) -- ($ (C1) + (-0.1,-0.1) $);
\draw[thick] ($ (C1) + (0.1,-0.1) $) -- ($ (C1) + (-0.1,0.1) $);
\end{tikzpicture}}}} &  \parbox{1.8cm}{${}$\\[-0cm]$\frac{(5-144\alpha_1)E_6}{2592(2\pi)^4}$\\[-0.3cm]} &   \parbox{3.6cm}{${}$\\[-0.1cm]$\frac{(20-576\alpha_1-9\alpha_{10}+72\alpha_3)E_4}{864(2\pi)^4}$\\[-0.3cm]}  &   \parbox{5cm}{${}$\\[-0.1cm]$\frac{(1152\alpha_1+288\alpha_2-144\alpha_3-27\alpha_9-16)E_4}{2593(2\pi)^4}$\\[-0.3cm]} &  \parbox{0.175cm}{${}$\\[-0.2cm]$0$\\[-0.2cm]} \\\cline{3-6}
& & \parbox{0.175cm}{${}$\\[-0cm]$0$\\[-0.3cm]} &  \parbox{0.175cm}{${}$\\[-0.1cm]$0$\\[-0.3cm]} &  \parbox{1cm}{${}$\\[-0.1cm]$-\frac{12\alpha_3}{(2\pi)^4}$\\[-0.3cm]} &  \parbox{0.8cm}{${}$\\[-0.1cm]$\frac{\alpha_{10}}{(2\pi)^4}$\\[-0.3cm]} \\\cline{3-6}
& & \parbox{0.175cm}{${}$\\[-0cm]$0$\\[-0.3cm]} &  \parbox{2.5cm}{${}$\\[-0.1cm]$\frac{2(12\alpha_3-\alpha_{10}-8\alpha_2)}{(2\pi)^4}$\\[-0.3cm]} & \parbox{0.8cm}{${}$\\[-0.1cm]$\frac{\alpha_9}{(2\pi)^4}$\\[-0.3cm]} &  \parbox{0.175cm}{${}$\\[-0.1cm]$0$\\[-0.3cm]} \\\cline{3-6}
& & \parbox{0.175cm}{${}$\\[-0cm]$0$\\[-0.3cm]}  &  \parbox{3.9cm}{${}$\\[-0.1cm]$\frac{2(8+3\alpha_{10}+96\alpha_2-7 2\alpha_3-9\alpha_9)}{9(2\pi)^4}$\\[-0.3cm]} &  \parbox{0.175cm}{${}$\\[-0.1cm]$0$\\[-0.3cm]} &  \parbox{0.175cm}{${}$\\[-0.1cm]$0$\\[-0.3cm]} \\\cline{3-6}
& & \parbox{3cm}{${}$\\[-0cm]$\frac{2(48\alpha_3+9\alpha_9-96\alpha_2-8)}{27(2\pi)^4}$\\[-0.3cm]} &  \parbox{0.175cm}{${}$\\[-0.1cm]$0$\\[-0.3cm]} &  \parbox{0.175cm}{${}$\\[-0.1cm]$0$\\[-0.3cm]} &  \parbox{0.175cm}{${}$\\[-0.1cm]$\phantom{0}$\\[-0.3cm]} \\\hline\hline
%
&&&&&\\[-14pt]
\multirow{5}{*}{{\bf $3$}} & \multirow{5}{*}{\scalebox{1.2}{\parbox{1.4cm}{\begin{tikzpicture}
\draw[fill=black] (0,0.75) circle (0.05);
\draw[fill=black] (-0.65,-0.375) circle (0.05);
\draw[fill=black] (0.65,-0.375) circle (0.05);
\draw[thick,dashed]  (-0.65,-0.375) -- coordinate[pos=0.5] (C1) (0.65,-0.375);
\end{tikzpicture}}}} & \parbox{4.3cm}{${}$\\[-0cm]$\frac{(184-5184\alpha_1-9\alpha_{13}-864\alpha_{14})E_4^2}{82944(2\pi)^2}$\\[-0.2cm]}  &   \parbox{2.8cm}{${}$\\[-0cm]$\frac{(11-216\alpha_1+36\alpha_3)E_6}{216(2\pi)^2}$\\[-0.2cm]} &  \parbox{3.8cm}{${}$\\[-0.2cm]$\frac{(216E\alpha_1+72\alpha_2-36\alpha_3-5)E_6}{324(2\pi)^2}$\\[-0.2cm]} &   \parbox{3.7cm}{${}$\\[-0.2cm]$\frac{(36-288\alpha_1-\alpha_{12}+192\alpha_3)E_4}{96(2\pi)^2}$\\[-0.2cm]}  \\\cline{3-6}
& & \parbox{1cm}{${}$\\[-0cm]$\frac{\alpha_{14} E_4}{(2\pi)^2}$\\[-0.3cm]} &  \parbox{6cm}{${}$\\[-0.1cm]$\frac{(384\alpha_1+256(\alpha_2-\alpha_3)-\alpha_{11}-4\alpha_{13}-192\alpha_{14})E_4}{96(2\pi)^2}$\\[-0.3cm]} &  \parbox{0.175cm}{${}$\\[-0.2cm]$0$\\[-0.2cm]} & \parbox{0.175cm}{${}$\\[-0.2cm]$0$\\[-0.2cm]} \\\cline{3-6}
& & \parbox{6.3cm}{${}$\\[-0cm]$\frac{(768\alpha_3+576\alpha_{14}-1152\alpha_2-1536\alpha_2-3\alpha_{11}-16)E_4}{864(2\pi)^2}$\\[-0.3cm]}  &  \parbox{0.175cm}{${}$\\[-0.2cm]$0$\\[-0.2cm]} & \parbox{0.175cm}{${}$\\[-0.2cm]$0$\\[-0.2cm]} & \parbox{0.8cm}{${}$\\[-0.1cm]$\frac{\alpha_{12}}{(2\pi)^2}$\\[-0.3cm]} \\\cline{3-6}
& & \parbox{0.8cm}{${}$\\[-0cm]$\frac{\alpha_{13}}{(2\pi)^2}$\\[-0.3cm]}  &  \parbox{0.175cm}{${}$\\[-0.2cm]$0$\\[-0.2cm]} &  \parbox{0.9cm}{${}$\\[-0.1cm]$\frac{-2\alpha_{12}}{(2\pi)^2}$\\[-0.3cm]} &  \parbox{0.8cm}{${}$\\[-0.1cm]$\frac{\alpha_{11}}{(2\pi)^2}$\\[-0.3cm]} \\\cline{3-6}
& &\parbox{0.175cm}{${}$\\[-0cm]$0$\\[-0.3cm]}  &  \parbox{3cm}{${}$\\[-0.1cm]$\frac{2(\alpha_{12}-4-3\alpha_{11}-6\alpha_{13})}{3(2\pi)^2}$\\[-0.3cm]} &  \parbox{2.2cm}{${}$\\[-0.1cm]$\frac{16+3\alpha_{11}+12\alpha_{13}}{9(2\pi)^2}$\\[-0.3cm]} & \parbox{0.175cm}{${}$\\[-0.1cm]$\phantom{0}$\\[-0.3cm]}\\\hline
\end{tabular}}
}}
\end{center}
\caption{Coefficient functions $c^{(1)}_{k;i_1,i_2;i_3,i_4}(\rho)$ for $a=1$.}
\label{Tab:Coefsa1}
\end{table}

\noindent
Similarly, for $a=2$, we have four different $\mathcal{K}_2^{k}(\widehat{a}_{1,2},\rho)$ (for which we use the same condensed graphical representation) such that $k\in\{1,2,3,4\}$. The coefficients $c^{(2)}_{k;i_1,i_2;i_3,i_4}(\rho)$ are tabulated in Table~\ref{Tab:Coefsa2}. Finally, for $a=3$, there are four different $\mathcal{K}_3^{k}(\widehat{a}_{1,2},\rho)$ (for which we use the same condensed graphical representation) and therefore $k\in\{1,2,3,4\}$. The coefficients $c^{(3)}_{k;i_1,i_2;i_3,i_4}(\rho)$ are tabulated in Table~\ref{Tab:Coefsa3}. The parameters $\alpha_{1,\ldots,19}$ appearing in eq.~(\ref{CoefsN3a0}) and Tables~\ref{Tab:Coefsa1}, \ref{Tab:Coefsa2} and \ref{Tab:Coefsa3} are undetermined constants and label the ambiguity in the decomposition of~$P^{(r=2)}_{3,(0)}$.

\begin{table}[htbp]
\begin{center}
\parbox{21.2cm}{\begin{tabular}{|c||c||c|c|c|c|}\hline
&&&&&\\[-16pt]
\multirow{3}{*}{{\bf $k$}} & \multirow{3}{*}{$\mathcal{K}_2^{k}(\widehat{a}_{1,2},\rho)$} & $c^{(2)}_{k;6,0;0,0}$ & $c^{(2)}_{k;4,1;0,0}$ & $c^{(2)}_{k;2,2;0,0}$ & $c^{(2)}_{k;0,3;0,0}$ \\[2pt]\cline{3-6}
& & $c^{(2)}_{k;4,0;2,0}$ & $c^{(2)}_{k;3,1;1,0}$ & $c^{(2)}_{k;1,2;1,0}$ & $c^{(2)}_{k;2,1;2,0}$  \\[2pt]\cline{3-6}
& & $c^{(2)}_{k;5,0;1,0}$ & $c^{(2)}_{k;4,0;0,1}$ & $c^{(3)}_{k;2,1;0,1}$ & $c^{(2)}_{k;3,0;1,1}$\\[2pt]  \hline\hline
&&&&&\\[-14pt]
\multirow{3}{*}{{\bf $1$}} & \multirow{3}{*}{\scalebox{1.2}{\parbox{1.4cm}{${}$\\\begin{tikzpicture}
\draw[fill=black] (0,0.75) circle (0.05);
\draw[fill=black] (-0.65,-0.375) circle (0.05);
\draw[fill=black] (0.65,-0.375) circle (0.05);
\draw[thick,dashed]  (-0.65,-0.375) -- coordinate[pos=0.3] (C1) coordinate[pos=0.69] (C2)  (0,0.75);
\draw[thick,dashed]  (0.65,-0.375) -- coordinate[pos=0.5] (B) (0,0.75);
\fill[black] (C1) circle (1.5pt);
\draw[thick,rotate=60] ($ (C1) + (0.1,0.1) $) -- ($ (C1) + (-0.1,-0.1) $);
\draw[thick,rotate=60] ($ (C1) + (0.1,-0.1) $) -- ($ (C1) + (-0.1,0.1) $);
\fill[black] (C2) circle (1.5pt);
\draw[thick,rotate=60] ($ (C2) + (0.1,0.1) $) -- ($ (C2) + (-0.1,-0.1) $);
\draw[thick,rotate=60] ($ (C2) + (0.1,-0.1) $) -- ($ (C2) + (-0.1,0.1) $);
\fill[black] (B) circle (1.5pt);
\draw[thick,rotate=-60] ($ (B) + (0.1,0.1) $) -- ($ (B) + (-0.1,-0.1) $);
\draw[thick,rotate=-60] ($ (B) + (0.1,-0.1) $) -- ($ (B) + (-0.1,0.1) $);
\end{tikzpicture}}}
} & \parbox{1.4cm}{${}$\\[-0cm]$\frac{1-24\alpha_1}{288(2\pi)^{10}}$\\[-0.2cm]} & \parbox{0.175cm}{${}$\\[-0.2cm]$0$\\[-0.2cm]}  & \parbox{0.175cm}{${}$\\[-0.2cm]$0$\\[-0.2cm]} & \parbox{0.175cm}{${}$\\[-0.2cm]$0$\\[-0.2cm]} \\\cline{3-6}
& & \parbox{0.175cm}{${}$\\[-0cm]$0$\\[-0.2cm]} &\parbox{0.175cm}{${}$\\[-0cm]$0$\\[-0.2cm]} & \parbox{0.175cm}{${}$\\[-0cm]$0$\\[-0.2cm]} & \parbox{0.175cm}{${}$\\[-0cm]$0$\\[-0.2cm]}   \\\cline{3-6}
& & \parbox{0.175cm}{${}$\\[-0.2cm]$0$\\[-0.2cm]} & \parbox{0.175cm}{${}$\\[-0.2cm]$0$\\[-0.2cm]}  & \parbox{0.175cm}{${}$\\[-0cm]$0$\\[-0.2cm]} & \parbox{0.175cm}{${}$\\[-0cm]$0$\\[-0.2cm]} \\ \hline\hline
&&&&&\\[-14pt]
\multirow{3}{*}{{\bf $2$}} & \multirow{3}{*}{\scalebox{1.2}{\parbox{1.4cm}{${}$\\\begin{tikzpicture}
\draw[fill=black] (0,0.75) circle (0.05);
\draw[fill=black] (-0.65,-0.375) circle (0.05);
\draw[fill=black] (0.65,-0.375) circle (0.05);
\draw[thick,dashed]  (-0.65,-0.375) -- coordinate[pos=0.5] (C1)  (0,0.75);
\draw[thick,dashed]  (0.65,-0.375) -- coordinate[pos=0.5] (B) (0,0.75);
\fill[black] (C1) circle (1.5pt);
\draw[thick,rotate=60] ($ (C1) + (0.1,0.1) $) -- ($ (C1) + (-0.1,-0.1) $);
\draw[thick,rotate=60] ($ (C1) + (0.1,-0.1) $) -- ($ (C1) + (-0.1,0.1) $);
\fill[black] (B) circle (1.5pt);
\draw[thick,rotate=-60] ($ (B) + (0.1,0.1) $) -- ($ (B) + (-0.1,-0.1) $);
\draw[thick,rotate=-60] ($ (B) + (0.1,-0.1) $) -- ($ (B) + (-0.1,0.1) $);
\end{tikzpicture}}}
} & \parbox{0.175cm}{${}$\\[-0.2cm]$0$\\[-0.2cm]}  & \parbox{2.4cm}{${}$\\[-0cm]$\frac{\frac{25}{216}-\frac{5\alpha_1-2\alpha_2+\alpha_3}{3}}{288(2\pi)^{10}}$\\[-0.3cm]} & \parbox{0.175cm}{${}$\\[-0.2cm]$0$\\[-0.2cm]}  & \parbox{0.175cm}{${}$\\[-0.2cm]$0$\\[-0.2cm]} \\\cline{3-6}
& & \parbox{0.175cm}{${}$\\[-0cm]$0$\\[-0.2cm]} & \parbox{0.175cm}{${}$\\[-0cm]$0$\\[-0.2cm]} & \parbox{0.175cm}{${}$\\[-0cm]$0$\\[-0.2cm]} & \parbox{0.175cm}{${}$\\[-0cm]$0$\\[-0.2cm]}   \\\cline{3-6}
& & \parbox{1.8cm}{${}$\\[-0.1cm]$\frac{45\alpha_1+9\alpha_3-2}{18(2\pi)^{10}}$\\[-0.3cm]}  & \parbox{0.175cm}{${}$\\[-0.2cm]$0$\\[-0.2cm]}  & \parbox{0.175cm}{${}$\\[-0cm]$0$\\[-0.2cm]} & \parbox{0.175cm}{${}$\\[-0cm]$0$\\[-0.2cm]}
\\ \hline\hline
&&&&&\\[-14pt]
\multirow{3}{*}{{\bf $3$}} & \multirow{3}{*}{\scalebox{1.2}{\parbox{1.4cm}{${}$\\\begin{tikzpicture}
\draw[fill=black] (0,0.75) circle (0.05);
\draw[fill=black] (-0.65,-0.375) circle (0.05);
\draw[fill=black] (0.65,-0.375) circle (0.05);
\draw[thick,dashed]  (-0.65,-0.375) -- coordinate[pos=0.5] (C1)  (0,0.75);
\draw[thick,dashed]  (0.65,-0.375) -- coordinate[pos=0.5] (B) (0,0.75);
\fill[black] (C1) circle (1.5pt);
\draw[thick,rotate=60] ($ (C1) + (0.1,0.1) $) -- ($ (C1) + (-0.1,-0.1) $);
\draw[thick,rotate=60] ($ (C1) + (0.1,-0.1) $) -- ($ (C1) + (-0.1,0.1) $);
\end{tikzpicture}}}
} & \parbox{2.6cm}{${}$\\[-0cm]$\frac{(10-288\alpha_1-9\alpha_5)E_4}{864(2\pi)^{6}}$\\[-0.2cm]}  & \parbox{0.175cm}{${}$\\[-0.2cm]$0$\\[-0.2cm]}  & \parbox{3cm}{${}$\\[-0.2cm]$\frac{2(2+48\alpha_2-24\alpha_3+3\alpha_5)}{9(2\pi)^{6}}$\\[-0.2cm]}  & \parbox{0.175cm}{${}$\\[-0.2cm]$0$\\[-0.2cm]} \\\cline{3-6}
& & \parbox{1.2cm}{${}$\\[-0.1cm]$-\frac{12\alpha_3}{(2\pi)^{6}}$\\[-0.3cm]} & \parbox{2.2cm}{${}$\\[-0.1cm]$\frac{2(8\alpha_3-8\alpha_2+\alpha_5)}{(2\pi)^{6}}$\\[-0.3cm]} & \parbox{0.175cm}{${}$\\[-0cm]$0$\\[-0.2cm]} & \parbox{0.175cm}{${}$\\[-0cm]$0$\\[-0.2cm]}   \\\cline{3-6}
& & \parbox{0.175cm}{${}$\\[-0.2cm]$0$\\[-0.2cm]}   &\parbox{0.8cm}{${}$\\[-0.2cm]$\frac{\alpha_5}{(2\pi)^{6}}$\\[-0.2cm]} & \parbox{0.175cm}{${}$\\[-0cm]$0$\\[-0.2cm]} & \parbox{0.175cm}{${}$\\[-0cm]$0$\\[-0.2cm]}
\\ \hline\hline
&&&&&\\[-14pt]
\multirow{3}{*}{{\bf $4$}} & \multirow{3}{*}{\scalebox{1.2}{\parbox{1.4cm}{${}$\\\begin{tikzpicture}
\draw[fill=black] (0,0.75) circle (0.05);
\draw[fill=black] (-0.65,-0.375) circle (0.05);
\draw[fill=black] (0.65,-0.375) circle (0.05);
\draw[thick,dashed]  (-0.65,-0.375) -- coordinate[pos=0.5] (C1)  (0,0.75);
\draw[thick,dashed]  (0.65,-0.375) -- coordinate[pos=0.5] (B) (0,0.75);
\end{tikzpicture}}}
} & \parbox{2cm}{${}$\\[-0cm]$\frac{(11-216\alpha_1)E_6}{864(2\pi)^{4}}$\\[-0.2cm]}  & \parbox{4cm}{${}$\\[-0cm]$\frac{(96\alpha_1+64\alpha_2-32\alpha_3-\alpha_6-4)E_4}{96(2\pi)^{4}}$\\[-0.2cm]}  & \parbox{0.175cm}{${}$\\[-0.2cm]$0$\\[-0.2cm]}  & \parbox{1.3cm}{${}$\\[-0.2cm]$\frac{2(8+3\alpha_6)}{9(2\pi)^{4}}$\\[-0.2cm]} \\\cline{3-6}
& & \parbox{0.175cm}{${}$\\[-0cm]$0$\\[-0.2cm]} &\parbox{0.175cm}{${}$\\[-0cm]$0$\\[-0.2cm]} & \parbox{1.9cm}{${}$\\[-0.1cm]$\frac{2(\alpha_7-3\alpha_6-2)}{3(2\pi)^{6}}$\\[-0.3cm]} & \parbox{1.2cm}{${}$\\[-0.1cm]$-\frac{2\alpha_7}{(2\pi)^{6}}$\\[-0.3cm]}   \\\cline{3-6}
& & \parbox{3.4cm}{${}$\\[-0.1cm]$\frac{(18-144\alpha_1+48 \alpha_3-\alpha_7)E_4}{96(2\pi)^{4}}$\\[-0.3cm]} &\parbox{0.175cm}{${}$\\[-0.2cm]$0$\\[-0.2cm]} & \parbox{0.8cm}{${}$\\[-0.2cm]$\frac{\alpha_6}{(2\pi)^{6}}$\\[-0.3cm]} & \parbox{0.8cm}{${}$\\[-0.2cm]$\frac{\alpha_7}{(2\pi)^{6}}$\\[-0.2cm]}
\\ \hline
\end{tabular}}
\end{center}
\caption{Coefficient functions $c^{(2)}_{k;i_1,i_2;i_3,i_4}(\rho)$ for $a=2$.}
\label{Tab:Coefsa2}
\end{table}



\begin{table}[htbp]
\begin{center}
\begin{tabular}{|c||c||c|c|c|c|c|c|}\hline
&&&&&&&\\[-16pt]
{\bf $k$} & $\mathcal{K}_3^{k}(\widehat{a}_{1,2},\rho)$ & $c^{(3)}_{k;6,0;0,0}$ & $c^{(3)}_{k;4,1;0,0}$ & $c^{(3)}_{k;2,2;0,0}$ & $c^{(3)}_{k;5,0;1,0}$ & $c^{(3)}_{k;3,1;1,0}$ & $c^{(3)}_{k;4,0;0,1}$  \\[2pt] \hline\hline
&&&&&&&\\[-14pt]
$1$ & \scalebox{1.2}{\parbox{1.4cm}{\begin{tikzpicture}
\draw[fill=black] (0,0.75) circle (0.05);
\draw[fill=black] (-0.65,-0.375) circle (0.05);
\draw[fill=black] (0.65,-0.375) circle (0.05);
\draw[thick,dashed]  (0.65,-0.375) -- coordinate[pos=0.5] (A) (0,0.75);
\draw[thick,dashed]  (-0.65,-0.375) -- coordinate[pos=0.5] (B) (0,0.75);
\draw[thick,dashed]  (-0.65,-0.375) -- coordinate[pos=0.3] (C1) coordinate[pos=0.69] (C2) (0.65,-0.375);
\fill[black] (C1) circle (1.5pt);
\draw[thick] ($ (C1) + (0.1,0.1) $) -- ($ (C1) + (-0.1,-0.1) $);
\draw[thick] ($ (C1) + (0.1,-0.1) $) -- ($ (C1) + (-0.1,0.1) $);
\fill[black] (C2) circle (1.5pt);
\draw[thick] ($ (C2) + (0.1,0.1) $) -- ($ (C2) + (-0.1,-0.1) $);
\draw[thick] ($ (C2) + (0.1,-0.1) $) -- ($ (C2) + (-0.1,0.1) $);
\end{tikzpicture}}}
& $\frac{\alpha_1}{(2\pi)^{10}}$ & $0$ & $0$ & $0$ & $0$ & $0$\\[18pt] \hline
&&&&&&&\\[-14pt]
$2$ & \scalebox{1.2}{\parbox{1.4cm}{\begin{tikzpicture}
\draw[fill=black] (0,0.75) circle (0.05);
\draw[fill=black] (-0.65,-0.375) circle (0.05);
\draw[fill=black] (0.65,-0.375) circle (0.05);
\draw[thick,dashed]  (0.65,-0.375) -- coordinate[pos=0.5] (A) (0,0.75);
\draw[thick,dashed]  (-0.65,-0.375) -- coordinate[pos=0.5] (B) (0,0.75);
\draw[thick,dashed]  (-0.65,-0.375) -- coordinate[pos=0.3] (C1) coordinate[pos=0.69] (C2) (0.65,-0.375);
\fill[black] (A) circle (1.5pt);
\draw[thick,rotate=-60] ($ (A) + (0.1,0.1) $) -- ($ (A) + (-0.1,-0.1) $);
\draw[thick,rotate=-60] ($ (A) + (0.1,-0.1) $) -- ($ (A) + (-0.1,0.1) $);
\fill[black] (B) circle (1.5pt);
\draw[thick,rotate=60] ($ (B) + (0.1,0.1) $) -- ($ (B) + (-0.1,-0.1) $);
\draw[thick,rotate=60] ($ (B) + (0.1,-0.1) $) -- ($ (B) + (-0.1,0.1) $);
\end{tikzpicture}}}
& $\frac{45\alpha_1-2}{9(2\pi)^{10}}$ & $0$ & $0$ & $0$ & $0$ & $0$\\[18pt] \hline
&&&&&&&\\[-14pt]
$3$ & \scalebox{1.2}{\parbox{1.4cm}{\begin{tikzpicture}
\draw[fill=black] (0,0.75) circle (0.05);
\draw[fill=black] (-0.65,-0.375) circle (0.05);
\draw[fill=black] (0.65,-0.375) circle (0.05);
\draw[thick,dashed]  (0.65,-0.375) -- coordinate[pos=0.5] (A) (0,0.75);
\draw[thick,dashed]  (-0.65,-0.375) -- coordinate[pos=0.5] (B) (0,0.75);
\draw[thick,dashed]  (-0.65,-0.375) -- coordinate[pos=0.5] (C1) (0.65,-0.375);
\fill[black] (C1) circle (1.5pt);
\draw[thick] ($ (C1) + (0.1,0.1) $) -- ($ (C1) + (-0.1,-0.1) $);
\draw[thick] ($ (C1) + (0.1,-0.1) $) -- ($ (C1) + (-0.1,0.1) $);
\end{tikzpicture}}}
& $0$ & $\frac{8(\alpha_3-2\alpha_2)}{(2\pi)^{8}}$ & $0$ & $-\frac{12\alpha_3}{(2\pi)^8}$ & $0$ & $0$\\[18pt] \hline
&&&&&&&\\[-14pt]
$4$ & \scalebox{1.2}{\parbox{1.4cm}{\begin{tikzpicture}
\draw[fill=black] (0,0.75) circle (0.05);
\draw[fill=black] (-0.65,-0.375) circle (0.05);
\draw[fill=black] (0.65,-0.375) circle (0.05);
\draw[thick,dashed]  (0.65,-0.375) -- coordinate[pos=0.5] (A) (0,0.75);
\draw[thick,dashed]  (-0.65,-0.375) -- coordinate[pos=0.5] (B) (0,0.75);
\draw[thick,dashed]  (-0.65,-0.375) -- coordinate[pos=0.5] (C1) (0.65,-0.375);
\end{tikzpicture}}}
& $\frac{12-96\alpha_1-\alpha_4)}{96(2\pi)^{6}}\,E_4$ & $\frac{2(3\alpha_4-4)}{9(2\pi)^6}$ & $0$ & $0$ & $-\frac{2\alpha_4}{(2\pi)^6}$ & $\frac{\alpha_4}{(2\pi)^6}$\\[18pt] \hline
\end{tabular}
\end{center}
\caption{Coefficient functions $c^{(3)}_{k;i_1,i_2;i_3,i_4}(\rho)$ for $a=3$.}
\label{Tab:Coefsa3}
\end{table}

Despite this structural ambiguity, we see that (similar to the case of $N=2$), a decomposition is possible which lends itself to an interpretation in terms of $n$-point functions (with $n\leq 6$) where the external states are given by $\buildH{r}{0}$ and $\buildW{r}{0}$ for $r\in\{1,2\}$. Determining precise coupling functions, however, is very difficult, due to the many undetermined parameters $\alpha_i$. We remark, however, that the fact that all of the $\mathcal{K}_a^{k=1}(\widehat{a}_{1,2},\rho)$ can again be written entirely as combinations of 'decorated' scalar Greens functions, is in line with the interpretation we have given in the case of $N=2$. In particular, just as before, these 'decorations' again consist of differential operators $D_{\widehat{a}}^{2}$ or of multiplication with holomorphic Eisenstein series that can be interpreted as dihedral modular graph forms $\mathcal{C}\big[{}^{2k\,0}_{\,\,0\,\,0}\big](\rho)$ in (\ref{ModGraphFormHol}) and can thus be interpreted as disconnected contributions.
\section{Example $N=4$}\label{Sect:ExN4}
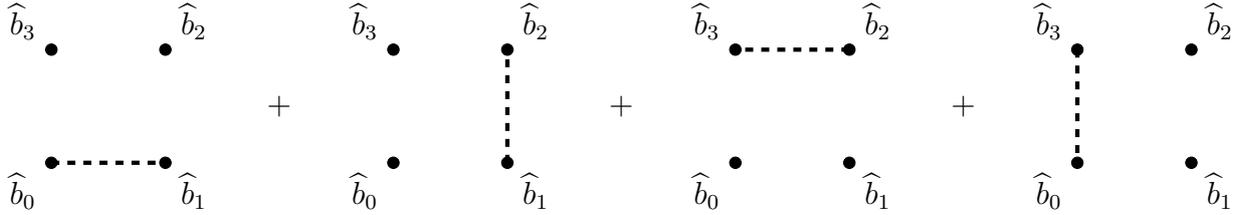
\begin{figure}[htbp]
\begin{center}
\scalebox{1}{\parbox{16.2cm}{\begin{tikzpicture}[scale = 1.50]
\draw[fill=black] (0,0) circle (0.05);
\draw[fill=black] (1,0) circle (0.05);
\draw[fill=black] (0,1) circle (0.05);
\draw[fill=black] (1,1) circle (0.05);
\draw[ultra thick,dashed]  (0,0) -- (1,0);
\node at (-0.25,-0.25) {$\widehat{b}_0$};
\node at (1.25,-0.25) {$\widehat{b}_1$};
\node at (1.25,1.25) {$\widehat{b}_2$};
\node at (-0.25,1.25) {$\widehat{b}_3$};
\node at (2,0.5) {$+$};
\begin{scope}[xshift=3cm]
\draw[fill=black] (0,0) circle (0.05);
\draw[fill=black] (1,0) circle (0.05);
\draw[fill=black] (0,1) circle (0.05);
\draw[fill=black] (1,1) circle (0.05);
\draw[ultra thick,dashed]  (1,0) -- (1,1);
\node at (-0.25,-0.25) {$\widehat{b}_0$};
\node at (1.25,-0.25) {$\widehat{b}_1$};
\node at (1.25,1.25) {$\widehat{b}_2$};
\node at (-0.25,1.25) {$\widehat{b}_3$};
\end{scope}
\node at (5,0.5) {$+$};
\begin{scope}[xshift=6cm]
\draw[fill=black] (0,0) circle (0.05);
\draw[fill=black] (1,0) circle (0.05);
\draw[fill=black] (0,1) circle (0.05);
\draw[fill=black] (1,1) circle (0.05);
\draw[ultra thick,dashed]  (1,1) -- (0,1);
\node at (-0.25,-0.25) {$\widehat{b}_0$};
\node at (1.25,-0.25) {$\widehat{b}_1$};
\node at (1.25,1.25) {$\widehat{b}_2$};
\node at (-0.25,1.25) {$\widehat{b}_3$};
\end{scope}
\node at (8,0.5) {$+$};
\begin{scope}[xshift=9cm]
\draw[fill=black] (0,0) circle (0.05);
\draw[fill=black] (1,0) circle (0.05);
\draw[fill=black] (0,1) circle (0.05);
\draw[fill=black] (1,1) circle (0.05);
\draw[ultra thick,dashed]  (0,1) -- (0,0);
\node at (-0.25,-0.25) {$\widehat{b}_0$};
\node at (1.25,-0.25) {$\widehat{b}_1$};
\node at (1.25,1.25) {$\widehat{b}_2$};
\node at (-0.25,1.25) {$\widehat{b}_3$};
\end{scope}
\end{tikzpicture}}}
\caption{\sl Diagrammatical representation of the coupling $\coup{4}{1}$ in (\ref{CouplingsN4012}). The dashed line represent the two-point function $\mathbb{G}''(\widehat{b}_\ell-\widehat{b}_j)+\frac{2\pi i}{\rho-\bar{\rho}}$.}
\label{Fig:N4CouplingO1}
\end{center}
\end{figure}

\begin{figure}[htbp]
\begin{center}
\scalebox{1}{\parbox{16.2cm}{\begin{tikzpicture}[scale = 1.50]
\draw[fill=black] (0,0) circle (0.05);
\draw[fill=black] (1,0) circle (0.05);
\draw[fill=black] (0,1) circle (0.05);
\draw[fill=black] (1,1) circle (0.05);
\draw[ultra thick,dashed]  (0,0) -- (1,0);
\draw[ultra thick,dashed]  (0,0) -- (1,1);
\node at (-0.25,-0.25) {$\widehat{b}_0$};
\node at (1.25,-0.25) {$\widehat{b}_1$};
\node at (1.25,1.25) {$\widehat{b}_2$};
\node at (-0.25,1.25) {$\widehat{b}_3$};
\node at (2.5,0.5) {$+$};
\begin{scope}[xshift=4cm]
\draw[fill=black] (0,0) circle (0.05);
\draw[fill=black] (1,0) circle (0.05);
\draw[fill=black] (0,1) circle (0.05);
\draw[fill=black] (1,1) circle (0.05);
\draw[ultra thick,dashed]  (0,0) -- (1,0);
\draw[ultra thick,dashed]  (0,0) -- (0,1);
\node at (-0.25,-0.25) {$\widehat{b}_0$};
\node at (1.25,-0.25) {$\widehat{b}_1$};
\node at (1.25,1.25) {$\widehat{b}_2$};
\node at (-0.25,1.25) {$\widehat{b}_3$};
\end{scope}
\node at (6.5,0.5) {$+$};
\begin{scope}[xshift=8cm]
\draw[fill=black] (0,0) circle (0.05);
\draw[fill=black] (1,0) circle (0.05);
\draw[fill=black] (0,1) circle (0.05);
\draw[fill=black] (1,1) circle (0.05);
\draw[ultra thick,dashed]  (0,0) -- (1,1);
\draw[ultra thick,dashed]  (0,0) -- (0,1);
\node at (-0.25,-0.25) {$\widehat{b}_0$};
\node at (1.25,-0.25) {$\widehat{b}_1$};
\node at (1.25,1.25) {$\widehat{b}_2$};
\node at (-0.25,1.25) {$\widehat{b}_3$};
\end{scope}
\node at (-1.25,-2.5) {$+$};
\begin{scope}[yshift=-3cm]
\draw[fill=black] (0,0) circle (0.05);
\draw[fill=black] (1,0) circle (0.05);
\draw[fill=black] (0,1) circle (0.05);
\draw[fill=black] (1,1) circle (0.05);
\draw[ultra thick,dashed]  (0,0) -- (1,0);
\draw[ultra thick,dashed]  (1,0) -- (0,1);
\node at (-0.25,-0.25) {$\widehat{b}_0$};
\node at (1.25,-0.25) {$\widehat{b}_1$};
\node at (1.25,1.25) {$\widehat{b}_2$};
\node at (-0.25,1.25) {$\widehat{b}_3$};
\node at (2.5,0.5) {$+$};
\begin{scope}[xshift=4cm]
\draw[fill=black] (0,0) circle (0.05);
\draw[fill=black] (1,0) circle (0.05);
\draw[fill=black] (0,1) circle (0.05);
\draw[fill=black] (1,1) circle (0.05);
\draw[ultra thick,dashed]  (1,0) -- (0,0);
\draw[ultra thick,dashed]  (1,0) -- (1,1);
\node at (-0.25,-0.25) {$\widehat{b}_0$};
\node at (1.25,-0.25) {$\widehat{b}_1$};
\node at (1.25,1.25) {$\widehat{b}_2$};
\node at (-0.25,1.25) {$\widehat{b}_3$};
\end{scope}
\node at (6.5,0.5) {$+$};
\begin{scope}[xshift=8cm]
\draw[fill=black] (0,0) circle (0.05);
\draw[fill=black] (1,0) circle (0.05);
\draw[fill=black] (0,1) circle (0.05);
\draw[fill=black] (1,1) circle (0.05);
\draw[ultra thick,dashed]  (1,0) -- (1,1);
\draw[ultra thick,dashed]  (1,0) -- (0,1);
\node at (-0.25,-0.25) {$\widehat{b}_0$};
\node at (1.25,-0.25) {$\widehat{b}_1$};
\node at (1.25,1.25) {$\widehat{b}_2$};
\node at (-0.25,1.25) {$\widehat{b}_3$};
\end{scope}

\end{scope}
\node at (-1.25,-5.5) {$+$};
\begin{scope}[yshift=-6cm]
\draw[fill=black] (0,0) circle (0.05);
\draw[fill=black] (1,0) circle (0.05);
\draw[fill=black] (0,1) circle (0.05);
\draw[fill=black] (1,1) circle (0.05);
\draw[ultra thick,dashed]  (1,1) -- (0,1);
\draw[ultra thick,dashed]  (1,1) -- (0,0);
\node at (-0.25,-0.25) {$\widehat{b}_0$};
\node at (1.25,-0.25) {$\widehat{b}_1$};
\node at (1.25,1.25) {$\widehat{b}_2$};
\node at (-0.25,1.25) {$\widehat{b}_3$};
\node at (2.5,0.5) {$+$};
\begin{scope}[xshift=4cm]
\draw[fill=black] (0,0) circle (0.05);
\draw[fill=black] (1,0) circle (0.05);
\draw[fill=black] (0,1) circle (0.05);
\draw[fill=black] (1,1) circle (0.05);
\draw[ultra thick,dashed]  (1,1) -- (0,1);
\draw[ultra thick,dashed]  (1,1) -- (1,0);
\node at (-0.25,-0.25) {$\widehat{b}_0$};
\node at (1.25,-0.25) {$\widehat{b}_1$};
\node at (1.25,1.25) {$\widehat{b}_2$};
\node at (-0.25,1.25) {$\widehat{b}_3$};
\end{scope}
\node at (6.5,0.5) {$+$};
\begin{scope}[xshift=8cm]
\draw[fill=black] (0,0) circle (0.05);
\draw[fill=black] (1,0) circle (0.05);
\draw[fill=black] (0,1) circle (0.05);
\draw[fill=black] (1,1) circle (0.05);
\draw[ultra thick,dashed]  (1,1) -- (0,0);
\draw[ultra thick,dashed]  (1,1) -- (1,0);
\node at (-0.25,-0.25) {$\widehat{b}_0$};
\node at (1.25,-0.25) {$\widehat{b}_1$};
\node at (1.25,1.25) {$\widehat{b}_2$};
\node at (-0.25,1.25) {$\widehat{b}_3$};
\end{scope}

\end{scope}

\node at (-1.25,-8.5) {$+$};
\begin{scope}[yshift=-9cm]
\draw[fill=black] (0,0) circle (0.05);
\draw[fill=black] (1,0) circle (0.05);
\draw[fill=black] (0,1) circle (0.05);
\draw[fill=black] (1,1) circle (0.05);
\draw[ultra thick,dashed]  (0,1) -- (0,0);
\draw[ultra thick,dashed]  (0,1) -- (1,0);
\node at (-0.25,-0.25) {$\widehat{b}_0$};
\node at (1.25,-0.25) {$\widehat{b}_1$};
\node at (1.25,1.25) {$\widehat{b}_2$};
\node at (-0.25,1.25) {$\widehat{b}_3$};
\node at (2.5,0.5) {$+$};
\begin{scope}[xshift=4cm]
\draw[fill=black] (0,0) circle (0.05);
\draw[fill=black] (1,0) circle (0.05);
\draw[fill=black] (0,1) circle (0.05);
\draw[fill=black] (1,1) circle (0.05);
\draw[ultra thick,dashed]  (0,1) -- (0,0);
\draw[ultra thick,dashed]  (0,1) -- (1,1);
\node at (-0.25,-0.25) {$\widehat{b}_0$};
\node at (1.25,-0.25) {$\widehat{b}_1$};
\node at (1.25,1.25) {$\widehat{b}_2$};
\node at (-0.25,1.25) {$\widehat{b}_3$};
\end{scope}
\node at (6.5,0.5) {$+$};
\begin{scope}[xshift=8cm]
\draw[fill=black] (0,0) circle (0.05);
\draw[fill=black] (1,0) circle (0.05);
\draw[fill=black] (0,1) circle (0.05);
\draw[fill=black] (1,1) circle (0.05);
\draw[ultra thick,dashed]  (0,1) -- (1,1);
\draw[ultra thick,dashed]  (0,1) -- (1,0);
\node at (-0.25,-0.25) {$\widehat{b}_0$};
\node at (1.25,-0.25) {$\widehat{b}_1$};
\node at (1.25,1.25) {$\widehat{b}_2$};
\node at (-0.25,1.25) {$\widehat{b}_3$};
\end{scope}

\end{scope}

\end{tikzpicture}}}
\caption{\sl Diagrammatical representation of the coupling $\coup{4}{2}$ in (\ref{CouplingsN4012}). The dashed lines represent the two-point function $\mathbb{G}''(\widehat{b}_\ell-\widehat{b}_j)+\frac{2\pi i}{\rho-\bar{\rho}}$.}
\label{Fig:N4CouplingO2}
\end{center}
\end{figure}
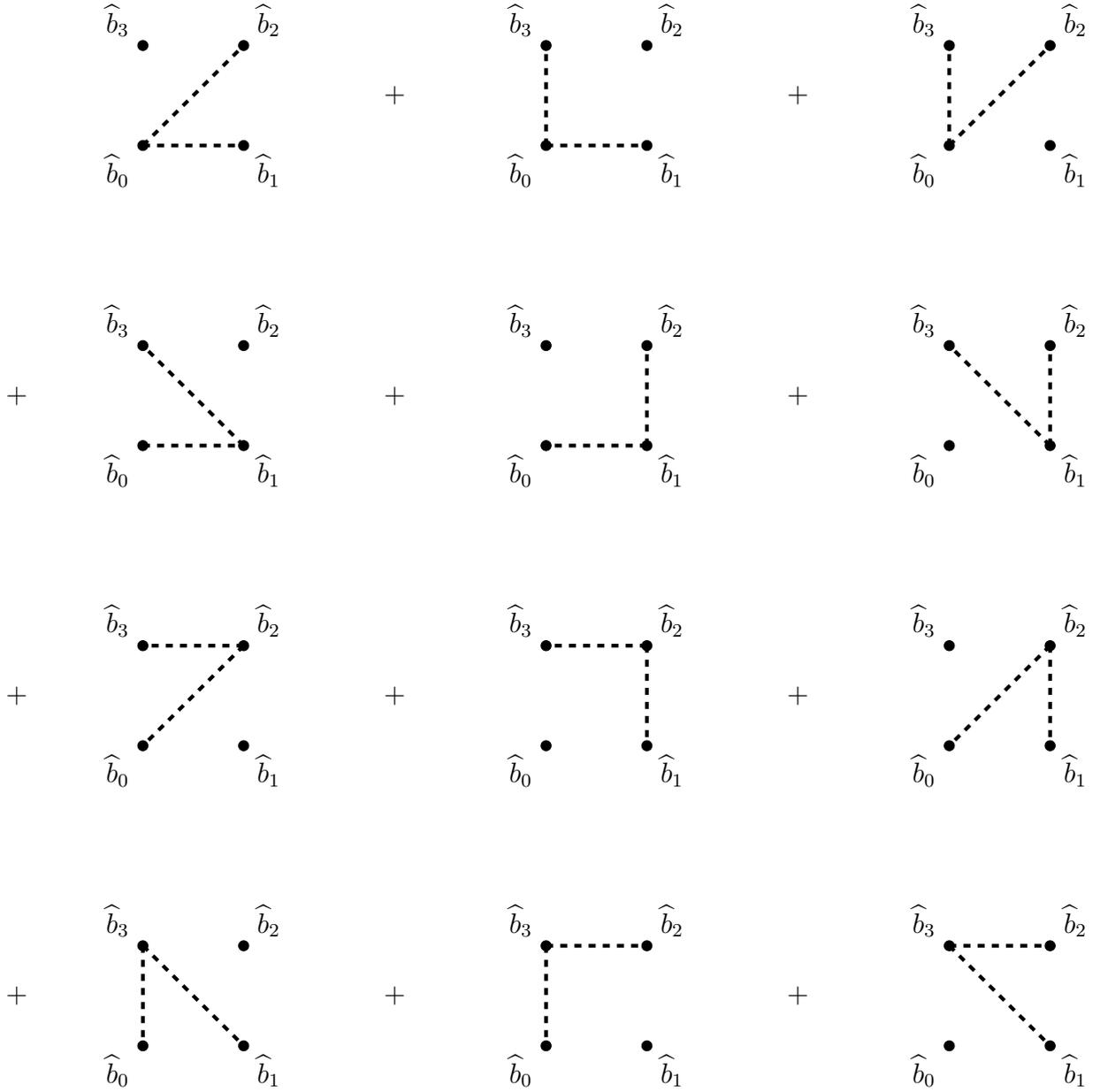

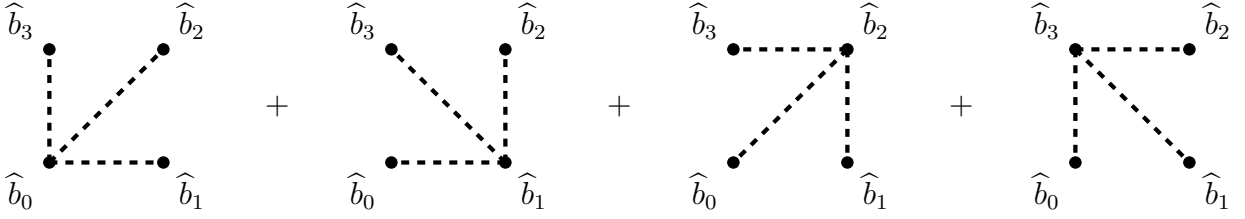
\begin{figure}[htbp]
\begin{center}
\scalebox{1}{\parbox{16.2cm}{\begin{tikzpicture}[scale = 1.50]
\draw[fill=black] (0,0) circle (0.05);
\draw[fill=black] (1,0) circle (0.05);
\draw[fill=black] (0,1) circle (0.05);
\draw[fill=black] (1,1) circle (0.05);
\draw[ultra thick,dashed]  (0,0) -- (1,0);
\draw[ultra thick,dashed]  (0,0) -- (1,1);
\draw[ultra thick,dashed]  (0,0) -- (0,1);
\node at (-0.25,-0.25) {$\widehat{b}_0$};
\node at (1.25,-0.25) {$\widehat{b}_1$};
\node at (1.25,1.25) {$\widehat{b}_2$};
\node at (-0.25,1.25) {$\widehat{b}_3$};
\node at (2,0.5) {$+$};
\begin{scope}[xshift=3cm]
\draw[fill=black] (0,0) circle (0.05);
\draw[fill=black] (1,0) circle (0.05);
\draw[fill=black] (0,1) circle (0.05);
\draw[fill=black] (1,1) circle (0.05);
\draw[ultra thick,dashed]  (1,0) -- (0,0);
\draw[ultra thick,dashed]  (1,0) -- (0,1);
\draw[ultra thick,dashed]  (1,0) -- (1,1);
\node at (-0.25,-0.25) {$\widehat{b}_0$};
\node at (1.25,-0.25) {$\widehat{b}_1$};
\node at (1.25,1.25) {$\widehat{b}_2$};
\node at (-0.25,1.25) {$\widehat{b}_3$};
\end{scope}
\node at (5,0.5) {$+$};
\begin{scope}[xshift=6cm]
\draw[fill=black] (0,0) circle (0.05);
\draw[fill=black] (1,0) circle (0.05);
\draw[fill=black] (0,1) circle (0.05);
\draw[fill=black] (1,1) circle (0.05);
\draw[ultra thick,dashed]  (1,1) -- (0,0);
\draw[ultra thick,dashed]  (1,1) -- (0,1);
\draw[ultra thick,dashed]  (1,1) -- (1,0);
\node at (-0.25,-0.25) {$\widehat{b}_0$};
\node at (1.25,-0.25) {$\widehat{b}_1$};
\node at (1.25,1.25) {$\widehat{b}_2$};
\node at (-0.25,1.25) {$\widehat{b}_3$};
\end{scope}
\node at (8,0.5) {$+$};
\begin{scope}[xshift=9cm]
\draw[fill=black] (0,0) circle (0.05);
\draw[fill=black] (1,0) circle (0.05);
\draw[fill=black] (0,1) circle (0.05);
\draw[fill=black] (1,1) circle (0.05);
\draw[ultra thick,dashed]  (0,1) -- (0,0);
\draw[ultra thick,dashed]  (0,1) -- (1,0);
\draw[ultra thick,dashed]  (0,1) -- (1,1);
\node at (-0.25,-0.25) {$\widehat{b}_0$};
\node at (1.25,-0.25) {$\widehat{b}_1$};
\node at (1.25,1.25) {$\widehat{b}_2$};
\node at (-0.25,1.25) {$\widehat{b}_3$};
\end{scope}
\end{tikzpicture}}}
\caption{\sl Diagrammatical representation of the proposed coupling $\coup{4}{3}$ in (\ref{CouplingsN43}). The dashed line represent the two-point function $\mathbb{G}''(\widehat{b}_\ell-\widehat{b}_j)+\frac{2\pi i}{\rho-\bar{\rho}}$.}
\label{Fig:N4CouplingO3}
\end{center}
\end{figure}

Repeating the above discussion for the case $N=4$ is much more difficult due to the increased complexity of the free energy even for $r=1$. Nevertheless we can report some non-trivial results to order $\mathcal{O}(Q_R)$, which are in line with the general picture advocated above: as was argued in \cite{Hohenegger:2019tii} to this order the free energy can be decomposed as
\begin{align}
P_{4,(s)}^{(r=1)}(\widehat{a}_{1,2,3},\rho,S)&=\buildH{1}{s}\,\left(\buildW{1}{s}\right)^3\,\coup{4}{0}+\left(\buildH{1}{s}\right)^2\,\left(\buildW{1}{s}\right)^2\,\coup{4}{1}(\widehat{a}_{1,2,3},\rho)\nonumber\\
&\hspace{0.5cm}+\left(\buildH{1}{s}\right)^3\,\buildW{1}{s} \,\coup{4}{2}(\widehat{a}_{1,2,3},\rho)+\left(\buildH{1}{s}\right)^4 \,\coup{4}{3}(\widehat{a}_{1,2,3},\rho)\,.
\end{align}
Implicit expressions for the couplings $\coup{4}{\alpha}$ for $\alpha=0,1,2,3$ as infinite series can be inferred from the expansions of the free energy presented in \cite{Bastian:2019wpx}. These series representations can be analysed with the same methods outlined in appendix~\ref{SeriesExpansionSimCoup}. For $\alpha=0,1,2$ we have checked up to order $\mathcal{O}(\widehat{a}_{1,2,3}^6)$ that they are compatible with 
\begin{align}
\coup{4}{0}&=4\,,\nonumber\\
\coup{4}{1}(\widehat{a}_{1,2,3},\rho)&=\frac{1}{(2\pi)^2}\sum_{\ell=1}^4\sum_{j\neq \ell}\left(\mathbb{G}''(\widehat{b}_\ell-\widehat{b}_j)+\frac{2\pi i}{\rho-\bar{\rho}}\right)\,,\nonumber\\
\coup{4}{2}(\widehat{a}_{1,2,3},\rho)&=\frac{1}{(2\pi)^4}\sum_{\ell=1}^4\sum_{{j_1\neq j_2\atop j_1\neq\ell\neq j_2}}\left(\mathbb{G}''(\widehat{b}_\ell-\widehat{b}_{j_1})+\frac{2\pi i}{\rho-\bar{\rho}}\right)\,\left(\mathbb{G}''(\widehat{b}_\ell-\widehat{b}_{j_2})+\frac{2\pi i}{\rho-\bar{\rho}}\right)\,,\label{CouplingsN4012}
\end{align}
where we have introduced the points
\begin{align}
&\widehat{b}_0=0\,,&&\widehat{b}_1=\widehat{a}_1\,,&&\widehat{b}_2=\widehat{a}_1+\widehat{a}_2\,,&&\widehat{b}_3=\widehat{a}_1+\widehat{a}_2+\widehat{a}_3\,.
\end{align}
For $\coup{4}{3}$, it was remarked in \cite{Bastian:2019wpx} that the coefficients for the free energy for the case $N=4$ have been matched up to the maximal order $\mathcal{O}(Q_\rho^5)$, but may receive additional corrections beyond that. This order allows us to unambiguously only determine the leading singularity of $\coup{4}{3}$, for which we find after a lengthy computation\footnote{$\coup{4}{3}$ receives contributions from all coefficients listed in \cite{Bastian:2019wpx}, which involve up to 3 infinite series with various powers of $Q_{\widehat{a}_{1,2,3}}$. We refrain from presenting the details of this calculation.}
\begin{align}
\coup{4}{3}(\widehat{a}_{1,2,3},\rho)=\frac{1}{(2\pi)^6}\bigg[&\frac{1}{\widehat{a}_1^2\,(\widehat{a}_1+\widehat{a}_2)^2\,(\widehat{a}_1+\widehat{a}_2+\widehat{a}_3)^2}+\frac{1}{\widehat{a}_1^2\,\widehat{a}_2^2\,(\widehat{a}_2+\widehat{a}_3)^2}\nonumber\\
&+\frac{1}{\widehat{a}_2^2\,\widehat{a}_3^2\,(\widehat{a}_1+\widehat{a}_2)^2}+\frac{1}{\widehat{a}_3^2\,(\widehat{a}_2+\widehat{a}_3)^2\,(\widehat{a}_1+\widehat{a}_2+\widehat{a}_3)^2}\bigg]+\mathcal{O}(\widehat{a}_{1,2,3}^{-4})\,.
\end{align}
This pole-structure is compatible with the closed form expression
\begin{align}
\coup{4}{3}(\widehat{a}_{1,2,3},\rho)&=\frac{1}{(2\pi)^6}\sum_{\ell=1}^4\sum_{{j_1< j_2<j_3\atop j_{1,2,3}\neq\ell}}\prod_{a=1}^3\left(\mathbb{G}''(\widehat{b}_\ell-\widehat{b}_{j_a})+\frac{2\pi i}{\rho-\bar{\rho}}\right)\,.\label{CouplingsN43}
\end{align}
The form of the couplings $\coup{4}{\alpha}$ in (\ref{CouplingsN4012}) and (\ref{CouplingsN43}) is compatible with the general form (\ref{StructureCouplingR1}). The (non-trivial) couplings $\coup{4}{1,2,3}$ can also be graphically presented: up to numerical factors $\coup{4}{1}$ in \figref{Fig:N4CouplingO1} corresponds to all possible ways to contract two out of the 4 points $\widehat{b}_{0,1,2,3}$ with a single two-point function $\mathbb{G}''(\widehat{b}_\ell-\widehat{b}_j)+\frac{2\pi i}{\rho-\bar{\rho}}$. The coupling $\coup{4}{2}$ in \figref{Fig:N4CouplingO2} corresponds to all possible ways to contract one out of the 4 points $\widehat{b}_{0,1,2,3}$ with two distinct other points through one two-point function respectively. Finally, the coupling $\coup{4}{3}$ in \figref{Fig:N4CouplingO3} corresponds to all possible ways to contract one of the 4 points with all other points through two-point functions.

\noindent 
Due to the high complexity and the large amount of intrinsic ambiguity, we refrain from discussing a decomposition of the $N=4$ free energy to higher orders in $\mathcal{O}(Q_R)$. We leave this problem for further work.
\section{Conclusions}\label{Sect:Conclusions}
In this paper we have continued and extended the study of \cite{Hohenegger:2019tii} to decompose the free energy of LSTs of type $A_N$. In contrast to \cite{Hohenegger:2019tii}, we have considered the full free energy that counts all BPS states (including multi particle states), which does not impact the results of \cite{Hohenegger:2019tii} to leading instanton order (from the perspective of the low energy $U(N)$ gauge theory), but streamlines some of the results to higher order in $Q_R$. We have studied the examples $N=2,3$ and $4$ which exhibit clear repeating patterns, that we thus conjecture to hold in general: to leading order $\mathcal{O}(Q_R)$, it was already argued in \cite{Hohenegger:2019tii} that the free energy can be presented in a way that resembles a Feynman diagrammatic expansion, which is schematically shown in~\figref{Fig:FeynmanR1}. The external states are given by the building blocks $\buildH{1}{s}$ and $\buildW{1}{0}$, which are the expansion coefficients of the free energy for $N=1$ as well as the leading term in the expansion of the quasi-Jacobi form $W(\rho,S,\epsilon_1)$ defined in (\ref{DefWFunct}). In the current paper, we have analysed in detail the effective couplings $\coup{N}{\alpha}$ (for $N=2,3$ and $4$) appearing in this decomposition and have provided evidence that they can be entirely written as combinations of (derivatives of the) scalar two-point function of a free scalar field $\phi$ on the torus with the general conjectured form given in (\ref{StructureCouplingR1}). The latter can in fact be also written as a correlation function in the following form 
\begin{align}
\coup{N}{\alpha}(\widehat{a}_{1,\ldots,N-1},\rho)&=\frac{(-1)^\alpha}{(2\pi)^{2\alpha}\alpha!}\sum_{\ell=0}^{N-1}\sum_{{\mathcal{S}\subset\{0,\ldots,N-1\}\setminus\{\ell\}}\atop {|\mathcal{S}|=\alpha}}\left\langle :(\partial\phi)^\alpha\,(\widehat{b}_\ell):\,:\prod_{j\in \mathcal{S}}\partial\phi(\widehat{b}_j):\right\rangle\nonumber\\
&=\sum_{\ell=0}^{N-1}\sum_{{\mathcal{S}\subset\{0,\ldots,N-1\}\setminus\{\ell\}}\atop {|\mathcal{S}|=\alpha}}\left\langle :\text{exp}\left(-\frac{\lambda}{(2\pi)^2}\,\partial\phi\,(\widehat{b}_\ell)\right):\,:\prod_{j\in \mathcal{S}}\partial\phi(\widehat{b}_j):\right\rangle\bigg|_{\lambda^{\alpha}}\,.\label{CorrelatorRewriteR1}
\end{align}
In the last line we have introduced the counting parameter $\lambda$ and it is understood to extract the coefficient of $\lambda^\alpha$. Furthermore $:\ldots:$ denotes normal ordering to prevent self-contractions.

We also note that the form (\ref{StructureCouplingR1}) (or equivalently (\ref{CorrelatorRewriteR1})) also implies a recursive structure, that allows to obtain $\coup{N}{\alpha}(\widehat{a}_{1,\ldots,N-1},\rho)$ from $\coup{N+1}{\alpha+1}(\widehat{a}_{1,\ldots,N},\rho)$ by contour integration of~$\widehat{a}_{N}$ 
\begin{align}
\coup{N}{\alpha}(\widehat{a}_{1,\ldots,N-1},\rho)=-i\pi \oint_{\mathcal{C}}d\widehat{a}_N\,\left(\sum_{i=0}^{N-2}(\widehat{a}_N-p_i)\right)\,\coup{N+1}{\alpha+1}(\widehat{a}_{1,\ldots,N},\rho)\,.\label{ContourPrescript}
\end{align}
Here it is understood that $\widehat{a}_{1,\ldots,N}$ and $\rho$ are independent variables. Furthermore, the details of the choice of the contour $\mathcal{C}$ are not important, except that it encircles all $N-1$ poles of $\widehat{a}_N$
\begin{align}
&p_0=0\,,&&\text{and} &&p_i=-\sum_{j=i}^{N-2}\widehat{a}_j\hspace{0.5cm}\forall i=1,\ldots,N-2\,.
\end{align}
The contour integration in (\ref{ContourPrescript}) is designed to extract the pole of a single scalar two-point function in the decomposition of $\coup{N+1}{\alpha+1}(\widehat{a}_{1,\ldots,N},\rho)$, while the additional factor of $\tfrac{1}{2}$ takes into account that for a given pole $p_i$, there are exactly two diagrams of the type \figref{Fig:ScalarGenCorrelatorR1} that provide such a singularity.

Finally, to orders $\mathcal{O}(Q_R^r)$ for $r>1$, we have found that the free energy still affords a decomposition in the basic building blocks $\buildH{r}{s}$ and $\buildW{r}{s}$. The tentative coupling functions are again composed of combinations of second derivatives of the scalar Greens function on the torus, which, however, can be decorated in two different ways: either through the action of a differential operator $D_{\widehat{a}_i}^{2n}$ (for $n\in\mathbb{N}\cup\{0\}$) or through multiplication with combinations of holomorphic Eisenstein series. We have argued that the latter precisely correspond to (dihedral) modular graph forms with bivalent vertices as defined in \cite{DHoker:2015wxz} and thus lend themselves to be interpreted as disconnected contributions from a Feynman diagrammatic point of view. This suggests that also higher instanton orders allow a decomposition in terms of graphs. However, an inherent ambiguity in the decomposition of the free energy (due to the fact that $\buildH{r}{s}$ and $\buildW{r}{s}$ are an overcomplete basis) prevents us from making this statement more precise: in the future it will therefore be important to get a better understanding of the origin of this diagrammatic expansion.

From a physical perspective, the observations made in this paper suggest that non-perturbative information about the LSTs of A-type can be obtained using simple, purely perturbative ingredients, namely two-point correlation functions of a free scalar field on the torus and the BPS counting function of a single M5-brane on a torus. It would be interesting to analyse if similar statements can also be made for other LSTs and/or supersymmetric gauge theories. Similarly, it would be interesting to extend the current study to quantities other than the free energy. Yet another question is to understand if the decomposition of the free energy introduced in \cite{Hohenegger:2019tii} and further elaborated in the current paper reveals new symmetries of the underlying LSTs, which might for example be linked to the conformal symmetry of the scalar two-point function we have encountered in the effective couplings $\coup{N}{\alpha}$. We leave these questions for future work.
\section*{Acknowledgements}
I am deeply indebted to Oliver Schlotterer, for a carefully reading of the draft and many valuable comments and suggestions. I would also like to warmly thank Pierre Vanhove for several interesting exchanges on modular graph functions as well as Amer Iqbal for many inspiring discussions and collaboration on related topics. 
\appendix
\section{Modular Toolkit}
Throughout this paper we are using various different modular objects. This appendix serves to define them and present various different of their properties along with other concepts, which are useful for the discussion in the main body of this article.
\subsection{Jacobi Forms and Eisenstein Series}\label{App:JacobiForms}
Many of the objects discussed in this paper are (related to) \emph{Jacobi forms}: Let $\mathbb{H}$ be the upper half-plane. Following \cite{EichlerZagier}, a weak Jacobi form $\phi(\rho,z)$ of index $m\in\mathbb{Z}$ and weight $w$ of $SL(2,\mathbb{Z})$ is a holomorphic function $\phi:\,\mathbb{H}\times \mathbb{C}\rightarrow\mathbb{C}$ that satisfies the following properties
\begin{align}
&\phi\left(\frac{a\rho+b}{c\rho+d},\frac{z}{c\rho+d}\right)=(c\rho+d)^{w}\,e^{\frac{2\pi i mc\,z^2}{c\rho+d}}\,\phi(\rho,z)\,,&&\forall\,\left(\begin{array}{cc}a & b \\ c & d\end{array}\right)\in SL(2,\mathbb{Z})\,,\nonumber\\
&\phi(\rho,z+k_1\,\rho+k_2)=e^{-2\pi im(k_1^2\rho+2 k_1z)}\,\phi(\rho,z)\,,&&\forall k_{1,2}\in\mathbb{N}\,,
\end{align}
and which affords a Fourier series 
\begin{align}
&\phi(z,\rho)=\sum_{n=0}^\infty\sum_{k\in\mathbb{Z}}c(n,k)\,Q_\rho^n\,e^{2\pi iz k}\,,&&\text{with} &&c(n,k)=(-1)^w\,c(n,-k)\,.
\end{align}
Further symmetries of Jacobi forms follow from the fact that two Fourier coefficients $c(n,\ell)$ and $c(n',\ell')$ (for fixed $n,n'$ and $\ell,\ell'$) are identical if
\begin{align}
&\ell^2-4mn=(\ell')^2-4mn'\,,&&\text{and} &&\ell=\ell'(\text{mod }2m)\,.
\end{align}
The Jacobi forms we shall encounter in the main body of this article can be written as homogeneous polynomials (both with regards to their index as well as weight) of the following two standard Jacobi forms
\begin{align}
&\phi_{-2,1}(\rho,z)=\frac{\theta_1^2(z,\rho)}{\eta^6(\rho)}\,,&&\text{and} &&\phi_{0,1}(\rho,z)=8\sum_{i=2}^4\left(\frac{\theta_i(z,\rho)}{\theta_i(0,\rho)}\right)^2\,,\label{DefStandardJacobiForms}
\end{align}
of weight and index $(-2,1)$ and $(0,1)$ respectively, where $\theta_{i}(\rho,z)$ (for $i=1,2,3,4$) are the Jacobi theta functions and $\eta$ is the Dedekind eta-function. Furthermore, the coefficients of these polynomials in $\phi_{-2,1}$ and $\phi_{0,1}$ themselves are homogeneous polynomials (with regards to their weight) of (holomorphic) Eisenstein series \cite{EichlerZagier,Lang,Stein}
\begin{align}
E_{2n}(\rho)=1-\frac{4n}{B_{2n}}\sum_{k=1}^\infty \sigma_{2n-1}(k)\,Q_\rho^k\,,\label{DefEisenstein}
\end{align}
where $B_{2n}$ are the Bernoulli numbers and $\sigma_n(k)$ is the divisor sigma function. In certain cases, we shall also use the notation $G_{2n}(\rho)=2\zeta(2n) E_{2n}(\rho)$. The $E_{2n}$ for $n>1$ are holomorphic modular forms of weight $2n$, while $E_{2}$ transforms in the following fashion under $SL(2,\mathbb{Z})$
\begin{align}
&E_2\left(\frac{a\rho+b}{c\rho+d}\right)=(c\rho+d)^2\,E_2(\rho)-\frac{3}{\pi}\,\frac{c}{\text{Im}(\rho)}\,,&&\forall\,\left(\begin{array}{cc}a & b \\ c & d\end{array}\right)\in SL(2,\mathbb{Z})\,.
\end{align}
Thus $E_2$ can be completed into the following quasi-modular form of weight $2$
\begin{align}
\widehat{E}_2(\rho,\bar{\rho})=E_2(\rho)-\frac{3}{\pi\,\text{Im}(\rho)}\,.
\end{align}
Furthermore, the holomorphic Eisenstein series display a ring structure in the sense that any $E_{2n}$ for $n>1$ can be written as a combination (of powers) of $E_4$ and $E_6$.

We shall sometimes also encounter derivatives $\mathfrak{d}=Q_\rho\,\frac{d}{dQ_\rho}$ of the Eisenstein series $E_{2n}$ (for $n>1$), which can again be expressed in terms of combinations of Eisenstein series with combined weight $2n+2$, which, however, are linear in $E_2$
\begin{align}
&\mathfrak{d}E_4=\frac{1}{3}\,(E_2 E_4-E_6)\,,\hspace{1.2cm}\mathfrak{d}E_6=-\frac{1}{2}\,(E_4^2-E_2 E_6)\,,\hspace{1.2cm}\mathfrak{d}E_8=\frac{2}{3}\,E_4\,(E_2E_4-E_6)\,,\nonumber\\
&\mathfrak{d} E_{10}=\frac{1}{6}\left(5E_2 E_4 E_6-3E_4^3-2E_6^2\right)\,,\hspace{0.5cm}\mathfrak{d}E_{12}=\frac{1}{691}\left(441 E_2 E_4^3-691 E_4^2 E_6+250E_2 E_6^2\right)\,.\label{DerivativeE4}
\end{align}
Finally, we remark that (quasi)Jacobi\footnote{For the purpose of this paper, we understand quasi-Jacobi forms as homogeneous polynomials of $\phi_{-2,1}$ and $\phi_{0,1}$, whose coefficients also depend on the Eisenstein series $E_2$. For a more rigorous definition we refer to \cite{Libgober} (see also \cite{Bastian:2019wpx}).} forms can be related to one-another through \emph{Hecke transformations}. Let $\mathcal{J}_{w,m}$ be the space of Jacobi forms of weight $w$ and index $m$ and let $n\in\mathbb{N}$, then
\begin{align}
\mathcal{H}_n:\,\mathcal{J}_{w,m}&\longrightarrow \mathcal{J}_{w,nm}\nonumber\\
\phi(\rho,z)&\longmapsto \mathcal{H}_n(f)=n^{w-1}\sum_{d|n\atop b\text{ mod }d}d^{-w}f\left(\frac{n\rho+bd}{d^2},\frac{nz}{d}\right)\,.\label{DefHecke}
\end{align} 
\subsection{Weierstrass' Elliptic Function and Scalar Two-Point Function}
A class of infinite series \cite{Bastian:2019wpx,Hohenegger:2019tii} that are useful in the discussion of the free energy in the case of $N=2$ is defined as (with $\widehat{a}\in\mathbb{C}$)
 \begin{align}
 \mathcal{I}_k(\rho,\widehat{a})=\sum_{n=1}^\infty \frac{n^{2k+1}}{1-Q_\rho^n}\left(Q_{\widehat{a}}^n+\frac{Q_\rho^n}{Q_{\widehat{a}}^n}\right)\,,&&\text{with}&&Q_{\widehat{a}}=e^{2\pi i \widehat{a}}\,, &&\forall k\in\mathbb{N}\cup\{0\}\,.\label{DefIalpha}
 \end{align}
The series for generic $k$ can be written as derivatives of the generating function $\mathcal{I}_0$
\begin{align}
&\mathcal{I}_{k}(\rho,\widehat{a})=D_{\widehat{a}}^{2k}\,\mathcal{I}_0(\rho,\widehat{a})\,,&&\text{with} &&D_{\widehat{a}}=\frac{1}{2\pi i}\,\frac{\partial}{\partial \widehat{a}}=Q_{\widehat{a}}\,\frac{\partial}{\partial Q_{\widehat{a}}}\,.\label{I0Generating}
\end{align}
Furthermore, it was shown in \cite{Bastian:2019wpx} that the generating function $\mathcal{I}_0$ can be related to Weierstrass' elliptic function $\wp$
\begin{align}
\mathcal{I}_0(\rho,\widehat{a})=\frac{1}{(2\pi i)^2}\,\left[G_2(\rho)+\wp(\widehat{a};\rho)\right]\,,\label{DefI0Weierstrass}
\end{align}
where the Eisenstein series $G_2=2\zeta(2)\,E_2$ is defined in (\ref{DefEisenstein}) and
\begin{align}
\wp(z;\rho)=\frac{1}{z^2}+\sum_{k=1}^\infty (2k+1)\,G_{2k+2}(\rho)\,z^{2k}\,.\label{WeierstrassDef}
\end{align}
A proof of (\ref{DefI0Weierstrass}) (which is slightly complementary to the argument presented in \cite{Bastian:2019wpx}) can be found in appendix~\ref{TechnicalCoupO21}. Finally, it was observed in \cite{Hohenegger:2019tii} that via the relation \cite{Eguchi:1986sb,Kuzenko:1991vu}
\begin{align}
\wp(\widehat{a};\rho)=\mathbb{G}''(\widehat{a};\rho) -\frac{\pi^2}{3}\,\widehat{E}_2(\rho)\,,
\end{align}
the infinite series $\mathcal{I}_k$ in (\ref{DefIalpha}) can be related to derivatives of the two-point function of a free scalar field $\phi$ on the torus
\begin{align}
\mathbb{G}(\widehat{a};\rho)=\langle \phi(\widehat{a})\,\phi(0)\rangle=-\ln\left|\frac{\theta_1(\widehat{a};\rho)}{\theta'_1(0,\rho)}\right|^2-\frac{\pi}{2\text{Im}\rho}\,(\widehat{a}-\bar{\widehat{a}})^2\,,\label{ScalarGreensFunction}
\end{align}
in particular 
\begin{align}
\mathcal{I}_0(\rho,\widehat{a})=\frac{1}{(2\pi i)^2}\left[\mathbb{G}''(\widehat{a};\rho)+\frac{2\pi i}{\rho-\bar{\rho}}\right]\,.\label{RelI0Greens}
\end{align}
For later use, we also exhibit another way of presenting this scalar two-point function: let $\Lambda=\mathbb{Z}\oplus \rho\,\mathbb{Z}$ be a two-dimensional lattice and define
\begin{align}
\langle \cdot|\cdot\rangle:\,&\Lambda\times \mathbb{C}\longrightarrow\mathbb{C}\nonumber\\
&(p,\widehat{a})=(m\rho+n,\alpha_1\rho+\alpha_2)\longmapsto\langle p|\widehat{a}\rangle=m\alpha_2-n\alpha_1\,.
\end{align}
Eq.(\ref{ScalarGreensFunction}) can then be written as
\begin{align}
\mathbb{G}(\widehat{a};\rho)=\frac{\text{Im}\rho}{\pi}\sum'_{p\in\Lambda}\frac{e^{2\pi i \langle p|\widehat{a}\rangle}}{|p|^2}\,,\label{GreensFunctionLattice}
\end{align}
where the summation excludes the origin of $\Lambda$. 

Furthermore, as is evident from (\ref{DefI0Weierstrass}), the generating function $\mathcal{I}_0$ is not holomorphic and modular, due to the presence of $G_2$. However, by multiplying $\mathcal{I}_0$ with holomorphic Eisenstein series $E_{2k}$ (for $k>1$) and subtracting a judicious derivative of the same Eisenstein series (as listed in (\ref{DerivativeE4})) one obtains an object whose Taylor series expansion in $\widehat{a}$ is purely holomorphic and modular: in addition to the example (\ref{ExpandE4I0}) discussed in the main body of this paper, we also find the following expressions relevant up to $s=4$ 
\begin{align}
\mathfrak{d}E_6+6 E_6\,\mathcal{I}_0(\rho,\widehat{a})&=-\frac{3 E_6}{2 \pi ^2 \widehat{a}^2}-\frac{E_4^2}{2}-\frac{6E_6}{4\pi^2}\sum_{k=1}^\infty (2k+1)G_{2k+2}(\rho)\,\widehat{a}^{2k}\,,\nonumber\\
\mathfrak{d}E_8+8 E_4^2\,\mathcal{I}_0(\rho,\widehat{a})&=-\frac{2 E_4^2}{\pi ^2 \widehat{a}^2}-\frac{2 E_4 E_6}{3}-\frac{8E_4^2}{4\pi^2}\sum_{k=1}^\infty (2k+1)G_{2k+2}(\rho)\,\widehat{a}^{2k}\,,\nonumber\\
\mathfrak{d}(E_{4} E_6)+10 E_4 E_6\,\mathcal{I}_0(\rho,\widehat{a})&=-\frac{5 E_4 E_6}{2 \pi ^2 \widehat{a}^2}-\frac{E_4^3}{2}-\frac{E_6^2}{3}-\frac{10E_4 E_6}{4\pi^2}\sum_{k=1}^\infty (2k+1)G_{2k+2}(\rho)\,\widehat{a}^{2k}\,,\nonumber\\
\mathfrak{d}(E_{12})+12 E_{12}\,\mathcal{I}_0(\rho,\widehat{a})&=-\frac{3 E_{12}}{\pi ^2 \widehat{a}^2}-E_6 E_4^2-3E_{12}\sum_{k=1}^\infty (2k+1)G_{2k+2}(\rho)\,\widehat{a}^{2k}\,.
\end{align}

\subsection{Graph Functions and Graph Forms}\label{App:GraphFunctions}
In this appendix we review so-called modular graph forms. Our conventions and presentation

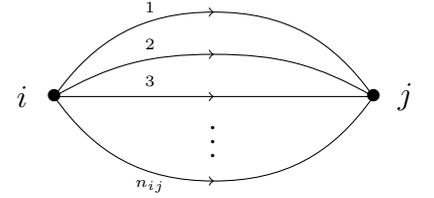
\begin{wrapfigure}{r}{0.31\textwidth}
\begin{center}
\vspace{-1cm}
\scalebox{1}{\parbox{5.5cm}{\begin{tikzpicture}[scale = 1.40]
\node at (0,0) {$\bullet$};
\node at (3,0) {$\bullet$};
\begin{scope}[ultra thick,decoration={markings,mark=at position 0.5 with {\arrow{>}}}] 
\draw[thin,postaction={decorate}] (0,0) to [out=55,in=180] (1.5,0.8) to [out=0,in=125](3,0);
\draw[thin,postaction={decorate}] (0,0) to [out=30,in=180] (1.5,0.4) to [out=0,in=150](3,0);
\draw[thin,postaction={decorate}] (0,0) -- (3,0);
\draw[thin,postaction={decorate}] (0,0) to [out=-55,in=180] (1.5,-0.8) to [out=0,in=235](3,0);
\end{scope}
\node at (0.9,0.85) {\tiny $1$};
\node at (0.9,0.5) {\tiny $2$};
\node at (0.9,0.15) {\tiny $3$};
\node at (0.9,-0.85) {\tiny $n_{ij}$};
\node[rotate=90] at (1.5,-0.4) {$\cdots$}; 
\node at (-0.3,0) {$i$};
\node at (3.3,0) {$j$};
\end{tikzpicture}}}
\caption{\sl $n_{ij}$ oriented edges connecting vertex $i$ to $j$.}
\label{Fig:UndecoratedGraph}
\end{center}
${}$\\[-2cm]
\end{wrapfigure} 

\noindent
mostly follow \cite{DHoker:2016mwo} as well as \cite{Gerken:2019cxz}. \emph{Modular graph functions} have first been introduced in \cite{DHoker:2015wxz} (see also \cite{DHoker:2016mwo,DHoker:2017pvk,Zerbini:2018sox,Zerbini:2018hgs,Gerken:2018jrq,Gerken:2019cxz,Gerken:2020yii}): Let $\Gamma$ be a graph of $N$ vertices (labelled by $i,j=1,\ldots,N$) with $r_{ij}$ oriented edges connecting vertex $i$ to vertex $j$. An example of $n$ edges connecting the vertices $i$ and $j$ is shown in \figref{Fig:UndecoratedGraph}. The weight $w$ of the graph is defined as the total number of all edges, \emph{i.e.} $w=\sum_{1\leq i<j\leq N}r_{ij}$. The modular graph function associated with the graph $\Gamma$ is then defined as the following integral over the positions $z_{i}$ (with $i=1,\ldots,N$) of the $N$ vertices on the torus $\Sigma_\rho$ that is parametrised by the modular parameter~$\rho$
\begin{align}
\mathcal{C}_\Gamma(\rho)=\prod_{k=1}^N\int_{\Sigma_\rho}\frac{d^2z_k}{\text{Im}\rho}\,\prod_{1\leq i<j\leq N}\mathbb{G}(z_i-z_j;\rho)^{r_{ij}}\,.\label{DefModularGraphFunction}
\end{align}
Due to the properties of the scalar Greens functions, $\mathcal{C}_\Gamma$ is invariant under $SL(2,\mathbb{Z})$ acting on $\rho$. Notice, however, that it is a non-holomorphic modular function, which can be made more manifest by using the presentation (\ref{GreensFunctionLattice}) of $\mathbb{G}(z;\rho)$, to write\footnote{Since it will turn out more convenient for the discussion in the main body of this paper, our normalisation here follows \cite{Gerken:2019cxz} and is missing a factor of $\left(\frac{\text{Im}\rho}{\pi}\right)^w$ relative to \cite{DHoker:2016mwo}. In (\ref{DefModularGraphFunction}) this factor has been accounted for by using the normalisation $\int_\Sigma \frac{d^2z}{\text{Im}\rho}=1$.}
\begin{align}
&\mathcal{C}_\Gamma(\rho)=\sum'_{p_1,\ldots,p_w\in\Lambda}\prod_{a=1}^w\frac{1}{ |p_a|^2}\,\prod_{i=1}^N\,\delta_{\text{K}}\left(\sum_{b=1}^w\Gamma_{ib}\,p_b\right)\,,\label{DefGraphFunctLit}
\end{align}
where we used the shorthand notation
\begin{align}
\delta_{\text{K}}\left(\sum_{b=1}^w\Gamma_{ib}p_b\right) =\delta_{\text{K}}\left(\sum_{a=1}^w\Gamma_{ia}\,m_a\right)\,\delta_{\text{K}}\left(\sum_{b=1}^w\Gamma_{ib}\,n_b\right)\,, &&\text{for} && p_a=m_a+\rho n_a\,,
\end{align}
and we have defined
\begin{align}
&\delta_{\text{K}}(x)=\left\{\begin{array}{lcl}1 & \text{if} & x=0\,,\\ 0 & \text{if} & x\neq 0\,,\end{array}\right.&&\text{and} &&\Gamma_{ia}=\left\{\begin{array}{lcl}1 & \text{if} &\parbox{4.7cm}{\small edge $a$ ends on vertex $i$ and\\[-6pt] points into vertex $i$}\,,\\[8pt]
-1 & \text{if} &\parbox{4.7cm}{\small edge $a$ ends on vertex $i$ and\\[-6pt] points out of vertex $i$}\,,\\[8pt]
0 & & \text{else} \end{array}\right.
\end{align}
This definition of modular graph functions has been generalised in \cite{DHoker:2016mwo} to so-called \emph{modular}

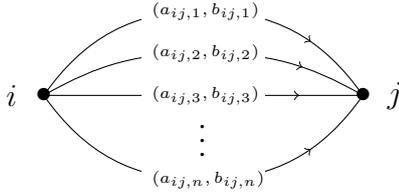
\begin{wrapfigure}{l}{0.29\textwidth}
\begin{center}
\vspace{-0.75cm}
\scalebox{1}{\parbox{5.5cm}{\begin{tikzpicture}[scale = 1.40]
\node at (0,0) {$\bullet$};
\node at (3,0) {$\bullet$};
\begin{scope}[ultra thick,decoration={markings,mark=at position 0.8 with {\arrow{>}}}] 
\draw[thin,postaction={decorate}] (0,0) to [out=55,in=180] (1.5,0.8) to [out=0,in=125](3,0);
\draw[thin,postaction={decorate}] (0,0) to [out=30,in=180] (1.5,0.4) to [out=0,in=150](3,0);
\draw[thin,postaction={decorate}] (0,0) -- (3,0);
\draw[thin,postaction={decorate}] (0,0) to [out=-55,in=180] (1.5,-0.8) to [out=0,in=235](3,0);
\end{scope}
\node[fill=white] at (1.5,0.8) {\parbox{1.35cm}{\tiny $(a_{ij,1},b_{ij,1})$}};
\node[fill=white] at (1.5,0.4) {\parbox{1.35cm}{\tiny $(a_{ij,2},b_{ij,2})$}};
\node[fill=white] at (1.5,0) {\parbox{1.35cm}{\tiny$(a_{ij,3},b_{ij,3})$}};
\node[fill=white] at (1.5,-0.8) {\parbox{1.35cm}{\tiny$(a_{ij,n},b_{ij,n})$}};
\node[rotate=90] at (1.5,-0.4) {$\cdots$}; 
\node at (-0.3,0) {$i$};
\node at (3.3,0) {$j$};
\end{tikzpicture}}}
\caption{\sl $n$ oriented and decorated edges connecting vertex $i$ to $j$.}
\label{Fig:DecoratedGraph}
\end{center}
${}$\\[-2cm]
\end{wrapfigure} 

\noindent
\emph{graph forms}. To describe the latter, we first generalise the graph $\Gamma$ by decorating each edge by two integers: instead of the oriented edges in \figref{Fig:UndecoratedGraph} between two vertices $i$ and $j$, we now consider \figref{Fig:DecoratedGraph} with labels $a_{ij,a}$ and $b_{ij,a}$ (for $a=1,\ldots,n$). We then define\footnote{As in (\ref{DefModularGraphFunction}), we follow the normalisation of \cite{Gerken:2019cxz}, which differs by a factor $(\text{Im}\rho/\pi)^{\frac{a_r+b_r}{2}}$ relative to \cite{DHoker:2016mwo}.}
\begin{align}
\mathcal{C}\big[{}^{a_{ij,r}}_{b_{ij,r}}\big](\rho)=\sum'_{p_1,\ldots,p_w\in\Lambda}\prod_{r}\frac{1}{p_r^{a_r}\bar{p}_r^{b_r}}\prod_{i=1}^N\delta_{\text{K}}\left(\sum_{s}\Gamma_{is}p_s\right),\label{DefGraphFormGeneral}
\end{align}
where the sums over $r$ and $s$ run over all possible labels $(ij,a)$. Since in the current paper only the modular graph forms with two vertices are important, we shall not make the notation in (\ref{DefGraphFormGeneral}) precise for generic $N$\footnote{For this purpose, we refer the reader to the original literature \cite{DHoker:2016mwo} (see also \cite{Gerken:2019cxz}).}, but focus on graphs with $N=2$. Such graphs are called \emph{dihedral graphs} and for $n$ edges connecting the two vertices, we arrange the decorations $(a_\alpha,b_\alpha)$ (for $\alpha=1,\ldots,n$) into the matrix $\big[{}^{A}_{B}\big]=\big[{}^{a_1\,a_2\,\ldots\,a_n}_{b_1\,b_2\,\ldots\,b_n}\big]
$. We then write for (\ref{DefGraphFormGeneral})
\begin{align}
\mathcal{C}\big[{}^{A}_{B}\big](\rho)=\sum'_{p_1,\ldots,p_n\in\Lambda}\prod_{r=1}^n\frac{1}{p_r^{a_r}\bar{p}_r^{b_r}}\,\delta\left(\sum_{s=1}^N p_s\right)\,.
\end{align}
Dihedral graph forms with a single edge vanish
\begin{align}
&\mathcal{C}\big[{}^{a_1}_{b_1}\big](\rho)=0\,,&&\forall a_1,b_1\in\mathbb{N}\,.
\end{align}
In this paper we shall exclusively encounter dihedral graph forms with two edges, for which numerous identities and relations are known. For example, in \cite{DHoker:2016mwo} it was shown that
\begin{align}
&\nabla^n\,(\text{Im}\rho)^k\,\mathcal{C}\big[{}^{k\,0}_{k\,0}\big](\rho)=(\text{Im}\rho)^{k+n}\,\frac{(k+n-1)!}{(k-1)!}\,\mathcal{C}\big[{}^{k+n\,\,0}_{k-n\,\,0}\big](\rho)\,,&&\text{with} &&\nabla=2i\,(\text{Im}\rho)^2\,\frac{\partial}{\partial\rho}\,,\label{GraphFunctionSplit}
\end{align}
as well as
\begin{align}
\mathcal{C}\big[{}^{a_1\,a_2}_{\,b_1\,b_2}\big](\rho)=(-1)^{a_2+b_2}\,\mathcal{C}\big[{}^{a_1+a_2\,\,0}_{b_1+b_2\,\,\,0}\big](\rho)\,.\label{GraphFunctionShift}
\end{align}
Furthermore, the following explicit expressions have been given in \cite{DHoker:2016mwo} \begin{align}
&\mathcal{C}\big[{}^{k\,0}_{k\,0}\big](\rho)=\left(\frac{\pi}{\text{Im}\rho}\right)^k\,\mathcal{E}_k(\rho)\,,&&\forall k>1\,,
\end{align}
where $\mathcal{E}_k(\rho)$ is the so-called non-holomorphic Eisenstein series
\begin{align}
\mathcal{E}_k(\rho)=\sum_{(m,n)\neq (0,0)}\frac{1}{|m+n\rho|^{2k}}\,.
\end{align}
Finally, the modular graph forms (along with the dihedral graph with two bi-valent vertices) relevant for the current paper are 
\begin{align}
&\mathcal{C}\big[{}^{2k\,0}_{\,\,0\,\,0}\big](\rho)=\left\{\begin{array}{lcl}2\zeta(2)\,\widehat{E}_2(\rho,\bar{\rho}) & \text{if} & k=1\,, \\ 2\zeta(2k)\,E_{2k}(\rho) & \text{if} & k>1\,.\end{array}\right.\label{ModGraphFormHol}
\end{align}
\subsection{Generating Functions of Multiple Divisor Sums}\label{App:DivisorSums}
In \cite{Bachmann:2013wba} the following generating function was introduced
\begin{align}
T(z_1,\ldots,z_k;\rho)=\sum_{n_1,\ldots,n_k>0}\prod_{j=1}^k\frac{e^{2\pi in_j z_j}\,Q_\rho^{n_1+\ldots+n_j}}{1-Q_\rho^{n_1+\ldots+n_j}}\,.\label{GenFunctionsT}
\end{align}
From a mathematical perspective, these functions can be related to multiple zeta values and are expressable in terms of (reduced) polylogarithms. These, in turn, have made appearances in the physics literature, related to (loop) amplitudes in (supersymmetric) string theories, \emph{e.g.} \cite{Schlotterer:2012ny,Broedel:2013aza,Broedel:2013tta,Stieberger:2013wea,Broedel:2014vla,DHoker:2015wxz,MatthesPhD,Brown1,Brown2,Broedel:2018izr,Zerbini:2018sox,Mafra:2019ddf,Mafra:2019xms,DHoker:2019xef,Zagier:2019eus}.  For the current work, we remark that (\ref{GenFunctionsT}) affords the following Taylor series expansion
\begin{align}
T(z_1,\ldots,z_k;\rho)=\sum_{s_1,\ldots,s_k>0}[s_1,\ldots,s_k;\rho]\,(2\pi i z_1)^{s_1-1}\ldots (2\pi i z_k)^{s_k-1}\,. \label{TaylorSeriesT}
\end{align}
Here $[s_1,\ldots,s_k;\rho]$ are called brackets of length $k\in\mathbb{N}$ in \cite{Bachmann:2013wba}
\begin{align}
&[s_1,\ldots,s_k;\rho]=\frac{\sum_{n>0}Q_\rho^n\,\sigma_{s_1-1,\ldots,s_k-1}(n)}{(s_1-1)!\ldots(s_k-1)!}\,,&&s_j\in\mathbb{N}\hspace{0.3cm}\forall\,j\in\{1,\ldots,k\}\,,
\end{align}
are generating functions of multiple divisor sums
\begin{align}
&\sigma_{p_1,\ldots,p_k}(n)=\sum_{{u_1v_1+\ldots+u_kv_k=n}\atop{u_1>\ldots>u_k>0}}v_1^{p_1}\ldots v_k^{p_k}\,,&&\begin{array}{l}p_1,\ldots,p_k\in\mathbb{N}\cup\{0\}\,,\\[4pt] k,n\in\mathbb{N}\,.\end{array}
\end{align}
\section{Building Blocks}\label{App:BuildingBlocks}
The decompositions of the LST free energy studied in the main body of this paper use two classes of functions as their basic building blocks: the expansion coefficients of the free energy for the case $N=1$ and the expansion coefficients of a specific function that governs the BPS counting of a configuration of a single M5-brane with M2-branes ending on it on either side. In this appendix we review both of these building blocks and also provide basic expansions for the convenience of the reader.
\subsection{Free Energy for $N=1$}\label{App:BuildingN1}
We first consider the expansion coefficients of the free energy for $N=1$ in the unrefined limit. For $N=1$, the decomposition (\ref{DecompositionFourierModes}) of the Fourier modes $P_{N,(s)}^{(r)}$ of the free energy $\mathcal{F}_{N,1}$, as introduced in eq.~(\ref{FourierModesFreeEnergy}), is simply\footnote{Notice, for $N=1$ we have $\rho=\widehat{a}_1$.}
\begin{align}
P_{N=1,(s)}^{(r)}(\rho,S)=H_{(s)}^{(r),\{0\}}(\rho,S)\,.
\end{align}
The functions $H_{N,(s)}^{(r),\{0\}}(\rho,S)$ are in fact (quasi) Jacobi forms of weight $2s-2$ and index $r$. For $r=1$, they are related to the basic Jacobi forms $\phi_{0,1}$ and $\phi_{-2,1}$ in (\ref{DefStandardJacobiForms})
\begin{align} 
\buildH{1}{s}(\rho,S)=\left\{\begin{array}{lcl}-\phi_{-2,1}(\rho,S) & \text{if} & s=0\,, \\ \frac{\phi_{0,1}(\rho,S)}{24} &\text{if} & s=1\,, \\ \frac{(-1)^s B_{2s}\,E_{2s}(\rho)}{(2s-3)!!(2s)!!}\,\phi_{-2,1}(\rho,S) & \text{if} & s>1\,,\end{array}\right.
\end{align}
where $E_{2s}(\rho)$ are the (holomorphic) Eisenstein series and $B_{2s}$ the Bernoulli numbers. Higher orders in $r$ are related to $\buildH{1}{s}$ through the action of the Hecke operator
\begin{align}
&\buildH{r}{s}(\rho,S)=\mathcal{H}_r\left(\buildH{1}{s}(\rho,S)\right)\,,&&\forall s\geq 0\,,\label{DefBuildBlockHr}
\end{align}
where $\mathcal{H}_r$ is the Hecke operator defined in (\ref{DefHecke}).
\subsection{The Function $W(\rho,S)$}\label{App:BuildingW}
The second class of building blocks that we shall use in the main body of this paper are the expansion coefficients (in powers of $\epsilon$) of the quasi-Jacobi form $W$ that was first introduced in \cite{Hohenegger:2015btj,Ahmed:2017hfr}
\begin{align}
W(\rho,S,\epsilon)=\frac{\theta^2_1(\rho,S)-\theta_1(\rho,S+\epsilon)\theta_1(\rho,S-\epsilon)}{\theta^2_1(\rho,\epsilon)}=\sum_{s=0}^\infty \epsilon^{2s}\,\buildW{1}{s}(\rho,S)\,.\label{DefWFunct}
\end{align}
The $\buildW{1}{s}$ are quasi Jacobi forms of weight $2s$ and index $1$ and the first few of them are
\begin{align}
&\buildW{1}{0}=\frac{1}{24}(\phi_{0,1}+2E_2\phi_{-2,1})\,,&&\buildW{1}{1}=\frac{E_2^2-E_4}{288}\,\phi_{-2,1}\,,&&\buildW{1}{2}=\frac{5E_2^3+3E_2 E_4-8E_6}{51840}\,\phi_{-2,1}\,.\nonumber
\end{align}
Following \cite{IqbalHohenegger} , similar to (\ref{DefBuildBlockHr}), we can define more general building blocks suitable also for $r>1$ through Hecke transformations
\begin{align}
\buildW{r}{s}(\rho,S)=\mathcal{H}_r\left(\buildW{1}{s}(\rho,S)\right)\,.
\end{align}
For further convenience, we present $\buildW{r}{s=0}$ for $r=2$ and $r=3$
\begin{align}
\buildW{2}{0}(\rho,S)&=\frac{1}{384}\,\left(\phi_{0,1}^2+4E_2\phi_{0,1}\phi_{2,1}+4 E_4\,\phi_{-2,1}\right)\,,\nonumber\\
\buildW{3}{0}(\rho,S)&=\frac{1}{18\cdot 24^2}\left(\phi_{0,1}^3+6 E_2\phi_{0,1}^2\phi_{-2,1}+12 E_4\phi_0 \phi_{-2,1}^2+8(9E_2 E_4-8E_6)\phi_{-2,1}^3\right)\,.
\end{align}
\section{Series Expansion of Coupling Functions}\label{SeriesExpansionSimCoup}
In this appendix we present some technical details of the Taylor series expansions of various couplings~$\coup{N}{\alpha}$.
\subsection{Coupling $\coup{2}{1}$}\label{TechnicalCoupO21}
The simplest non-trivial example is the coupling $\coup{2}{1}(\widehat{a}_1,\rho)$. Although eq.~(\ref{N2CoupR12}) has already been argued for in \cite{Hohenegger:2019tii}, we present a somewhat different approach here, which (in principle) is also applicable to some later (more complicated) examples. We start from (\ref{N2CoupR11}) which, as already remarked in \cite{Bastian:2019wpx}, can be written in terms of the generating functions (\ref{GenFunctionsT})
\begin{align}
\coup{2}{1}(\widehat{a}_1,\rho)=-2D_{\widehat{a}_1}\left[T(\widehat{a}_1-\rho;\rho)-T(-\widehat{a}_1;\rho)\right]\,.\label{CN2corrTform}
\end{align}
In principle (\ref{TaylorSeriesT}) affords a full series expansion in powers of $\widehat{a}_1$, except that care has to be taken since $T(\widehat{a}_1-\rho;\rho)$ has a singularity for $\widehat{a}_1=0$. To this end, we write
\begin{align}
T(\widehat{a}_1-\rho;\rho)=\sum_{n=1}^\infty Q_{\widehat{a}_1}\left[1+\frac{Q_\rho^n}{1-Q_\rho^n}\right]=T(\widehat{a}_1;\rho)+\sum_{n=1}^\infty Q_{\widehat{a}_1}^n\,.
\end{align}
The derivative of the last term can be written as
\begin{align}
D_{\widehat{a}_1}\sum_{n=1}^\infty Q_{\widehat{a}_1}^n=D_{\widehat{a}_1}\,\frac{Q_{\widehat{a}_1}}{1-Q_{\widehat{a}_1}}=\frac{1}{\left(e^{\pi i \widehat{a}_1}-e^{-i\pi \widehat{a}_1}\right)^2}=-\frac{1}{4\sin^2(\pi\widehat{a}_1)}\,, 
\end{align}
such that we can rewrite (\ref{CN2corrTform}) as
\begin{align}
\coup{2}{1}&=\frac{1}{2\sin^2(\pi\widehat{a}_1)}-2D_{\widehat{a}_1}\sum_{s=1}^\infty (1+(-1)^s)(2\pi i \widehat{a}_1)^{s-1}\,\sum_{n=1}^\infty Q_\rho^{n}\,\frac{\sigma_{s-1}(n)}{(s-1)!}\nonumber\\
&=\frac{1}{2\sin^2(\pi\widehat{a}_1)}-4\sum_{k=1}^\infty \frac{(2\pi i \widehat{a}_1)^{2k-2}}{(2k-2)!}\frac{B_{2k}}{4k}\left(1-E_{2k}(\rho)\right)\,.
\end{align}
Using furthermore the Laurent series expansion
\begin{align}
\frac{1}{2\sin^2(\pi\widehat{a}_1)}=\frac{1}{2\pi^2}\left[\frac{1}{\widehat{a}_1^2}+\frac{\pi^2}{4}+\sum_{k=1}^\infty (2k+1)\,2\zeta(2k+2)\,\widehat{a}_1^{2k}\right]\,,
\end{align}
along with the definition (\ref{WeierstrassDef}) we find
\begin{align}
\coup{2}{1}&=\frac{E_2}{6}+\frac{1}{2\pi^2}\left[ \frac{1}{\widehat{a}_1^2}+\sum_{k=1}^\infty (2k+1)\,2\zeta(2k+2)\,E_{2k+2}(\rho)\,\widehat{a}_1^{2k}\right]=\frac{E_2}{6}+\frac{1}{2\pi^2}\,\wp(\widehat{a}_1;\rho)\,,\label{ExplicitFormO21}
\end{align}
which is indeed (\ref{N2CoupR12}). 

For later use, we also remark that a (limited) series expansion in $Q_\rho$ to fixed order in $\widehat{a}_1$ can be obtained in a much simpler fashion. To this end, we recall that it was shown in \cite{Bastian:2019wpx} that $\coup{2}{1}$ can also be written as\footnote{Here it was assumed that $|Q_\rho|<1$ and $|Q_{\widehat{a}_1}|<1$, such that $\coup{2}{1}$ is absolutely convergent.}
\begin{align}
\coup{2}{1}(\widehat{a}_1,\rho)
=\frac{1}{2\pi^2}\left[\frac{\pi^2}{\sin^2(\pi\widehat{a}_1)}-4\pi^2\sum_{k=1}^\infty\sum_{n=1}^\infty n\,Q_\rho^{nk}\left(Q_{\widehat{a}_1}^n+Q_{\widehat{a}_1}^{-n}\right)\right]\,.
\end{align}
This expression is equivalent to the following series expansion in $Q_\rho$
\begin{align}
\coup{2}{1}(\widehat{a}_1,\rho)=\frac{1}{2\pi^2}\left[\frac{\pi^2}{\sin^2(\pi\widehat{a}_1)}-4\pi^2\sum_{n=1}^\infty Q_\rho^n\sum_{\ell|n} \ell\,\left(Q_{\widehat{a}_1}^\ell+Q_{\widehat{a}_1}^{-\ell}\right)\right]\,.\label{CoupO2InterResum}
\end{align}
Since the sum over $\ell$ is finite (for fixed $n$), we can extract limited expansions for fixed orders in $\widehat{a}_1$. For the first few orders we find
{\allowdisplaybreaks
\begin{align}
\mathcal{O}(\widehat{a}_1^{-2}):\,&\frac{1}{2\pi^2}\,,\nonumber\\
\mathcal{O}(\widehat{a}_1^{0}):\,&\frac{1}{6}\left(1-24\,Q_\rho-72\,Q_\rho^2-96\,Q_\rho^3-168\,Q_\rho^4+\mathcal{O}(Q_\rho^5)\right)\sim\frac{E_2(\rho)}{6}\,,\nonumber\\
\mathcal{O}(\widehat{a}_1^{2}):\,&\frac{\pi^2}{30}\left(1+240\, Q_\rho + 2160\, Q_\rho^2 + 6720 \,Q_\rho^3 + 17520\, Q_\rho^4 +\mathcal{O}(Q_\rho^5)\right)\sim\frac{\pi^2 E_4(\rho)}{30}\,,\nonumber\\
\mathcal{O}(\widehat{a}_1^{4}):\,&\frac{\pi^4}{189}\left(1 - 504\,Q_\rho - 16632\,Q_\rho^2 - 122976\,Q_\rho^3 - 532728\,Q_\rho^4+\mathcal{O}(Q_\rho^5)\right)\sim \frac{\pi^4 E_6(\rho)}{189}\,,\nonumber\\
\mathcal{O}(\widehat{a}_1^{6}):\,&\frac{\pi^6}{1350}\left(1 + 960\,Q_\rho + 354240\,Q_\rho^2 + 61543680\,Q_\rho^3 + 4858169280\,Q_\rho^4 +\mathcal{O}(Q_\rho^5)\right)
\sim \frac{\pi^6 E_4^2(\rho)}{1350}\,.\label{LimitEisenstein}
\end{align}}
Here the identification in terms of Eisenstein series follows if we assume that the coefficient of $\widehat{a}_1^{2k}$ in $\coup{2}{1}$ is a quasi-modular form of weight $2k+2$. These terms of low order in $\widehat{a}_1$ indeed match the precise expansion in (\ref{ExplicitFormO21}). In fact, we remark, that in the current case this can be made precise by extracting the coefficient of the term $\widehat{a}_1^{2s}$ (for $s\geq 0$) in a Laurent series expansion of (\ref{CoupO2InterResum}) (the terms $\widehat{a}_1^{2s+1}$ are identically zero) 
\begin{align}
&\frac{(2s+1)2\zeta(2s+2)}{2\pi^2}-2\sum_{n=1}^\infty\,Q_\rho^n\sum_{\ell|n}\ell\,\frac{(2\pi i \ell \widehat{a}_1)^{2s}}{(2s)!}=\frac{(2s+1)2\zeta(2s+2)}{2\pi^2}-\frac{4(2\pi i)^{2s}}{(2s)!}\sum_{n=1}^\infty\,Q_\rho^n\,\sigma_{2s+1}(n)\nonumber\\
&\hspace{0.3cm}=\frac{(2s+1)2\zeta(2s+2)}{2\pi^2}-\frac{4(2\pi i)^{2s}}{(2s)!}+\frac{B_{2s+2}}{4(s+1)}(1-E_{2s+2}(\rho))=\frac{(2s+1)2\zeta(2s+2)}{2\pi^2}\,E_{2s+2}\,.\nonumber
\end{align}
Combining with the $\widehat{a}_1^{-2}$ term stemming from $\frac{1}{2\sin^2(\pi\widehat{a}_1)}$, we obtain
\begin{align}
\coup{2}{1}(\widehat{a}_1,\rho)=\frac{1}{2\pi^2}\left[\widehat{a}_{1}^{-2}+\frac{E_2(\rho)}{6}+\sum_{s=1}^\infty (2s+1)\,G_{2s+2}(\rho)\,\widehat{a}_1^{2s}\right]\,,
\end{align}
which is indeed (\ref{ExplicitFormO21}), thus confirming once more the result we have inferred from analysing the limited expansions (\ref{LimitEisenstein}) and assuming modularity of the final result.
\subsection{Coupling $\coup{3}{2}$}\label{App:ExpansionCoupO32}
The coupling $\coup{3}{2}$ is defined in (\ref{N3R1CoupsPre}). Due to the complexity of this expression, a purely analytic approach (as discussed in the first part of the previous subsection) is much more difficult and we therefore resort to studying limited expansions in powers of $Q_\rho$. To leading orders in $\widehat{a}_{1,2}$, we can match the latter to combinations of Eisenstein series, which in turn we can compare to combinations of scalar two-point functions.\footnote{While this procedure does not constitute a rigorous proof, the fact that we find agreement for many orders in $\widehat{a}_{1,2}$ leads us to believe that (\ref{CouplingO32R1}) is indeed the complete result.} More precisely, we start from the following (schematic) presentation of the coupling 
\begin{align}
\coup{3}{2}(\widehat{a}_1,\widehat{a}_2,\rho)=\sum_{k=-2}^\infty\sum_i\,\mathfrak{r}^{(2k)}_i(\widehat{a}_1,\widehat{a}_2)\,\mathfrak{p}^{(2k)}_i(\rho)\,,\label{SeriesExpandN3Coup2}
\end{align}
where $\mathfrak{r}^{(2k)}_i(\widehat{a}_1,\widehat{a}_2)$ are rational functions in $\widehat{a}_{1,2}$, which are homogeneous of order $2k$ and $\mathfrak{p}^{(2k)}_i(\rho)$ are series expansions in $Q_\rho$. We shall then work out the first orders of the latter and match them to combinations of Eisenstein series of weight $4+2k$. From the ensuing pattern, we will be able to present a closed form expression that fits to all orders that we were able to compute.

We start with the terms in the first line of (\ref{N3R1CoupsPre}), which are governed by the quotient $\frac{n^2}{(1-Q_\rho^n)^2}$. In order to sum up their contribution, there are three conceptually different terms
\begin{align}
J_\ell(x)=&\sum_{n=1}^\infty\frac{n^2}{(1-Q_\rho^n)^2}\,Q_\rho^{n \ell}\, e^{2\pi i n x}\,,&&\text{for} && \ell\in\{0,1,2\}\,,
\end{align}
where $x$ can stand for various linear combinations of $\widehat{a}_{1,2}$ (with $|e^{2\pi ix}|<1$). We can further write this as
\begin{align}
J_\ell(x)&=\sum_{n=1}^\infty n^2\sum_{k=1}^\infty k\,Q_\rho^{n(k+\ell-1)}\,e^{2\pi i nx}\nonumber\\
&=\left\{\begin{array}{lcl}\sum_{n=1}^\infty n^2 e^{2\pi i nx}+\sum_{k=1}^\infty (k+1)\sum_{n=1}^\infty n^2Q_\rho^{nk}\,e^{2\pi i nx} & \text{for} &\ell=0\,,\\ \sum_{k=1}^\infty k\,\sum_{n=1}^\infty n^2 Q_\rho^{nk}\,e^{2\pi inx} & \text{for} & \ell=1\,,\\
\sum_{k=2}^\infty(k-1)\,\sum_{n=1}^\infty n^2 Q_\rho^{nk}\,e^{2\pi inx} & \text{for} &\ell=2\,.\end{array}\right.\nonumber
\end{align}
Notice that the first term for $\ell=0$ diverges for $x=0$, thus leading to a pole in the free energy. Furthermore, in order to obtain a (limited) series expansion in powers of $Q_\rho$ it is useful to write this expression in the form
\begin{align}
J_\ell(x)=\left\{\begin{array}{lcl}\frac{e^{2\pi i x}(1+e^{2\pi i x})}{(1-e^{2\pi ix})^3}+\sum_{m=1}^\infty Q_\rho^m\sum_{k|m}(m+k)k\,e^{2\pi ikx} & \text{for} &\ell=0\,,\\\sum_{m=1}^\infty Q_\rho^m\sum_{k|m}\,m\,k\, e^{2\pi ikx} & \text{for} & \ell=1\,,\\\sum_{m=1}^\infty Q_\rho^m\sum_{k|m}(m-k)k\,e^{2\pi i k x }&\text{for} &\ell=2\,.\end{array}\right.
\end{align}  
Since (for fixed $m$) the sum over $k$ is finite, we can derive a limited series expansion in $Q_\rho$ for a given power of $x$.

In order to analyse the terms in the second and third line of (\ref{N3R1CoupsPre}), we consider
\begin{align}
&J_{\ell_1,\ell_2}(x_1,x_2)=\sum_{n_1,n_2=1}^\infty\left(\frac{n_2(2n_1+n_2)}{(1-Q_\rho^{n_1})(1-Q_\rho^{n_2})}+\frac{(n_1+n_2)(n_1-n_2)}{(1-Q_\rho^{n_1})(1-Q_\rho^{n_1+n_2})}\right)\,Q_{\rho}^{\ell_1 n_1+\ell_2 n_2}\,e^{2\pi i (n_1 x_1+n_2 x_2)}\,,\nonumber
\end{align}
for $\ell_1,\ell_2\in\{0,1\}$. We can start by writing (with $Q_{x_1}=e^{2\pi i x_1}$ and $Q_{x_2}=e^{2\pi i x_2}$)
\begin{align}
J_{\ell_1,\ell_2}(x_1,x_2)=\sum_{n_1,n_2=1}^\infty\sum_{k_1,k_2=0}^\infty\left[n_2(2n_1+n_2) +(n_1^2-n_2^2)Q_\rho^{n_1 k_2}\right]\,Q_\rho^{n_1(k_1+\ell_1)+n_2(k_2+\ell_2)}\,\,Q_{x_1}^{n_1}\,Q_{x_2}^{n_2}\,.\nonumber
\end{align}
The terms in which $(k_1+\ell_1)=0$ and/or $(k_2+\ell_2)=0$ lead to divergent sums over $n_{1,2}$ and need to be considered separately. In order to do so, we need to distinguish the various cases for $\ell_{1,2}$:
\begin{itemize}
\item $\ell_1=0$ and $\ell_2=0$: 
{\allowdisplaybreaks
\begin{align}
J_{0,0}(x_1,x_2)&=\sum_{n_1,n_2=1}n_1(n_1+2n_2)\,Q_{x_1}^{n_1}\,Q_{x_2}^{n_2}+\sum_{k_1=1}^\infty \sum_{n_1=1}^\infty Q_\rho^{n_1 k_1}\sum_{n_2=1}^\infty n_1(n_1+2n_2)\,Q_{x_1}^{n_1}\,Q_{x_2}^{n_2}\nonumber\\
&+\sum_{k_2=1}^\infty \sum_{n_1,n_2=1}^\infty\left[n_2(2n_1+n_2) +(n_1^2-n_2^2)Q_\rho^{n_1 k_2}\right]\,Q_\rho^{n_2 k_2}\,Q_{x_1}^{n_1}\,Q_{x_2}^{n_2}\nonumber\\
&+\sum_{k_1,k_2=1}^\infty\sum_{n_1,n_2=1}^\infty\left[n_2(2n_1+n_2) +(n_1^2-n_2^2)Q_\rho^{n_1 k_2}\right]\,Q_\rho^{n_1k_1+n_2k_2}\,Q_{x_1}^{n_1}\,Q_{x_2}^{n_2}\,,\nonumber
\end{align}}
which can be cast into the form 
\begin{align}
&J_{0,0}(x_1,x_2)=D_{x_1}(D_{x_1}+2 D_{x_2})\,\frac{1}{(1-Q_{x_1})(1-Q_{x_2})}+\sum_{m=1}^\infty Q_\rho^m\sum_{k=1}^\infty\sum_{n_1,n_2=1\atop{(n_1+n_2)k=m}}(n_1^2-n_2^2)\,Q_{x_1}^{n_1}Q_{x_2}^{n_2}\nonumber\\
&\hspace{0.5cm}+\sum_{m=1}^\infty Q_\rho^m\sum_{k|m} \left[\frac{Q_{x_2}k\left(k(1-Q_{x_2})+2\right)}{(1-Q_{x_2})^2}\,Q_{x_1}^k+\frac{k Q_{x_1}(k(1-Q_{x_1})+2)}{(1-Q_{x_1})^2}\,Q_{x_2}^k\right]\nonumber\\
&\hspace{0.5cm}+\sum_{m=1}^\infty Q_\rho^m\sum_{k_1,k_2=1}^{m}\left[\sum_{n_1,n_2=1\atop{n_1k_1+n_2k_2=m}}^m n_2(2n_1+n_2)\,Q_{x_1}^{n_1}Q_{x_2}^{n_2}+\sum_{n_1,n_2=1\atop{n_1(k_1+k_2)+n_2k_2=m}}^m(n_1^2-n_2^2)\,Q_{x_1}^{n_1}Q_{x_2}^{n_2}\right]\,.\nonumber
\end{align}
\item $\ell_1=1$ and $\ell_2=0$: 
\begin{align}
J_{1,0}(x_1,x_2)&=\sum_{k_1=1}^\infty\sum_{n_1,n_2=1}^\infty Q_\rho^{n_1k_1}n_1(n_1+2n_2)\,Q_{x_1}^{n_1}Q_{x_2}^{n_2}\nonumber\\
&\hspace{0.5cm}+\sum_{k_1,k_2=1}^\infty\sum_{n_1,n_2=1}^\infty Q_\rho^{n_1k_1+n_2 k_2}\left[n_2(2n_1+n_2)+(n_1^2-n_2^2)Q_\rho^{n_1k_2}\right]\,Q_{x_1}^{n_1}Q_{x_2}^{n_2}\,,\nonumber
\end{align}
which can be recast into the form
{\allowdisplaybreaks
\begin{align}
&J_{1,0}(x_1,x_2)=\sum_{m=1}^\infty Q_\rho^m\sum_{k|m}\frac{kQ_{x_2}(2+k(1-Q_{x_2}))}{(1-Q_{x_2})^2}\,Q_{x_1}^k\nonumber\\
&\hspace{0.5cm}+\sum_{m=1}^\infty Q_\rho^m\sum_{k_1,k_2=1}^{m}\left[\sum_{n_1,n_2=1\atop{n_1k_1+n_2k_2=m}}^m n_2(2n_1+n_2)\,Q_{x_1}^{n_1}Q_{x_2}^{n_2}+\sum_{n_1,n_2=1\atop{n_1(k_1+k_2)+n_2k_2=m}}^m(n_1^2-n_2^2)\,Q_{x_1}^{n_1}Q_{x_2}^{n_2}\right]
\end{align}}
\item $\ell_1=0$ and $\ell_2=1$: 
{\allowdisplaybreaks
\begin{align}
J_{0,1}(x_1,x_2)&=\sum_{k_2=0}^\infty\sum_{n_1,n_2=1}^\infty Q_\rho^{n_2(k_2+1)}\left[n_2(2n_1+n_2)+(n_1^2-n_2^2)Q_\rho^{n_1k_2}\right]\,Q_{x_1}^{n_1}Q_{x_2}^{n_2}\nonumber\\
&\hspace{0.5cm}+\sum_{k_1,k_2=1}^\infty\sum_{n_1,n_2=1}^\infty Q_\rho^{n_1k_1+n_2 k_2}\left[n_2(2n_1+n_2)+(n_1^2-n_2^2)Q_\rho^{n_1(k_2-1)}\right]\,Q_{x_1}^{n_1}Q_{x_2}^{n_2}\,,\nonumber
\end{align}}
which can be recast into the form
\begin{align}
&J_{0,1}(x_1,x_2)=\sum_{m=1}^\infty Q_\rho^m\sum_{k|m}\frac{kQ_{x_1}(2+k(1-Q_{x_1}))}{(1-Q_{x_1})^2}\,Q_{x_2}^k+\sum_{m=1}^\infty Q_\rho^m\sum^m_{n_1,n_2,k_2=1\atop {n_1(k_2-1)+n_2k_2=m}}(n_1^2-n_2^2)\, Q_{x_1}^{k_1}\,Q_{x_2}^{k_2}\nonumber\\
&\hspace{0.5cm}+\sum_{m=1}^\infty Q_\rho^m\sum_{k_1,k_2=1}^{m}\left[\sum_{n_1,n_2=1\atop{n_1k_1+n_2k_2=m}}^m n_2(2n_1+n_2)\,Q_{x_1}^{n_1}Q_{x_2}^{n_2}+\sum_{n_1,n_2=1\atop{n_1(k_1+k_2-1)+n_2k_2=m}}^m(n_1^2-n_2^2)\,Q_{x_1}^{n_1}Q_{x_2}^{n_2}\right]\,.
\end{align}
\item $\ell_1=1$ and $\ell_2=1$: 
\begin{align}
J_{1,1}(x_1,x_2)&=\sum_{k_1,k_2=1}^\infty\sum_{n_1,n_2=1}^\infty Q_\rho^{n_1k_1+n_2 k_2}\left[n_2(2n_1+n_2)+(n_1^2-n_2^2)Q_\rho^{n_1(k_2-1)}\right]\,Q_{x_1}^{n_1}Q_{x_2}^{n_2}\,,\nonumber
\end{align}
which can be recast in the form
\begin{align}
&J_{1,1}(x_1,x_2)\nonumber\\
&=\sum_{m=1}^\infty Q_\rho^m\sum_{k_1,k_2=1}^{m}\left[\sum_{n_1,n_2=1\atop{n_1k_1+n_2k_2=m}}^m n_2(2n_1+n_2)\,Q_{x_1}^{n_1}Q_{x_2}^{n_2}+\sum_{n_1,n_2=1\atop{n_1(k_1+k_2-1)+n_2k_2=m}}^m(n_1^2-n_2^2)\,Q_{x_1}^{n_1}Q_{x_2}^{n_2}\right]\,.
\end{align}
\end{itemize}
From these expressions we can extract an expansion in $Q_\rho$ for any (fixed) power in the arguments $x_{1,2}$. Consequently, we can work out the leading terms in a Fourier series expansion of the $\mathfrak{p}^{(2k)}_i(\rho)$ in (\ref{SeriesExpandN3Coup2}) for low values of $k$. The latter precisely match the following combinations of Eisenstein series
\begin{align}
&\mathfrak{p}^{(-4)}_1=1\,,&&\mathfrak{p}^{(-2)}_1=E_2\,,&&\mathfrak{p}_1^{(0)}=E_2^2\,,&&\mathfrak{p}^{(0)}_2=E_4\,,&&\mathfrak{p}_1^{(2)}=E_2\,E_4\,,\nonumber\\
&\mathfrak{p}_2^{(2)}=E_6\,,&&\mathfrak{p}^{(4)}_1=E_4^2\,,&&\mathfrak{p}^{(4)}_2=E_2\,E_6\,,&&\mathfrak{p}^{(6)}_1=E_2\,E_4^2\,,&&\mathfrak{p}^{(6)}_2=E_4\,E_6\,,
\end{align}
with the following matching rational functions in $\widehat{a}_{1,2}$
{\allowdisplaybreaks
\begin{align}
&\mathfrak{r}_1^{(-4)}=\frac{\widehat{a}_1+\widehat{a}_1\widehat{a}_2+\widehat{a}_2^2}{8\pi^4\widehat{a}_1^2\widehat{a}_2(\widehat{a}_1+\widehat{a}_2)^2}\,,\hspace{2.3cm}
\mathfrak{r}_1^{(-2)}=\frac{(\widehat{a}_1+\widehat{a}_1\widehat{a}_2+\widehat{a}_2^2)^2}{24\pi^2\widehat{a}_1^2\widehat{a}_2(\widehat{a}_1+\widehat{a}_2)^2}\,,\hspace{2.3cm}\mathfrak{r}_1^{(0)}=\frac{1}{48}\,, \nonumber\\
&\mathfrak{r}_2^{(0)}=\frac{2 \widehat{a}_1^6+6 \widehat{a}_1^5 \widehat{a}_2+9 \widehat{a}_1^4 \widehat{a}_2^2+8 \widehat{a}_1^3 \widehat{a}_2^3+9 \widehat{a}_1^2 \widehat{a}_2^4+6
   \widehat{a}_1 \widehat{a}_2^5+2 \widehat{a}_2^6}{240 \widehat{a}_1^2 \widehat{a}_2^2 (\widehat{a}_1+\widehat{a}_2)^2}\,,\hspace{1.1cm}\mathfrak{r}_1^{(2)}=\frac{\pi^2(\widehat{a}_1+\widehat{a}_1\widehat{a}_2+\widehat{a}_2^2)}{180}\,,\nonumber\\
&\mathfrak{r}_2^{(2)}=\frac{\pi ^2 \left(\widehat{a}_1^2+\widehat{a}_1 \widehat{a}_2+\widehat{a}_2^2\right) \left(\widehat{a}_1^6+3 \widehat{a}_1^5 \widehat{a}_2+5 \widehat{a}_1^4
   \widehat{a}_2^2+5 \widehat{a}_1^3 \widehat{a}_2^3+5 \widehat{a}_1^2 \widehat{a}_2^4+3 \widehat{a}_1 \widehat{a}_2^5+\widehat{a}_2^6\right)}{756\, \widehat{a}_1^2
   \widehat{a}_2^2 (\widehat{a}_1+\widehat{a}_2)^2}\,,  \nonumber\\
&\mathfrak{r}_1^{(4)}=\frac{\pi ^4 \left(\widehat{a}_1^2+\widehat{a}_1 \widehat{a}_2+\widehat{a}_2^2\right)^2 \left(\widehat{a}_1^6+3 \widehat{a}_1^5 \widehat{a}_2+8 \widehat{a}_1^4
   \widehat{a}_2^2+11 \widehat{a}_1^3 \widehat{a}_2^3+8 \widehat{a}_1^2 \widehat{a}_2^4+3 \widehat{a}_1 \widehat{a}_2^5+\widehat{a}_2^6\right)}{5400\, \widehat{a}_1^2
   \widehat{a}_2^2 (\widehat{a}_1+\widehat{a}_2)^2}\,,\nonumber\\
&\mathfrak{r}_2^{(4)}=\frac{(\widehat{a}_1+\widehat{a}_1\widehat{a}_2+\widehat{a}_2)^2\pi^4}{1134}\,,\hspace{1cm}\mathfrak{r}_1^{(6)}=\frac{\pi ^6 \left(2 \widehat{a}_1^6+6 \widehat{a}_1^5 \widehat{a}_2+15 \widehat{a}_1^4 \widehat{a}_2^2+20 \widehat{a}_1^3 \widehat{a}_2^3+15 \widehat{a}_1^2
   \widehat{a}_2^4+6 \widehat{a}_1 \widehat{a}_2^5+2 \widehat{a}_2^6\right)}{16200}\,,\nonumber\\
&\mathfrak{r}_2^{(6)}=\frac{\pi ^6}{124740\, \widehat{a}_1^2 \widehat{a}_2^2 (\widehat{a}_1+\widehat{a}_2)^2} \big[3 \widehat{a}_1^{12}+18 \widehat{a}_1^{11} \widehat{a}_2+83 \widehat{a}_1^{10} \widehat{a}_2^2+250 \widehat{a}_1^9 \widehat{a}_2^3+509
   \widehat{a}_1^8 \widehat{a}_2^4+734 \widehat{a}_1^7 \widehat{a}_2^5\nonumber\\
   &\hspace{2cm}+817 \widehat{a}_1^6 \widehat{a}_2^6+734 \widehat{a}_1^5 \widehat{a}_2^7+509 \widehat{a}_1^4
   \widehat{a}_2^8+250 \widehat{a}_1^3 \widehat{a}_2^9+83 \widehat{a}_1^2 \widehat{a}_2^{10}+18 \widehat{a}_1 \widehat{a}_2^{11}+3
   \widehat{a}_2^{12}\big]\,.
\end{align}}
We have checked up to $2k=10$ that this expansion matches the following closed form expression
\begin{align}
\coup{3}{2}(\widehat{a}_1,\widehat{a}_2,\rho)&=\frac{1}{(2\pi )^4}\bigg[\left(\mathbb{G}''(\widehat{a}_1;\rho)+\frac{2\pi i}{\rho-\bar{\rho}}\right)\left(\mathbb{G}''(\widehat{a}_2;\rho)+\frac{2\pi i}{\rho-\bar{\rho}}\right)\nonumber\\
&\hspace{1.35cm}+\left(\mathbb{G}''(\widehat{a}_1;\rho)+\frac{2\pi i}{\rho-\bar{\rho}}\right)\left(\mathbb{G}''(\widehat{a}_1+\widehat{a}_2;\rho)+\frac{2\pi i}{\rho-\bar{\rho}}\right)\nonumber\\
&\hspace{1.35cm}+\left(\mathbb{G}''(\widehat{a}_2;\rho)+\frac{2\pi i}{\rho-\bar{\rho}}\right)\left(\mathbb{G}''(\widehat{a}_1+\widehat{a}_2;\rho)+\frac{2\pi i}{\rho-\bar{\rho}}\right)\bigg]\,.\label{CoupN32Calc}
\end{align}

\end{document}